\documentclass[aps,prc,twocolumn,superscriptaddress,showpacs,amssymb,amsmath,amsfonts,floatfix,nofootinbib]{revtex4-1}

\usepackage{graphicx}
\usepackage{natbib}
\usepackage[usenames]{color}
\usepackage{epsfig}
\usepackage{amssymb,amsmath}
\usepackage{subfig}
\usepackage{lineno,hyperref}
\usepackage{enumitem}

\definecolor{grey}{rgb}{0.9,0.9,0.9}
\definecolor{black}{rgb}{0,0,0}

\hypersetup{pdfpagemode=FullScreene}

\newcommand{\fpf}[2]{{F}_{#1}^{*}{F}_{#2}}

\def\sl#1{\slash{\hspace{-0.2 truecm}#1}}

\newcommand{\Tuzla}[1]
{\affiliation{University of Tuzla, Faculty of Natural Sciences and Mathematics, Univerzitetska 4, 75000 Tuzla, Bosnia and Herzegovina}}
\newcommand{\Mainz}[1]
{\affiliation{Institut f\"ur Kernphysik, Johannes Gutenberg-Universit\"at Mainz, D-55099
Mainz,Germany}}
\newcommand{\Zagreb}[1]
{\affiliation{Rudjer Bo\v{s}kovi\'{c} Institute, Bijeni\v{c}ka cesta 54, P.O. Box 180, 10002 Zagreb, Croatia}}

\begin{document}

\title{Eta and Etaprime Photoproduction on the Nucleon with the Isobar Model EtaMAID2018}

\author{L.~Tiator}\thanks{tiator@uni-mainz.de}\Mainz \\
\author{M.~Gorchtein}\Mainz\\
\author{V.~L.~Kashevarov}\Mainz\\
\author{K.~Nikonov}\Mainz\\
\author{M.~Ostrick}\Mainz\\
\author{M.~Had\v{z}imehmedovi\'{c}}\Tuzla \\
\author{R.~Omerovi\'{c}}\Tuzla \\
\author{H.~Osmanovi\'{c}}\Tuzla \\
\author{J.~Stahov}\Tuzla\\
\author{A.~\v{S}varc}\Zagreb\\

\vspace{5cm}
\date{\today}

\begin{abstract}
The isobar model EtaMAID has been updated with new and high
precision data for $\eta$ and $\eta^\prime$ photoproduction on
protons and neutrons from MAMI, ELSA, GRAAL and CLAS. The background
is described in a recently developed Regge-cut model, and for the
resonance part the whole list of nucleon resonances has been
investigated with 21 $N^*$ states contributing to $\eta$
photoproduction and 12 $N^*$ states contributing to $\eta^\prime$
photoproduction. A new approach is discussed to avoid double
counting in the overlap region of Regge and resonances. A comparison
is done among four newly updated partial waves analyses for
observables and partial waves. Finally, the possibility of a narrow
resonance near $W=1900$~MeV is discussed, that would be able to
explain unexpected energy and angular dependence of observables in
$p(\gamma,\eta^\prime)p$ near $\eta^\prime$ threshold.
\end{abstract}

\maketitle

\section{Introduction}\label{sec:intro}

The photoinduced production of $\eta$ and $\eta'$ mesons is a
selective probe to study excitations of the nucleon. These mesons
represent the isoscalar members of the fundamental
pseudoscalar-meson nonet and, in contrast to the isovector $\pi$,
excitations with isospin $I = 3/2$ ($\Delta$ resonances) do not
decay into $\eta N$ and $\eta' N$ final states. An overview of the
current status of nucleon resonances can be found in
Ref.~\cite{Tanabashi:2018} and of the experimental and
phenomenological progress in $\eta$ photoproduction can be found in
Ref.~\cite{Krusche2014}.

The isobar model EtaMAID is part of the Mainz MAID
project~\cite{Tiator:2018pjq,MAID,MAID07} with online programs
performing real-time calculations of observables, amplitudes and
partial waves (multipoles). EtaMAID was introduced in
2001~\cite{Chiang:2002vq} as a model with eight prominent nucleon
resonances: $N(1535)\frac{1}{2}^-(S_{11})$,
$N(1650)\frac{1}{2}^-(S_{11})$, $N(1710)\frac{1}{2}^+(P_{11})$,
$N(1720)\frac{3}{2}^+(P_{13})$, $N(1520)\frac{3}{2}^-(D_{13})$,
$N(1700)\frac{3}{2}^-(D_{13})$, $N(1675)\frac{5}{2}^-(D_{15})$, and
$N(1680)\frac{5}{2}^+(F_{15})$.\footnote{Throughout this paper we
will use two notations for nucleon resonances, the general notation
as $N(1535)\frac{1}{2}^-$, introduced by PDG in 2012 and the older
$\pi N$ notation as $S_{11}(1535)$.} The background was modeled with
Born terms and $t$-channel vector meson exchanges of $\omega$ and
$\rho$ mesons. The model was developed for photo- and
electroproduction on protons and neutrons and was well fitted to the
few available data in 2001. Since that time a lot of developments
occurred, first of all for the experimental data base.

There was a huge effort at several accelerator facilities to combine
high intensity polarized photon beams with modern 4$\pi$ detectors
and spin-polarized targets. In particular, the Crystal Ball/TAPS
setup at MAMI in Mainz (Germany) \cite{McNicoll:2010qk}, the Crystal
Barrel/TAPS at ELSA in Bonn (Germany) \cite{Crede:2009zzb}, and the
CLAS detector at JLab in Newport News (USA) \cite{Mecking:2003zu}
have reached this goal and provided new, valuable information about
photo-induced $\eta$ and $\eta^\prime$ production. At the GRAAL
facility in Grenoble (France) \cite{Ajaka:1998zi} and the LEPS
facility at SPring-8 in Osaka (Japan) \cite{Sumihama:2009gf}, photon
beams with high linear polarization are available via
laser-backscattering and also data from ELPH at Tohoku University in
Sendai (Japan) \cite{Miyahara:2007zz} became available. The CLAS
detector was using a magnetic field in order to reconstruct the
recoiling proton with high resolution. The final state neutral
mesons were identified via a missing mass analysis. The other
detectors used electromagnetic calorimeters with almost 4$\pi$
coverage to detect photons, pions, protons and neutrons. The $\gamma
N \to \eta N$ and $\gamma N \to \eta' N$ reactions were identified
via a combination of missing mass and invariant mass techniques.

Photoproduction of $\eta$ or $\eta^\prime$ on the nucleon has been
studied in various theoretical approaches, in quark
models~\cite{Li:1997gd,Saghai:2001yd,Golli:2016dlj}, Lagrangian
models~\cite{Feuster:1998cj,Davidson:1999in}, effective field
theory~\cite{Borasoy:2001pj,Ruic:2011wf}, dispersion theoretical
calculations~\cite{Aznauryan:2003zg,Nikonov:2018}, Regge
models~\cite{Sibirtsev:2010yj}, isobar
models~\cite{Knochlein:1995qz,Tiator:1999gr,Chiang:2001as,Chiang:2002vq,Aznauryan:2003zg,
Tryasuchev:2003st,Tryasuchev:2018pdq}, and combined analyses by
using the additional information from $NN$
interaction~\cite{Huang:2012xj,Nakayama:2008tg}. Most flexible and
successful have been isobar models, where nucleon resonances are
treated in $s$-channel Breit-Wigner parametrization with
energy-dependent widths due to the coupling with other decay
channels. The non-resonant background in those models is described
by $s$- and $u$-channel Born terms and $t$-channel vector meson
exchanges.

Besides single-channel investigations, a series of coupled-channel
partial wave analyses
(PWA)~\cite{Shklyar:2006xw,McNicoll:2010qk,Shrestha:2012ep,Kamano:2013iva,
Ronchen:2015vfa,Anisovich:2017} have been performed with multiple
channels as $\pi N$, $\sigma N$, $\pi \Delta$, $\eta N$, $K
\Lambda$, $K \Sigma$, $\rho N$, $\omega N$, and $\eta^\prime N$.
Within the last few months, new updates have been obtained by the
Bonn-Gatchina group~\cite{Anisovich:2018}, the J\"ulich-Bonn
group~\cite{Ronchen:2018} and the Kent State University
group~\cite{KSU2018}.

All these PWA are energy-dependent (ED) analyses, where an
underlying model determines the functional dependence on the energy
and provides continuity and, in an optimal case, also analyticity of
the partial wave amplitudes. An energy-independent or single-energy
(SE) PWA is free of such a model dependence but depends very much on
the availability of a `complete experiment'~\cite{Barker:75} and on
analyticity and unitarity constraints. This has been done very
successfully for $\pi N$ scattering and pion photoproduction. For
$\eta N$ photoproduction such constraints are mostly unavailable,
making a single-energy PWA much more difficult and can lead to
ambiguous solutions. In a very recent work, we have accomplished
such a SE PWA for $\eta$ photoproduction using constraints from
fixed-$t$ analyticity~\cite{Osmanovic:2017fwe}.

The last update of EtaMAID was done in 2003~\cite{Chiang:2002vq}
with a reggeized isobar model for $p(\gamma,\eta)p$ and an extension
to $p(\gamma,\eta^\prime)p$ was established for the threshold
region, when new data on differential cross sections became
available from SAPHIR at ELSA in Bonn~\cite{Plotzke:1998ua}.

Combining reggeized $t$-channel exchanges with resonances in the
direct channel is by no means a new idea, see e.g. the model of Ref.
\cite{Barger:1966zzd} for charge-exchange $\pi N$ scattering or
models for meson photoproduction, e.g. EtaMAID2003
\cite{Chiang:2002vq} or Regge-plus-resonance approach for $K\Lambda$
photoproduction \cite{Corthals:2005ce}. In these models, the Regge
amplitude is obtained from a fit to high energy data and continued
into the resonance region. For $\eta$ and especially $\eta^\prime$
production the Regge regime that sets in at $W\geq 2.5$~GeV is quite
close to the accessible part of the resonance region. Matching the
invariant amplitudes that are obtained from the low-energy fit onto
Regge amplitude thus represents a valuable physics constraint. The
advantage from the technical point of view is that it is not
necessary to introduce many free parameters which would have been
necessary to fix the non-resonant background amplitude, so only
resonance parameters are used as fit parameters.

However, it has been realized early on that when projected on the
$s$-channel partial waves, Regge amplitudes generate resonance-like
Schmid loops on the Argand diagram for each partial wave
\cite{Schmid:1968zz}, which leads to a general problem of double
counting in the extraction of resonance parameters. Collins et
al.~\cite{Collins:1969ab} pointed out that to state a correspondence
between Regge asymptotic and $s$-channel resonances, one would have
to invoke unitarity, as per finite energy sum rules (FESR), see e.g.
Ref.~\cite{Dolen:1967jr} for an early application to $\pi N$
scattering.

With these reservations in mind, we pursue here another method which
uses as background the Regge amplitude with kinematical suppression
factor applied in the resonance region. This damping factor is
needed to at least partially remove the double counting. To address
this double counting in detail the FESR is the most natural tool,
and we postpone this study to the upcoming work. Moreover, in view
of the ambiguity Regge-resonances we opt not to discuss the
Breit-Wigner resonance parameters returned by the fit in detail.

Independent whatever procedure is applied, the resonance
parametrization using Breit-Wigner amplitudes remains model
dependent. Generalized Breit-Wigner amplitudes have enough freedom
with the energy dependence of the widths and of the vertex
functions, that changes in the background can usually be absorbed by
the resonance contributions, therefore leading to sizeable model
uncertainties to masses, widths, branching ratios and photo
couplings. In careful treatments, and for resonances with widths
$\Gamma \lesssim 120$~MeV, the model dependence is rather mild.
Therefore, PDG~\cite{Tanabashi:2018} decided to keep such
traditional resonance parameters, even if the spread of values is
often quite large. First priority in newer PWA are the fundamental
$t$-matrix pole positions and residues of various elastic and
inelastic reactions involving nucleon resonance excitations. In an
upcoming work we will use our obtained partial waves and analyze
nucleon resonances by its pole position and residues with the
Laurent-plus-Pietarinen (L+P)
method~\cite{Svarc:2014sqa,Svarc:2013laa}.

The paper is organized as follows. In section~\ref{sect:Formalism}
we will first give the basic formalism for kinematics, amplitudes
and observables. In section~\ref{sect:IsobarModel} we present the
details of our isobar model. We shortly describe the Regge-cut model
which has already been published and give our formulation for
nucleon resonance excitations. In section~\ref{sect:Results} we
present our results on $\eta$ and $\eta^\prime$ photoproduction from
protons and neutrons with comparisons to the data and PWA from other
analysis groups. In section~\ref{sect:NarrowResonances} we discuss a
recent attempt to search for a narrow $N^*$ resonance near the
$\eta^\prime$ threshold. Partial waves are compared with recent
solutions by the Bonn-Gatchina, J{\"u}lich-Bonn and
Kent-State-University groups in section~\ref{sec:pwa}, before we
summarize our method and results in section~\ref{sec:conclusions}.
In an appendix we give the formulas for polarization observables
used in our analysis and tables of our background and Breit-Wigner
resonance parameters.

\section{Formalism}\label{sect:Formalism}

\subsection{\boldmath Kinematics in $\eta$ photoproduction}

For $\eta$ photoproduction on the nucleon, we consider the reaction
\begin{equation}
\gamma(k)+N(p_i)\rightarrow \eta(q)+N'(p_f)\,,
\end{equation}
where the variables in brackets denote the four-momenta of the
participating particles. These are $k^\mu=(k,\bold{k})$,
$q^\mu=(\omega,\bold{q})$ for photon and $\eta$ meson, and
$p_i^\mu=(E_i,\bold{p}_i)$, $p_f^\mu=(E_f,\bold{p}_f)$ for incoming
and outgoing nucleon, respectively. The familiar Mandelstam
variables are given as
\begin{equation}
s=W^2=(p_i+k)^2,\qquad t=(q-k)^2,\qquad u=(p_i-q)^2,
\end{equation}
the sum of the Mandelstam variables is given by the sum of the
external masses
\begin{equation}
s+t+u=2m_N^2+m_{\eta}^2\,,
\end{equation}
where $m_N$ and $m_{\eta}$ are masses of proton and $\eta$ meson,
respectively. The crossing symmetrical variable is
\begin{equation}
\nu=\frac{s-u}{4m_N}\,.
\end{equation}

In the $\eta N$ center-of-mass (c.m.) system, we have
$\bold{p}_i=-\bold{k}$, $\bold{p}_f=-\bold{q}$, and the energies and
momenta can be related to the Mandelstam variable $s$ by
\begin{equation}
k=|\bold{k}|=\frac{s-m_N^2}{2\sqrt{s}},\quad
\omega=\frac{s+m_{\eta}^2-m_N^2}{2\sqrt{s}}\,,
\end{equation}
\begin{equation}
q=|\bold{q}|=\left[\left(\frac{s-m_{\eta}^2+m_N^2}{2\sqrt{s}}\right)^2-m_N^2\right]^{\frac{1}{2}}\,,
\end{equation}
\begin{equation}
E_i=\frac{s+m_N^2}{2\sqrt{s}},\quad
E_f=\frac{s+m_N^2-m_{\eta}^2}{2\sqrt{s}}\,,
\end{equation}
$W=\sqrt{s}$ is the c.m. energy. Furthermore, we will also refer to
the lab energy of the photon, $E=(s-m_N^2)/(2m_N)$.

\subsection{Cross section and polarization observables}

\begin{figure*}   
\begin{center}
\includegraphics[width=12.0cm]{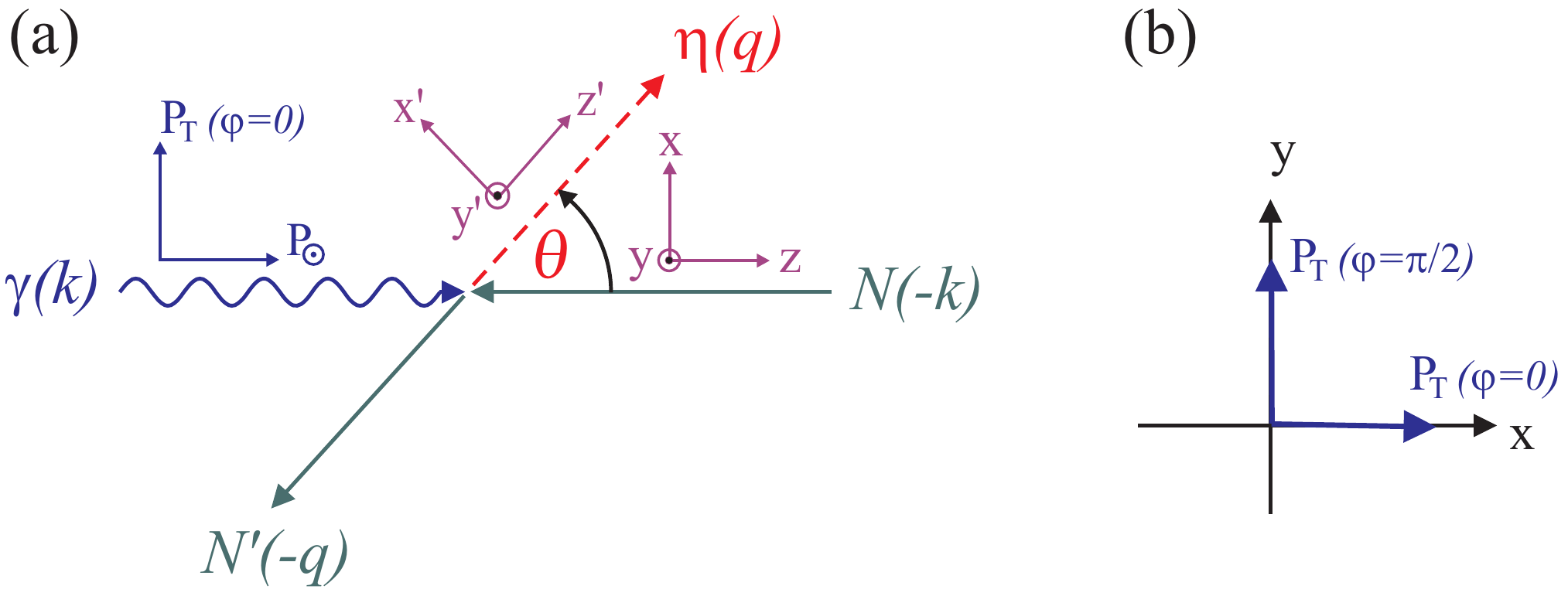} \vspace{3mm}
 \caption{ Kinematics of photoproduction and frames for polarization.
The frame $\{x,y,z\}$ is used for target polarization
$\{P_x,P_y,P_z\}$, whereas the recoil polarization
$\{P_{x'},P_{y'},P_{z'}\}$ is defined in the frame $\{x',y',z'\}$,
which is rotated around $y'=y$ by the polar angle $\theta$. The
azimuthal angle $\varphi$ is defined in the $\{x,y\}$ plane (b) and
is zero in the projection shown in the figure (a).}\label{fig:kin}
\end{center}
\end{figure*}

As depicted in Fig.~\ref{fig:kin}, the photon polarization can be
linear or circular. For a linear photon polarization $(P_T=1)$ along
the direction $\hat{\bold{x}}$ we define the azimuthal angle
$\varphi=0$, and perpendicular, in direction ${\hat{\bold{y}}}$, the
polarization angle is $\varphi=\pi/2$. For right-handed circular
polarization $P_{\odot}=+1$.

We may classify the differential cross sections by the three classes
of double polarization experiments and one class of triple
polarization experiments, which, however, do not give additional
information:
\begin{itemize}
\item polarized photons and polarized target
\end{itemize}
\begin{eqnarray}
\frac{d \sigma}{d \Omega} & = & \sigma_0
\left\{ 1 - P_T \Sigma \cos 2 \varphi \right. \nonumber \\
& & + P_x \left( - P_T H \sin 2 \varphi + P_{\odot} F \right)
\nonumber \\
& & + P_y \left( T - P_T P \cos 2 \varphi \right) \nonumber \\
& & \left. + P_z \left( P_T G \sin 2 \varphi - P_{\odot} E \right)
\right\} \, ,
\end{eqnarray}
\begin{itemize}
\item polarized photons and recoil polarization
\end{itemize}
\begin{eqnarray}
\frac{d \sigma}{d \Omega} & = & \sigma_0
\left\{ 1 - P_T \Sigma \cos 2 \varphi \right. \nonumber \\
& & + P_{x'} \left( -P_T O_{x'} \sin 2 \varphi - P_{\odot} C_{x'}
\right)
\nonumber \\
& & + P_{y'} \left( P - P_T T \cos 2 \varphi \right) \nonumber \\
& & \left. + P_{z'} \left( -P_T O_{z'} \sin 2 \varphi  - P_{\odot}
C_{z'} \right) \right\} \, ,
\end{eqnarray}
\begin{itemize}
\item polarized target and recoil polarization
\end{itemize}
\begin{eqnarray}
\frac{d \sigma}{d \Omega} & = & \sigma_0 \left\{ 1 + P_{y} T +
P_{y'} P
+ P_{x'} \left( P_x T_{x'} - P_{z} L_{x'} \right) \right. \nonumber \\
& & \left. + P_{y'} P_y \Sigma  + P_{z'}\left( P_x T_{z'} + P_{z}
L_{z'}\right) \right\}\,.
\end{eqnarray}

In these equations $\sigma_0$ denotes the unpolarized differential
cross section. Instead of asymmetries, in the following we will also
discuss the product of the unpolarized cross section with the
asymmetries and will use the notation
$\check{\Sigma}=\sigma_0\Sigma\,, \check{T}=\sigma_0T\,,\cdots\,$.
In appendix \ref{app:obs} we give expressions of the observables in terms of
CGLN amplitudes.

\subsection{\boldmath Invariant amplitudes}

The nucleon electromagnetic current for pseudoscalar meson
photoproduction can be expressed in terms of four invariant
amplitudes $A_i$~\cite{Chew:1957tf},
\begin{eqnarray}\label{eq:19}
J^\mu = \sum_{i=1}^4 A_i(\nu,t)\, M^\mu_i,
\end{eqnarray}
with the gauge-invariant four-vectors $M^\mu_i$ given by
\begin{eqnarray}
M^\mu_1&=&
-\frac{1}{2}i\gamma_5\left(\gamma^\mu\sl{k}-\sl{k}\gamma^\mu\right)\,
,
\nonumber\\
M^\mu_2&=&2i\gamma_5\left(P^\mu\, k\cdot(q-\frac{1}{2}k)-
(q-\frac{1}{2}k)^\mu\,k\cdot P\right)\, ,\nonumber\\
M^\mu_3&=&-i\gamma_5\left(\gamma^\mu\, k\cdot q
-\sl{k}q^\mu\right)\, ,\nonumber\\\
M^\mu_4&=&-2i\gamma_5\left(\gamma^\mu\, k\cdot P
-\sl{k}P^\mu\right)-2m_N \, M^\mu_1\, ,
 \label{eq:tensor}
\end{eqnarray}
where $P^\mu=(p_i^\mu+p_f^\mu)/2$, and the gamma matrices are
defined as in Ref.~\cite{Bjo65}.

The nucleon pole terms for $N(\gamma,\eta)N$, $A_i^{I,pole}$
($I=+,0$) are given by
\begin{eqnarray}\label{eq:Born}
A_1^{I,pole} & = & \ \ \ \frac{e\,g_{\eta N}}{2}
\left(\frac{1}{s-m_N^2}+\frac{1}{u-m_N^2}\right)\,,\nonumber \\
A_2^{I,pole} & = & -\frac{e\,g_{\eta N}}{t-m^2_\eta}
\left(\frac{1}{s-m_N^2}+\frac{1}{u-m_N^2}\right)\,,\nonumber \\
A_3^{I,pole} & = & -\frac{e\,g_{\eta N}}{2m_N}\frac{\kappa^{(I)}}{2}
\left(\frac{1}{s-m_N^2}-\frac{1}{u-m_N^2}\right)\,,\nonumber \\
A_4^{I,pole} & = & -\frac{e\,g_{\eta N}}{2m_N}\frac{\kappa^{(I)}}{2}
\left(\frac{1}{s-m_N^2}+\frac{1}{u-m_N^2}\right)\,,
\label{eq:a1-4pole}
\end{eqnarray}
with $\kappa^{(+)}= \kappa_p-\kappa_n$, and
$\kappa^{(0)}=\kappa_p+\kappa_n$, where $\kappa_p$ and $\kappa_n$
are the anomalous magnetic moments of the proton and the neutron,
respectively.

\subsection{CGLN amplitudes and multipoles}

For PWA the CGLN amplitudes $F_i(W,x)$~\cite{Chew:1957tf} are
conveniently used. They are defined in the c.m. frame using Coulomb
gauge. The matrix element ${\cal F}$ with the $e.m.$ current of
Eq.~(\ref{eq:19}) then reads
\begin{eqnarray}\label{cgln2}
\begin{split}
{\cal F} &= -\epsilon_\mu J_{}^\mu \\
         &= i\,({\vec {\sigma}}\cdot{ \hat{\epsilon}}) \,
{ F}_1 + ({\vec {\sigma}} \cdot\hat { {q}})\, ({\vec {\sigma}}
\times \hat{ {k}})\cdot{ \hat{\epsilon}}\,{ F}_2 \\ 
         &+ i\,({\hat{\epsilon}}\cdot\hat { {q}})\, ({\vec {\sigma}} 
\cdot\hat {{k}}) { F}_3 + i ({ \hat{\epsilon}} \cdot \hat{{q}})({\vec {\sigma}}
\cdot \hat { {q}}) \, { F}_4\,,
\end{split}
\end{eqnarray}
where $\epsilon^\mu=(0,\vec{\epsilon})$ and $\vec{\epsilon}\cdot
\vec{k} = 0$. In partial wave analysis of pseudoscalar meson
photoproduction it is convenient to work with CGLN amplitudes giving
simple representations in terms of electric and magnetic multipoles
and derivatives of Legendre polynomials
\begin{eqnarray}
\label{eq:multipoles}
\begin{split}
F_{1}(W,x) &=\sum_{l=0}^{\infty}[(lM_{l+}(W)+E_{l+}(W))P'_{l+1}(x) \\
           &+((l+1)M_{l-}(W)+E_{l-}(W))P'_{l-1}(x)] \,,  \\
F_{2}(W,x) &=\sum_{l=1}^{\infty}[(l+1)M_{l+}(W)+lM_{l-}(W)]P_{l}'(x)\,,\\
F_{3}(W,x) &=\sum_{l=1}^{\infty}[(E_{l+}(W)-M_{l+}(W))P''_{l+1} \\
           &+(E_{l-}(W)+M_{l-}(W))P''_{l-1}(x)]\,,\\
F_{4}(W,x) &=\sum_{l=2}^{\infty}[M_{l+}(W)-E_{l+}(W)-M_{l-}(W) \\
           &-E_{l-}(W)]P_{l}''(x)\,,
\end{split}
\end{eqnarray}
where $x=\cos\theta$ is the cosine of the scattering angle.
In appendix \ref{app:FtoA} we give relations between the CGLN and
the invariant amplitudes.

\section{The isobar model}\label{sect:IsobarModel}

In the isobar model the photoproduction amplitudes of $\eta$ and
$\eta^\prime$ mesons are written in terms of nucleon resonance
excitations in generalized Breit-Wigner forms and in non-resonant
background amplitudes. For simplicity we write all formulas in terms
of $(\gamma,\eta)$. For $(\gamma,\eta^\prime)$ all those formulas
and kinematical relations can easily be extended.

For a specific partial wave $\alpha=\alpha(\ell,j=\ell\pm 1/2,{\cal
M})$, where $\ell$ is the angular momentum of the $\eta N$ system in
the final state, $j$ is the total spin and ${\cal M}$ stands either
for an electric (E) or magnetic (M) transition. The total partial
wave amplitude can be written as a sum of a background amplitude
$t^{\alpha,b}$ and a resonance amplitude $t^{\alpha,r}$
\begin{equation}
t_{\gamma,\eta}^\alpha(W) = t_{\gamma,\eta}^{\alpha,b}(W) +
t_{\gamma,\eta}^{\alpha,r}(W)\,.
\end{equation}
In photoproduction we identify the partial wave amplitudes directly
with the electromagnetic multipoles $E_{\ell\pm}$ and $M_{\ell\pm}$.

\subsection{The non-resonant background}

Traditionally, the background amplitude is taken as a sum of Born
terms and $t$-channel meson exchange contributions
\begin{equation}
t_{\gamma,\eta}^{\alpha,b}(W) = t_{\gamma,\eta}^{\alpha,Born}(W)
 + t_{\gamma,\eta}^{\alpha,VM}(W)\,.
 \end{equation}
The Born terms for $\eta$ and $\eta^\prime$ photoproduction play a
minor role due to the small coupling constants. Whereas the $\pi NN$
coupling is very large, $g^2_{\pi NN}/4\pi\approx 14$, for $\eta$
and $\eta^\prime$ photoproduction $g^2_{\eta NN}/4\pi\sim
g^2_{\eta^\prime NN}/4\pi \lesssim 0.1$. This is a rather old
observation~\cite{Tiator:1994et} in contradiction to SU(3) symmetry,
where the coupling constants are predicted in the range of 1.

In all $\eta$ photoproduction analyses this suppression of the Born
terms has been confirmed and extensive studies have even found
$g^2_{\eta NN}/4\pi\le 10^{-3}$~\cite{Nakayama:2008tg}. For the
$\eta^\prime NN$ coupling our value is in agreement with a combined
analysis including also $NN\eta^\prime$~\cite{Huang:2012xj}.
Nevertheless, in interference terms and at high energies, the Born
terms can play some role, and similarly to our previous EtaMAID
models, the couplings are determined in the fits to the data. The
Born terms are most easily expressed in terms of invariant
amplitudes and in pseudoscalar coupling they are given by the
nucleon pole terms, Eq.~(\ref{eq:Born}).

As our goal in the 2018 update is a continuous description of
photoproduction from threshold up to the highest energies, where
experimental data exists ($W\sim 5$~GeV), we introduced an energy
dependence (damping) in order to suppress the strong rise of the
Born terms, and therefore a violation of unitarity at high energies
by
\begin{equation}\label{eq:Born_damp}
g_{\eta N} \rightarrow g_{\eta
N}\,\left(\frac{W_{thr}}{W}\right)^{\alpha_B}\,,
\end{equation}
where $\alpha_B$ will be found in the fit to the data. Ideally, a
correct high-energy behavior for the Born contribution should be
achieved by replacing the single nucleon exchange in the $u$-channel
by a Regge exchange of the nucleon trajectory. Such a modification
alone would, however, violate gauge invariance and a more elaborate
approach needs to be applied. We leave this to an upcoming work.

For $t$-channel exchanges the invariant amplitudes for vector and
axial-vector poles are given by
\begin{eqnarray}
A_1(t) & = & \frac{e\,\lambda_V\,g_V^{\mathfrak{t}}}{2m_\eta M_N}\;
\frac{t}{t-M_V^2}\, ,    \label{Eq:A1}\\
A_2'(t) & = & - \frac{e\,\lambda_A\,g_A^{\mathfrak{t}}}{2m_\eta M_N}
\;
\frac{t}{t-M_A^2}\, ,   \label{Eq:A2}\\
A_3(t)& = & \frac{e\,\lambda_A\,g_A^v}{m_\eta}\; \frac{1}{t-M_A^2}\, ,   \label{Eq:A3}\\
A_4(t) & = & \frac{-e\,\lambda_V\,g_V^v}{m_\eta}\;
\frac{1}{t-M_V^2}\,,\label{Eq:A4}
\end{eqnarray}
where $\lambda_{V(A)}$ denotes the electromagnetic coupling of the
vector ($V$) or axial ($A$) vector mesons with masses $M_{V(A)}$.
The constants $g_{V(A)}^{v(\mathfrak{t})}$ denote their vector $(v)$
or tensor $(\mathfrak{t})$ couplings to the nucleon. In order to
separate the vector and tensor contributions from individual mesons
we have used the amplitude
\begin{equation}\label{Eq:A2p}
A_2'(t)=A_1(t)+t\,A_2(t)\, ,
\end{equation}
which has only contributions from the tensor coupling of an axial
vector exchange.

Unlike in pion production, the physical region for $\eta$ and
especially $\eta^\prime$ production starts at considerably high
energy. It is generally expected that already at $\nu \sim 2$ GeV
the low-$t$ data are well-represented by Regge exchanges. At the
same time, a model with simple vector exchanges becomes inadequate
at high energy: a spin-1 exchange leads to a linearly increasing
amplitude which violates unitarity, so one is forced to introduce
phenomenological form factors to suppress this behavior.

To make use of all the data available for $\eta$ photoproduction we
propose here an alternative approach: A background function that is
smoothly joined onto a Regge amplitude at high energy, but is
modified in the resonance region to accommodate the nucleon
resonances by avoiding double counting.

For the Regge amplitudes we follow our recent work on Regge
phenomenology in $\pi^0$ and $\eta$ photoproduction,
Ref.~\cite{Kashevarov:2017vyl}. In that work we compared and
discussed four solutions, different in the Regge formulation and in
the data sets used in the fits. Here for EtaMAID we use our
preferred solution I, a Regge-cut model, where the full data set was
fitted.

\begin{figure}
\begin{center}
\resizebox{0.5\textwidth}{!}{\includegraphics{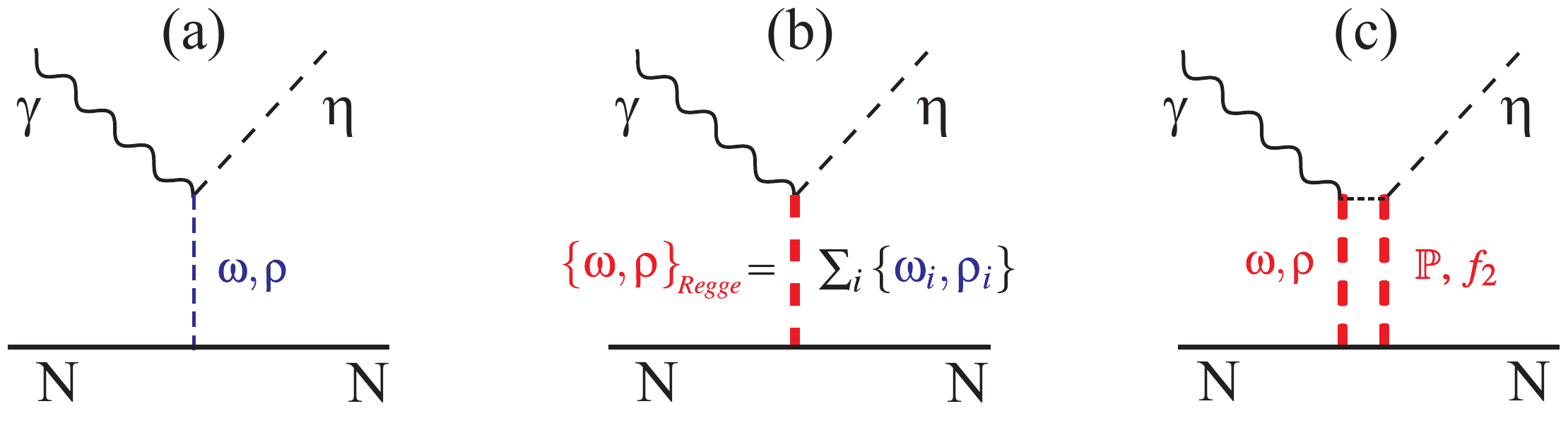}}
\caption{$t$-channel contributions to $\eta$ photoproduction from
single poles (a), Regge poles (b), and Regge cuts (c). An example
for $\rho$ and $\omega$ meson exchange and $\mathbb P$ (Pomeron) and
$f_2$ mesons for rescattering of two Reggeons.} \label{fig:regge}
\end{center}
\end{figure}

Technically, the $t$-channel exchange of Regge trajectories is done
by replacing the single meson propagator by the following expression
%
\begin{eqnarray}
\label{eq:Regge-1}
\begin{split}
&\frac{1}{t-M^2} \Rightarrow \\
&D(s,t)=(\frac{s}{s_0})^{\alpha(t)-1}\;
\frac{\pi\,\alpha^\prime}{\mbox{sin}[\pi\alpha(t)]}\; \frac{{\cal S}
+ e^{-i\pi\alpha(t)}}{2}\; \frac{1}{\Gamma(\alpha(t))}\,,
\end{split}
\end{eqnarray}
%
where $M$ is the mass of the Reggeon, $\cal S$ is the signature of
the Regge trajectory (${\cal S}=-1$ for vector and axial-vector
mesons), and $s_0$ is a mass scale factor, commonly set to
1~GeV$^2$. The Gamma function $\Gamma(\alpha(t))$ is introduced to
suppress additional poles of the propagator.

In addition Regge-cuts are added in our model. The Regge cuts can be
understood as a rescattering effect at high energies, e.g. an $\eta$
is produced via a vector or axial vector exchange at the first step,
and then rescattered via a Pomeron or tensor exchange. This effect
is shown in Fig.~\ref{fig:regge}~(c) as contracted box diagrams,
where two trajectories are exchanged consequently.

\begin{figure*}    
\begin{center}
\resizebox{0.6\textwidth}{!}{\includegraphics{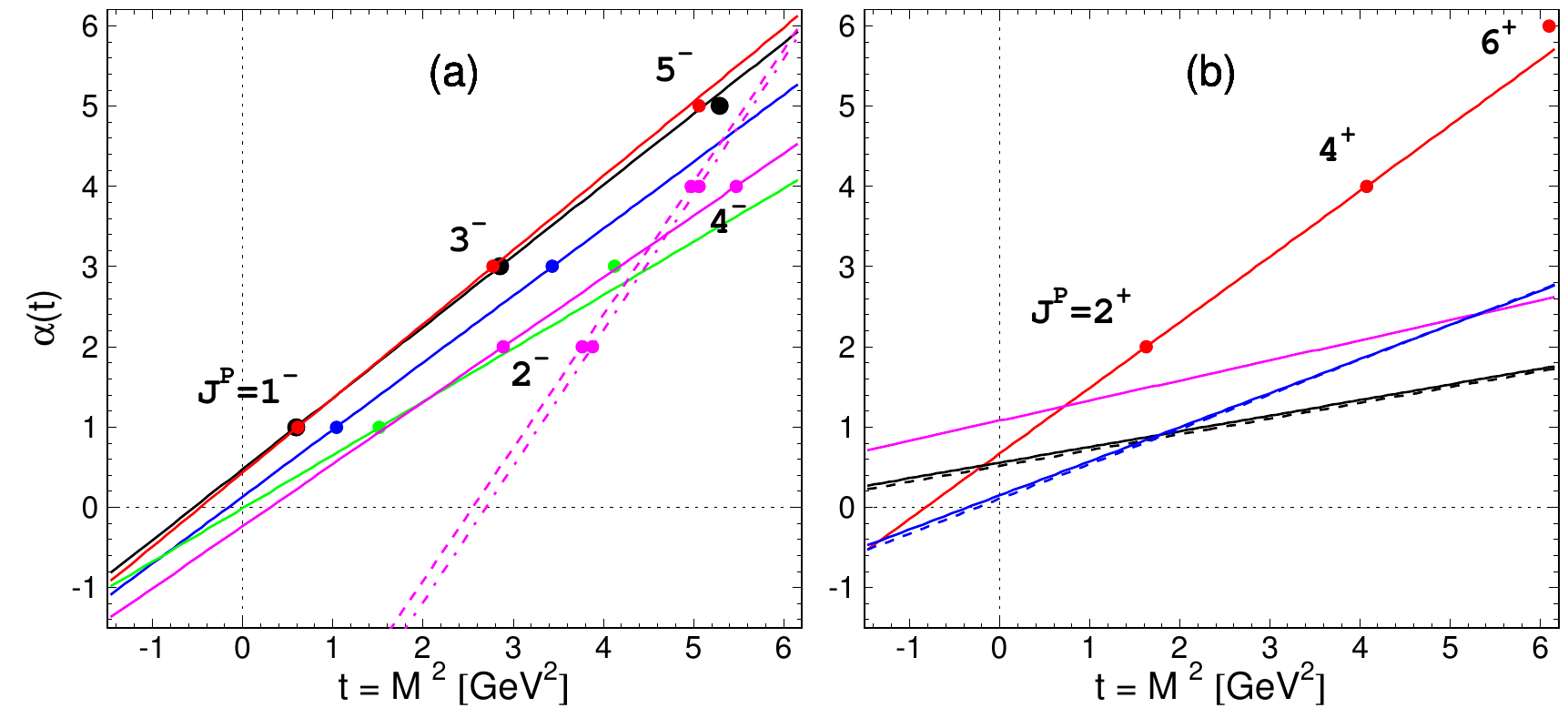}}
\caption{Regge trajectories: (a) $\rho$ black, $\omega$ red, $\phi$
blue, $b_1$ and $h_1$ green, $\rho_2$ and $\omega_2$ magenta; dashed
and dash-dotted magenta lines are $\rho_2$ and $\omega_2$ of
Ref.~\cite{Anisovich:2002-1,Anisovich:2002-2}; (b) $f_2$ red,
$\mathbb P$ magenta, $\rho f_2$ black solid, $\omega f_2$ blue
dashed, $\rho \mathbb P$ black solid, $\omega \mathbb P$ black
dashed. } \label{fig:traj}
\end{center}
\end{figure*}

The trajectories for $f_2$ and ${\mathbb P}$ are shown in
Fig.~\ref{fig:traj}~(b) together with four cut trajectories
$\rho{\mathbb P}$, $\omega{\mathbb P}$ (black solid and dashed
lines) and $\rho f_2$, $\omega f_2$ (blue solid and dashed lines).
Parameters of the Reggeon and cut trajectories used in the present
work are given in our previous paper~\cite{Kashevarov:2017vyl}.

All four Regge cuts can contribute to vector and axial vector
exchanges and can be written in the following
form~\cite{Donnachie:2015jaa}
\begin{equation}
D_{cut} =\left(\frac{s}{s_0}\right)^{\alpha_c(t)-1}\;
e^{-i\pi\alpha_c(t)/2}\; e^{d_c t} \,.
\end{equation}

In total, the vector meson propagators are replaced by
\begin{equation}
D_V \Rightarrow {D}_{V} + c_{V\mathbb P}\,{D}_{V\mathbb P} +
c_{Vf_2}\,D_{Vf_2},\; V=\rho,\omega
\end{equation}
and the axial vector meson propagators are replaced by
\begin{equation}
D_A \Rightarrow {D}_{A} + \sum_{V=\rho,\omega} ({\tilde c}_{V\mathbb
P}\,{D}_{V\mathbb P}
   + {\tilde c}_{Vf_2}\,D_{Vf_2}),\; A=b_1, h_1\,,
\end{equation}
where the coefficients $c_{V\mathbb P},c_{Vf_2}$ are for natural
parity cuts and ${\tilde c}_{V\mathbb P},{\tilde c}_{Vf_2}$ for
un-natural parity cuts and are obtained by a fit to the data.

In detail, the invariant amplitudes will be changed in the following
way
\begin{eqnarray}\label{Eq:cuts}
\begin{split}
\lambda_\rho\,g_\rho^{v,\mathfrak{t}}\;\frac{1}{t-M_\rho^2} &\rightarrow \lambda_\rho\,g_\rho^{v,\mathfrak{t}}\\
&\hspace{-1.8cm}[D_\rho(s,t) + c_{\rho \mathbb P}\,D_{\rho \mathbb P}(s,t) + c_{\rho f}\,D_{\rho f}(s,t)] \,,\\
\lambda_\omega\,g_\omega^{v,\mathfrak{t}}\;\frac{1}{t-M_\omega^2}
&\rightarrow
\lambda_\omega\,g_\omega^{v,\mathfrak{t}}\\
&\hspace{-1.8cm}[D_\omega(s,t) + c_{\omega \mathbb P}\,D_{\omega \mathbb P}(s,t) + c_{\omega f}\,D_{\omega f}(s,t)]\,,\\
\lambda_{b_1}\,g_{b_1}^{\mathfrak{t}}\;\frac{1}{t-M_{b_1}^2}
&\rightarrow \lambda_{b_1}\,g_{b_1}^{\mathfrak{t}}
D_{b_1}(s,t)\\
&\hspace{-1.8cm}+\lambda_\rho    \,g_\rho^{\mathfrak{t}}\,[{\tilde
c}_{\rho \mathbb P}  \,D_{\rho \mathbb P}(s,t)
 + {\tilde c}_{\rho f_2}  \,D_{\rho f_2}(s,t)] \\
&\hspace{-1.8cm}+\,\lambda_\omega\,g_\omega^{\mathfrak{t}}\,[{\tilde
c}_{\omega \mathbb P}\,D_{\omega \mathbb P}(s,t)
 + {\tilde c}_{\omega f_2}\,D_{\omega f_2}(s,t)]\,.
\end{split}
\end{eqnarray}

In practical calculations, it turns out that the axial vector Regge
pole contributions, proportional to $D_A$, can be neglected, but the
axial vector Regge cuts arising from $\rho$ and $\omega$ together
with $\mathbb P$ and $f_2$ are very important, in particular for
polarization observables, as the photon beam asymmetry $\Sigma$.

The Regge cuts also allow us to describe a long standing problem of
suitable candidates for an $A_3$ amplitude. While vector and
axial-vector single pole or Regge pole exchanges do not contribute,
Regge-cut exchanges $\rho{f_2}$ and $\omega{f_2}$ satisfy all
conservation law requirements. On the other hand, the $\rho{\mathbb
P}$ and $\omega{\mathbb P}$ cuts do not contribute to the $A_3$
amplitude.

The main aspect in EtaMAID is the exploration of nucleon resonance
excitation. Adding Regge amplitudes and resonances together, one
runs into the well-known double-counting problem. The duality
principle states that the full amplitude can be obtained  by summing
an infinite tower of either $s$- or $t$-channel resonances. In
isobar models only a finite number of nucleon resonances are
considered in the $s$-channel, still one cannot fully avoid this
problem. Various methods have been discussed in the literature to
treat with that problem. The so-called Regge-plus-Resonance models
simply ignore double counting. In another approach, applied e.g in
EtaMAID2003~\cite{Chiang:2002vq} and in the Bonn-Gatchina
model~\cite{Anisovich:2017}, the lowest partial waves, where
$s$-channel resonances are added, were projected out of the Regge
amplitudes. In models, where a lot of nucleon resonances are taken
into account, this would, however, lead to an almost completely
removed background amplitude in the resonance region. Recently, the
concept of finite-energy sum rules was discussed and applied to
$\pi^0$ and $\eta$
photoproduction~\cite{Nys:2016vjz,Mathieu:2018mjw}, where resonance
and Regge regions can be well separated and smoothly matched
together. Those applications for $\eta$ photoproduction are still in
progress.

Here we want to apply a further method, where the double counting is
removed by introducing a damping factor $F_d(W)$ to the Regge
amplitudes, which goes to zero at $\eta$ threshold and approaches
unity above some energy,
\begin{eqnarray}\label{eq:Regge_damp}
A_i^{Regge} &\rightarrow& A_i^{Regge}\cdot F_d(W)\\
\mbox{with}\quad F_d(W) &=& \left(
1-e^{-\frac{W-W_{thr}}{\Lambda_R}}\right) \theta(W-W_{thr})\,.
\end{eqnarray}
The scale $\Lambda_R$ describes at which energy Regge description
fully sets in and is obtained in the fit. For a very small
$\Lambda_R$ the damping factor introduced above is a step function,
whereas for large $\Lambda_R$ it only approaches unperturbed Regge
asymptotically. The way this damping factor cures the double
counting problem can be seen as follows. Assume that an exact dual
representation of the scattering amplitude $t$ is realized and
entails an infinite sum over the entire resonance spectrum in either
$s$- or $t$-channel,
\begin{eqnarray}
t=\sum_{i=1}^\infty t^{Res_i}_s=\sum_{i=1}^\infty t^{Res_i}_t.
\end{eqnarray}
At high $s$-channel energy, the $t$-channel sum can actually be
performed in terms of an exchange of a few leading Regge
trajectories $\alpha_i$, $t^{Regge}\sim \sum_i c_i\nu^{\alpha_i}$.
For the $s$-channel resonances, in turn, accounting for the full
spectrum is not possible, and we limit ourselves to explicitly
including only the lowest resonances up to $i=N$. We write,
\begin{eqnarray}
t&=&\sum_{i=1}^N t^{Res_i}_s+\left[ \sum_{i=1}^\infty t^{Res_i}_t-\sum_{i=1}^N t^{Res_i}_s\right]\nonumber\\
&\approx&\sum_{i=1}^N t^{Res_i}_s+ F_d(W)t^{Regge}.
\end{eqnarray}
The exact balance between the $s$-channel resonances and the part of
the Regge amplitude removed by the damping factor can be controlled
explicitly by the FESR. We will address these in an upcoming work.
Parameters for the background can be found in
table~\ref{tab:background} in appendix \ref{app:BG-BW}.

\subsection{Nucleon resonance excitations}

For a given partial wave $\alpha$, a set of $N_\alpha$ nucleon
resonances are added as generalized Breit-Wigner (BW) functions with
a unitary phase $\phi$ for each resonance,
\begin{equation}\label{eq:tres}
t_{\gamma,\eta}^{\alpha,r}(W)=\sum_{j=1}^{N_\alpha}\,
t_{\gamma,\eta}^{\alpha,BW,j}(W)\,e^{i\phi_j}\,.
\end{equation}
Due to the weakness of photoproduction, where the moduli of the
$t$-matrices are typically of the order $10^{-2}$ or smaller, a
simple addition of multiple resonances is sufficient and does not
violate unitarity. The phase $\phi_j$ introduced in
Eq.~(\ref{eq:tres}) is new for our EtaMAID models but was always
applied in pion photo- and electroproduction. Whereas in
$(\gamma,\pi)$ the Watson theorem determines the phase $\phi_j$ at
least below the $\pi\pi$ threshold, in $\eta$ and $\eta^\prime$
production we have no theoretical guideline and use $\phi_j$ as a
fit parameter. Furthermore, $\phi_j$ will be a constant in this
work, while in general it can be an energy-dependent function with
proper threshold behavior. The phase $\phi_j$ is often also called
the `background phase', because it is indirectly determined by the
background, which is different for the different channels $\eta p,
\eta n, \eta^\prime p, \eta^\prime n$ and also different for
electric and magnetic multipoles.

For a given partial wave $\alpha$, the relevant multipoles
$\mathcal{M}_{\ell\pm}$ ($E_{\ell\pm},\, M_{\ell\pm}$) are assumed
to have a Breit-Wigner energy dependence of the following form

\begin{eqnarray}\label{Eq:BWres}
\begin{split}
&t_{\gamma,\eta}^{\alpha,BW}(W)= \mathcal{M}_{\ell\pm}(W) \\ 
&= \bar{\mathcal{M}}_{\ell\pm}\,
 f_{\gamma N}(W)\,
 \frac{M_R \Gamma_\mathrm{tot}(W)}{M_R^2-W^2-i M_R \Gamma_\mathrm{tot}(W)}\,
 f_{\eta N}(W)\, C_{\eta N} \,,
\end{split}
\end{eqnarray}
where $f_{\eta N}(W)$ is the usual Breit-Wigner factor describing
the $\eta N$ decay of the $N^*$ resonance with total energy
dependent width $\Gamma_\mathrm{tot}(W)$, partial width
$\Gamma_{\eta N}(W)$ and spin $J$,
\begin{equation} \label{eq:fetaN_Maid}
 f_{\eta N}(W) = \zeta_{\eta N} \left[ \frac{1}{(2J+1)\pi}\,
 \frac{k(W)}{q_{\eta}(W)}\, \frac{M_N}{M_R}\,
 \frac{\Gamma_{\eta N}(W)}{\Gamma_\mathrm{tot}(W)^2} \right]^{1/2},
\end{equation}
with $k$ and $q_{\eta} = q$ the photon and $\eta$ meson momenta in
the c.m. system, and $\zeta_{\eta N} = \pm 1$ a relative sign
between the $N^* \rightarrow \eta N$ and $N^* \rightarrow \pi N$
couplings. $C_{\eta N}$ is an isospin factor, which is $-1$ for
$\eta N$ and $\eta^\prime N$ final states in the conventions used in
our work.

For the total widths of the resonances, we assume up to seven decay
channels, $\pi N$, $\pi\pi N$, $\eta N$, $K\Lambda$, $K\Sigma$,
$\omega N$, and $\eta^\prime N$,
\begin{equation}
 \Gamma_\mathrm{tot}(W) =
 \Gamma_{\pi N}(W) + \Gamma_{\pi\pi N}(W) + \Gamma_{\eta N}(W)+ \cdots\,.
\end{equation}
The threshold energies for the decays are listed in
table~\ref{tab:thresholds}.

\begin{table} [h]
\caption{Threshold energies in MeV of various $N^*$ decay
channels.\label{tab:thresholds}}
\begin{tabular}{|c|c|c|c|c|c|c|}
\hline
$\pi N$ & $\pi\pi N$ & $\eta N$ & $K \Lambda$ & $K \Sigma$ & $\omega N$ & $\eta^\prime N$ \\
\hline
1077.84 & 1217.41 & 1486.13 & 1609.36 & 1686.32 & 1720.92 & 1896.05\\
\hline
\end{tabular}
\end{table}

The energy dependence of the partial widths are given by
\begin{eqnarray}\label{eq:BW_widths_std}
 \Gamma_{\pi N}(W) &=& \beta_{\pi N }\,\Gamma_R
 \left(\frac{q_\pi(W)}{q_{\pi,R}}\right)^{2\ell+1}
 \left(\frac{X^2+q_{\pi,R}^2}{X^2+q_\pi(W)^2}\right)^\ell \,,\\
 \Gamma_{\eta N}(W) &=& \beta_{\eta N }\,\Gamma_R
 \left(\frac{q_\eta(W)}{q_{\eta,R}}\right)^{2\ell+1}
 \left(\frac{X^2+q_{\eta,R}^2}{X^2+q_\eta(W)^2}\right)^\ell \,,\\
 \Gamma_{\pi\pi N}(W) &=& \beta_{\pi\pi N }\,\Gamma_R
 \left(\frac{q_{2\pi}(W)}{q_{2\pi,R}}\right)^{2\ell+5}
 \left(\frac{X^2+q_{2\pi,R}^2}{X^2+q_{2\pi}(W)^2}\right)^{\ell+2}\,,\label{eq:BW_widths_std_2pi}
\end{eqnarray}
where $X$ is a cut-off parameter, which has been fixed in the
present work to $X=450$~MeV. The c.m. momenta of pion and eta are
denoted by $q_\pi$ and $q_\eta$, for the effective $2\pi$ channel we
use a mass of $2m_\pi$. All momenta, taken at the resonance
position, $W=M_R$, are denoted by an additional index $R$. All other
2-body channels are parameterized similarly as for $\pi N$ or $\eta
N$. In general the dynamics of 3-body decays as for the $\pi\pi N$
channel are rather complicated and have most extensively been
studied in the J{\"u}lich model. For single meson photoproduction
the effective 2-body treatment works very well.

For the energy dependence of the photon vertex, we assume the form
\begin{equation} \label{eq:fgammaN}
 f_{\gamma N}(W) =
 \left(\frac{k(W)}{k_R}\right)^2 \, \left(\frac{X_{\gamma}^2+k_R^2}{X_{\gamma}^2+k(W)^2}\right)^2\,,
\end{equation}
with the photon c.m. momentum $k$, which takes the value $k_R$ at
the resonance position. In EtaMAID2018 we found best fits for
$X_{\gamma}=0$.

The so-called reduced multipoles $\bar{\mathcal{M}}_{\ell\pm}$ are
related to the photon decay amplitudes $A_{1/2}$ and $A_{3/2}$ by
\begin{eqnarray}
\bar{M}_{\ell+} &=& -\frac{1}{\ell+1}\left(A_{1/2}^{\ell+}+\sqrt{\frac{\ell+2}{\ell}}A_{3/2}^{\ell+}\right)\,,\\
\bar{E}_{\ell+} &=& -\frac{1}{\ell+1}\left(A_{1/2}^{\ell+}-\sqrt{\frac{\ell}{\ell+2}}A_{3/2}^{\ell+}\right)\,,\\
\bar{M}_{\ell+1, -} &=& +\frac{1}{\ell+1}\left(A_{1/2}^{\ell+1, -}-\sqrt{\frac{\ell}{\ell+2}}A_{3/2}^{\ell+1, -}\right)\,,\\
\bar{E}_{\ell+1, -} &=& -\frac{1}{\ell+1}\left(A_{1/2}^{\ell+1,
-}+\sqrt{\frac{\ell+2}{\ell}}A_{3/2}^{\ell+1, -}\right)\,.
\end{eqnarray}
For specific resonances, see table~\ref{tab:amplitudes}.
\begin{table}[ht]
\caption{The reduced multipoles ${\bar{\cal M}}_{\alpha}$ in terms
of the photon decay amplitudes $A_\lambda$.\\}\label{tab:amplitudes}
\begin{center}
\begin{tabular}{|c|ccc|ccc|}
\hline
$N^{\ast}$        &      & $\bar{E}$  & & & $\bar{M}$  &    \\
\hline
$S_{11}$ &  & $-A_{1/2}$  &    &   & ---  &  \\
$P_{11}$ &  & ---  &    &   & $A_{1/2}$  & \\
$P_{13}$ &  & $\frac{1}{2}(\frac{1}{\sqrt{3}}A_{3/2}-A_{1/2})$ & & & $-\frac{1}{2}(\sqrt{3}A_{3/2}+A_{1/2})$ & \\
$D_{13}$ &  & $-\frac{1}{2}(\sqrt{3}A_{3/2}+A_{1/2})$ & & & $-\frac{1}{2}(\frac{1}{\sqrt{3}}A_{3/2}-A_{1/2})$ & \\
$D_{15}$&  & $\frac{1}{3}(\frac{1}{\sqrt{2}}A_{3/2}-A_{1/2})$ & & & $-\frac{1}{3}(\sqrt{2}A_{3/2}+A_{1/2})$ & \\
$F_{15}$ &  & $-\frac{1}{3}(\sqrt{2}A_{3/2}+A_{1/2})$ & & & $-\frac{1}{3}(\frac{1}{\sqrt{2}}A_{3/2}-A_{1/2})$ & \\
$F_{17}$ &  & $\frac{1}{4}(\sqrt{\frac{3}{5}}A_{3/2}-A_{1/2})$ & & & $-\frac{1}{4}(\sqrt{\frac{5}{3}}A_{3/2}+A_{1/2})$ & \\
$G_{17}$ &  & $-\frac{1}{4}(\sqrt{\frac{5}{3}}A_{3/2}+A_{1/2})$ & & & $-\frac{1}{4}(\sqrt{\frac{3}{5}}A_{3/2}-A_{1/2})$ & \\
$G_{19}$ &  & $\frac{1}{5}(\sqrt{\frac{2}{3}}A_{3/2}-A_{1/2})$ & & & $-\frac{1}{5}(\sqrt{\frac{3}{2}}A_{3/2}+A_{1/2})$ & \\
\hline
\end{tabular}
\end{center}
\end{table}

So far we assumed that the resonance mass $M_R$ is above all
considered decay channels. However, as nucleon resonances obtain
decay widths of the order of 100~MeV and more, also excitations of
resonances are very likely, if the nominal Breit-Wigner mass is only
a few MeV below threshold. But even for the Roper resonance, which
is about 50~MeV below $\eta N$ threshold, an excitation in $\eta$
photoproduction can be considered due to the large width of
$350$~MeV.

In such a case, however, the c.m. momentum $q_{a,R}$, which appears
in the parametrization of the partial width $\Gamma_a(W)$, is no
longer defined. In fact, one can analytically continue the momenta
below zero and obtains imaginary values.

In the literature, two different methods are discussed. The first
one takes a sharp cut-off with a $\theta$-function, giving a zero
value for the partial width below threshold. This is our EtaMAID
approach. The second one (Flatte's approach~\cite{Flatte:1976xu})
uses the analytical continuation of the momentum below threshold and
accepts the imaginary contribution of the width as a physical
contribution to the mass.

For both methods we can generalize the parametrization of a partial
width for arbitrary resonance masses
\begin{eqnarray}
 \Gamma_{a}(W) &=& g_a^2\;q_a(W)\,
  \left(\frac{|q_{a}^2(W)|}{X^2+|q_a^2(W)|}\right)^\ell\,.
\end{eqnarray}
The squared momenta $q_a^2(W)$ become negative below threshold and
could even produce singularities on the real axis in the physical
region. Therefore, we take the absolute values.

For resonances with masses larger than $W_{a,thr}$ this form can be
compared with the previous one, e.g. Eq.~(\ref{eq:BW_widths_std}),
and this gives the relation between the coupling constants $g_a$ and
the branching ratios $\beta_a$,
\begin{eqnarray}
 \beta_{a} &=& \frac{g_a^2\;q_a(M_{R})}{\Gamma_{R}\;
  (1+X^2/q_a^2(M_{R}))^\ell}\,,\\
 g_{a}^2 &=& \frac{\beta_a\,\Gamma_{R}}{q_a(M_{R})}\;
  (1+X^2/q_a^2(M_{R}))^\ell\,.
\end{eqnarray}

For the 3-body $2\pi$ channel we also make a small adjustment,
\begin{eqnarray}
 \Gamma_{\pi\pi}(W) &=& g_{\pi\pi}^2\;q_{2\pi}(W)\,
  \left(\frac{q_{2\pi}^2(W)}{X^2+q_{2\pi}^2(W)}\right)^{\ell+2}\,,
\end{eqnarray}
however, with a slightly different asymptotic behavior compared to
Eq.~(\ref{eq:BW_widths_std_2pi}).

For both $\pi N$ and $\pi\pi N$ channels, all nucleon resonances are
above threshold and the conventional definition of branching ratios
can be used. For $\eta N$ only the Roper resonance
$N(1440)\frac{1}{2}^+$ is below threshold. In the $K\Sigma$ channel
$N(1650)\frac{1}{2}^-$ and in the $\omega N$ channel
$N(1710)\frac{1}{2}^+$ are below threshold but with large couplings
that make significant contributions above threshold. Finally, in the
$\eta^\prime N$ channel we even find four states below threshold,
see table~\ref{tab:BWhadronic} in the appendix~\ref{app:BG-BW}.

\section{Results}\label{sect:Results}

\subsection{\boldmath Data base}
\label{sec:data-base}

In our analysis we only use modern data which cover a broad energy
and angular range. We prefer datasets with smallest statistical
uncertainties and we only combine data from different experiments if
they are in agreement in overlapping energy regions without
including additional scaling parameters. The unpolarized
differential cross section has been measured with by far highest
accuracy at MAMI. From several datasets we use those with the most
sophisticated reconstruction and error analysis
\cite{Kashevarov:2017}. The energy range of MAMI is limited to $W <
1970$~MeV. We used the differential cross section from the CLAS
Collaboration~\cite{Williams:2009yj} in this fit because of their
much smaller statistical errors, larger energy coverage, and better
agreement with the high statistics data from
A2MAMI~\cite{Kashevarov:2017} in an overlapping energy region than
CBELSA/TAPS data~\cite{Crede:2009zzb}. The angular-dependent
systematic uncertainty for results of Run-I and Run-II above
$W=1796$~MeV was evaluated as 3\%, for Run-III and for the $\eta'$
differential cross sections - as 5-6\%. These uncertainties were
added in quadrature to the statistical
uncertainties~\cite{Kashevarov:2017}. For other data, we use only
statistical uncertainties in the fit.

The photon beam asymmetry $\Sigma$ has been measured over the full
resonance region by GRAAL and CLAS. We include all polarized target
and beam-target asymmetries from modern experiments. Old data, in
particular an early target asymmetry measurement at ELSA
\cite{Bock1998}, cannot compete with regard to statistical and
systematic uncertainties and are not used in our analysis.

The differential cross sections cover the energy region from
threshold up to $W=2.8$~GeV. Polarization observables are from
threshold up to $W=1.85$~GeV for $T$ and $F$, up to $W=2.13$~GeV for
$E$, and up to $W=2.08$~GeV for $\Sigma$. These are five
polarization observables for the $\eta p$ channel with good energy
and angular coverage, which is however, still far away from a
complete experiment, that would require at least 8 observables
including those with recoil polarization detection. Therefore, some
ambiguities in the PWA can be expected.

Data sets for the other reactions are much more scarce than for
$\gamma p \to \eta p$. In the  $\eta n$ channel we have only three
observables, for $\eta^\prime p$ two and for $\eta^\prime n$ just
the differential cross section alone, see table~\ref{tab:expdata}.

In our fits to the data we have used a total of 208 parameters. For
the resonance sector with 21 $N^*$ resonances we have 112 parameters
for BW parametrization and 66 for unitarity phases. The background
is described with 20 parameters, mainly for the Regge
parametrization.

\begin{table*}[htbp]
\begin{center}
\caption{\label{tab:expdata} Experimental data on $\eta$ and
$\eta^\prime$ photoproduction. The column `used' shows the data that
were included in our fits and those that were ignored. $N$ is the
number of data points and $\chi^2$ is the total weighted deviation
from our standard 2018 solution for that dataset.}
\bigskip
\begin{tabular}{|c|c|c|c|c|c|c|c|}
\hline
Observable & Reaction  & used    &$W$~[MeV] &$N$ &$\chi^2$  &$\chi^2/N$  & Reference \\
\hline
$\sigma_0$ &$p(\gamma,\eta)p$        &  ---  &$1488 - 1870$ &2880 &9502&  3.3& A2MAMI-17 (Run I)~\cite{Kashevarov:2017}  \\
$\sigma_0$ &$p(\gamma,\eta)p$        &$\surd$&$1488 - 1891$ &2712 &4437&  1.6& A2MAMI-17 (Run II)~\cite{Kashevarov:2017}  \\
$\sigma_0$ &$p(\gamma,\eta)p$        &$\surd$&$1888 - 1957$ & 288 &329&   1.1& A2MAMI-17 (Run III)~\cite{Kashevarov:2017}  \\
$\sigma_0$ &$p(\gamma,\eta)p$        &$\surd$&$1965 - 2795$ & 634 &2276&  3.6& CLAS-09~\cite{Williams:2009yj} \\
$\sigma_0$ &$p(\gamma,\eta)p$        &  ---  &$1588 - 2370$ & 680 &8640&  13.& CBELSA/TAPS-09~\cite{Crede:2009zzb} \\
$\Sigma$   &$p(\gamma,\eta)p$        &$\surd$&$1496 - 1908$ & 150 &394&   2.6& GRAAL-07~\cite{Bartalini:2007fg} \\
$\Sigma$   &$p(\gamma,\eta)p$        &$\surd$&$1700 - 2080$ & 214 &617&   2.9& CLAS-17~\cite{Collins:2017} \\
$T$        &$p(\gamma,\eta)p$        &$\surd$&$1497 - 1848$ & 144 &246&   1.7& A2MAMI-14~\cite{Akondi:2014} \\
$F$        &$p(\gamma,\eta)p$        &$\surd$&$1497 - 1848$ & 144 &246&   1.7& A2MAMI-14~\cite{Akondi:2014} \\
$E$        &$p(\gamma,\eta)p$        &$\surd$&$1525 - 2125$ &  73 &155&   2.1& CLAS-16~\cite{Senderovich:2016} \\
$E$        &$p(\gamma,\eta)p$        &$\surd$&$1505 - 1882$ & 135 &255&   1.9& A2MAMI-17~\cite{Witthauer:2017} \\
\hline
$\sigma_0$ &$n(\gamma,\eta)n$        &$\surd$&$1492 - 1875$ & 880 &3079&  3.5& A2MAMI-14~\cite{Werthmueller:2014} \\
$\sigma_0$ &$n(\gamma,\eta)n$        &  ---  &$1505 - 2181$ & 322 &2986&  9.3& CBELSA/TAPS-11~\cite{Jaegle2:2011} \\
$\sigma_0$ &$n(\gamma,\eta)n$        &  ---  &$1588 - 2070$ & 317 &4992&  16.& CBELSA/TAPS-17~\cite{Witthauer:2017bonn} \\
$\Sigma$   &$n(\gamma,\eta)n$        &$\surd$&$1504 - 1892$ &  99 &177&   1.8& GRAAL-08~\cite{Fantini:2008} \\
$E$        &$n(\gamma,\eta)n$        &$\surd$&$1505 - 1882$ & 135 &209&   1.5& A2MAMI-17~\cite{Witthauer:2017} \\
\hline
$\sigma_0$ &$p(\gamma,\eta^\prime)p$ &$\surd$&$1898 - 1956$ & 120 &198&   1.7& A2MAMI-17~\cite{Kashevarov:2017}   \\
$\sigma_0$ &$p(\gamma,\eta^\prime)p$ &$\surd$&$1925 - 2795$ & 681 &2013&  3.0& CLAS-09~\cite{Williams:2009yj} \\
$\sigma_0$ &$p(\gamma,\eta^\prime)p$ &  ---  &$1934 - 2351$ & 200 &278&   1.4& CBELSA/TAPS-09~\cite{Crede:2009zzb} \\
$\Sigma$   &$p(\gamma,\eta^\prime)p$ &$\surd$&$1903 - 1913$ &  14 &35&    2.5& GRAAL-15~\cite{Sandri:2015} \\
$\Sigma$   &$p(\gamma,\eta^\prime)p$ &$\surd$&$1904 - 2080$ &  62 &85&    1.4& CLAS-17~\cite{Collins:2017} \\
\hline
$\sigma_0$ &$n(\gamma,\eta^\prime)n$ &$\surd$&$1936 - 2342$ & 170 &191&   1.1& CBELSA/TAPS-11~\cite{Jaegle:2011} \\

\hline
\end{tabular}
\end{center}
\end{table*}

\subsection{Total cross sections}

We begin the discussions of our results with the total cross
sections of the four channels considered in our work:
$p(\gamma,\eta)p$, $n(\gamma,\eta)n$, $p(\gamma,\eta^\prime)p$,
$n(\gamma,\eta^\prime)n$.

\begin{figure*}[!ht]
\begin{center}
\resizebox{0.7\textwidth}{!}{\includegraphics{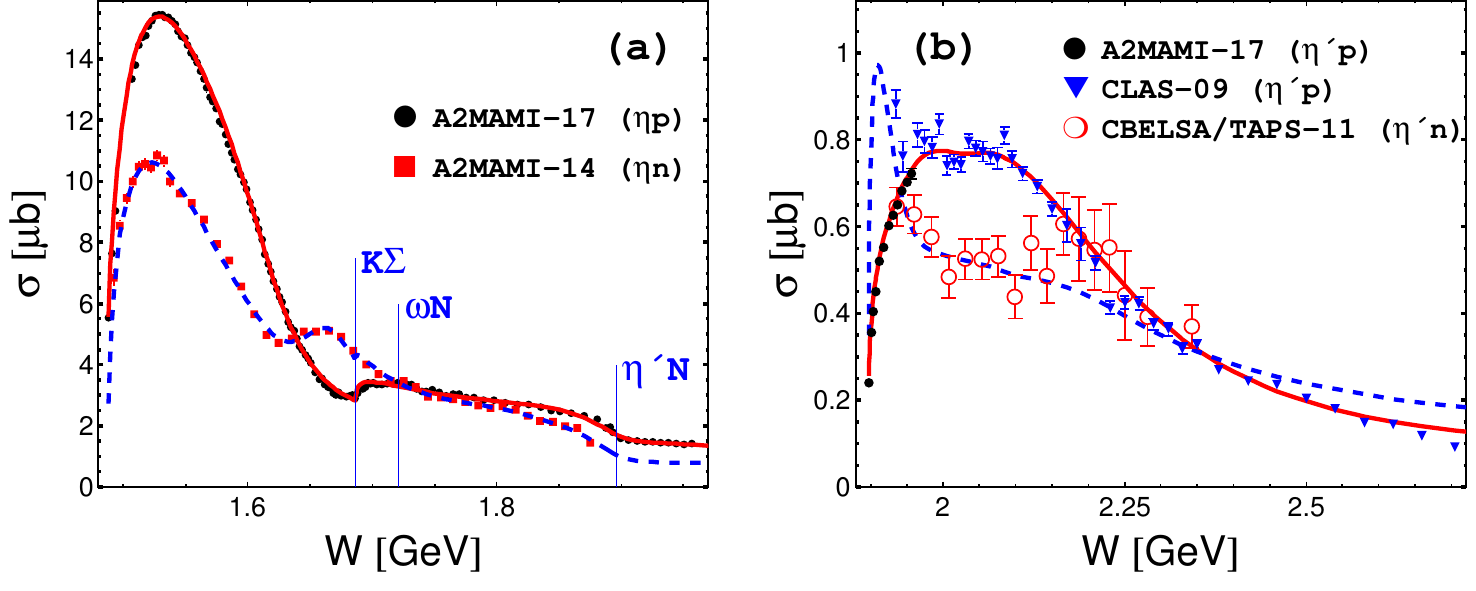}}
\caption{Total cross section for $(\gamma,\eta)$ (a) and
$(\gamma,\eta^\prime)$ (b) on protons and neutrons. The solid red
and dashed blue lines show our EtaMAID solution for proton and
neutron, respectively. } \label{fig:tcs}
\end{center}
\end{figure*}

\begin{figure*}[!h]
\begin{center}
\resizebox{0.7\textwidth}{!}{\includegraphics{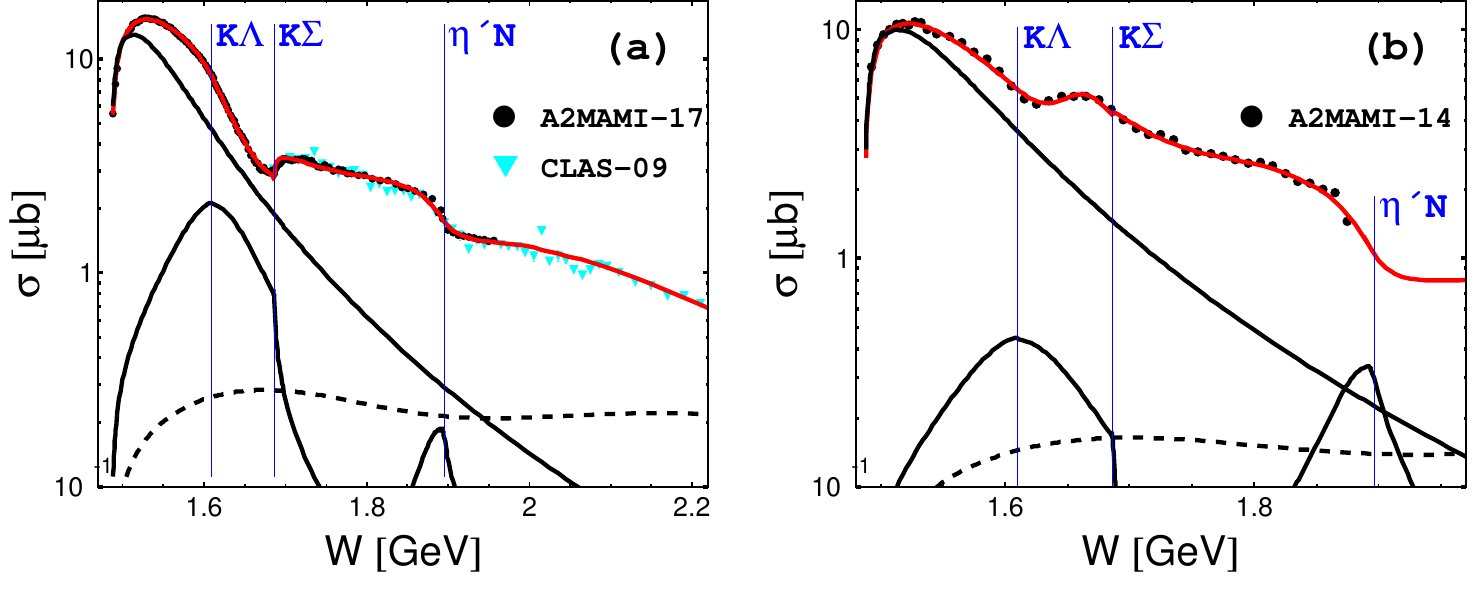}}
\caption{Partial contributions of the $S$-wave resonances to the total cross
section for $(\gamma,\eta)$ on protons (a) and neutrons (b) in comparison
with the non-resonant background. The
solid red lines show our full EtaMAID solution. The individual
contribution of $S_{11}(1535)$, $S_{11}(1650)$ and $S_{11}(1895)$
resonances are shown by solid black lines. The dashed line shows the
total background of Born and Regge contributions including the
damping factors. Vertical lines correspond to thresholds of
$K\Lambda$, $K\Sigma$, and $\eta^\prime N$ photoproduction. }
\label{fig:tcs_resonances_eta_log}
\end{center}
\end{figure*}

\begin{figure*}[!h]
\begin{center}
\resizebox{0.7\textwidth}{!}{\includegraphics{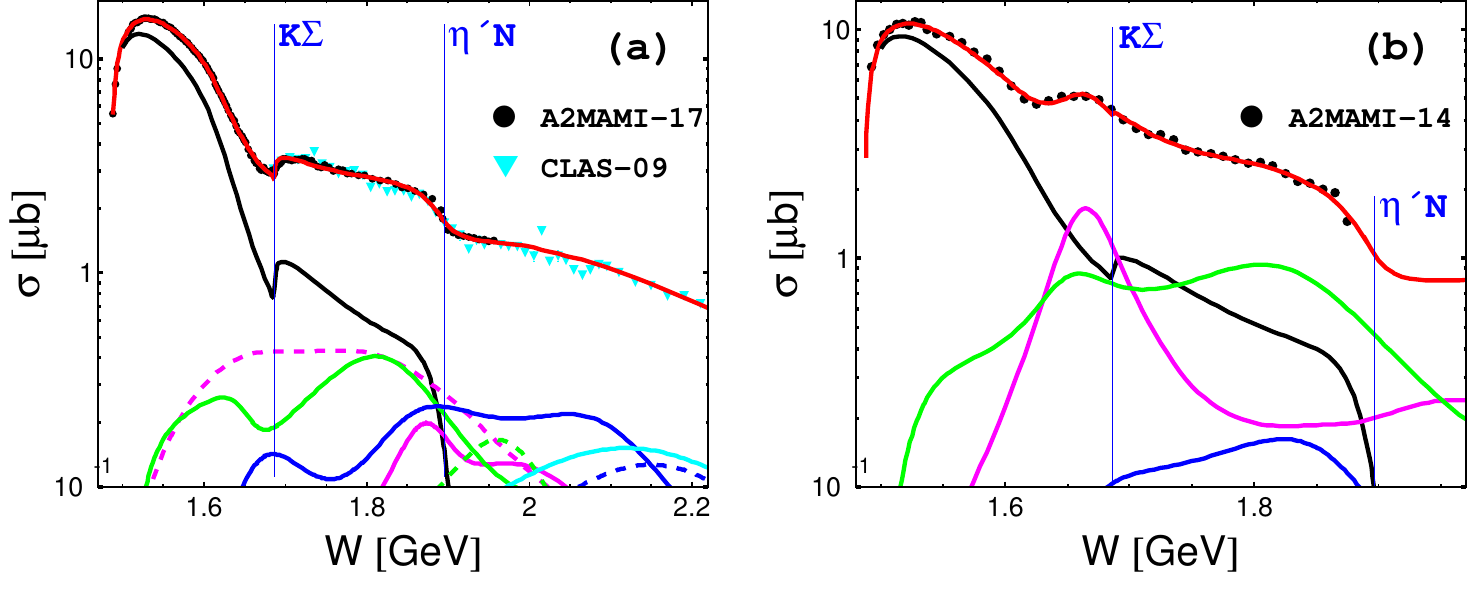}}
\caption{Resonance contributions of partial waves to the total cross
section for $(\gamma,\eta)$ on protons (a) and neutrons (b). The
solid red lines show our full EtaMAID solution including background.
The black solid lines are the sum of three $S_{11}(1535,1650,1895)$
resonances, magenta solid: four $P_{11}(1440,1710,1880,2100)$,
magenta dashed: two $P_{13}(1720,1900)$, green solid: four
$D_{13}(1520,1700,1875,2120)$, green dashed: two
$D_{15}(1675,2060)$, blue solid: three $F_{15}(1680,1860,2000)$,
blue dashed: $F_{17}(1990)$ and cyan solid: $G_{17}(2190)$. Vertical
lines correspond to thresholds of $K\Sigma$ and $\eta^\prime N$
photoproduction. } \label{fig:tcs_pw_eta_log}
\end{center}
\end{figure*}

\begin{figure*}[!ht]
\begin{center}
\resizebox{0.7\textwidth}{!}{\includegraphics{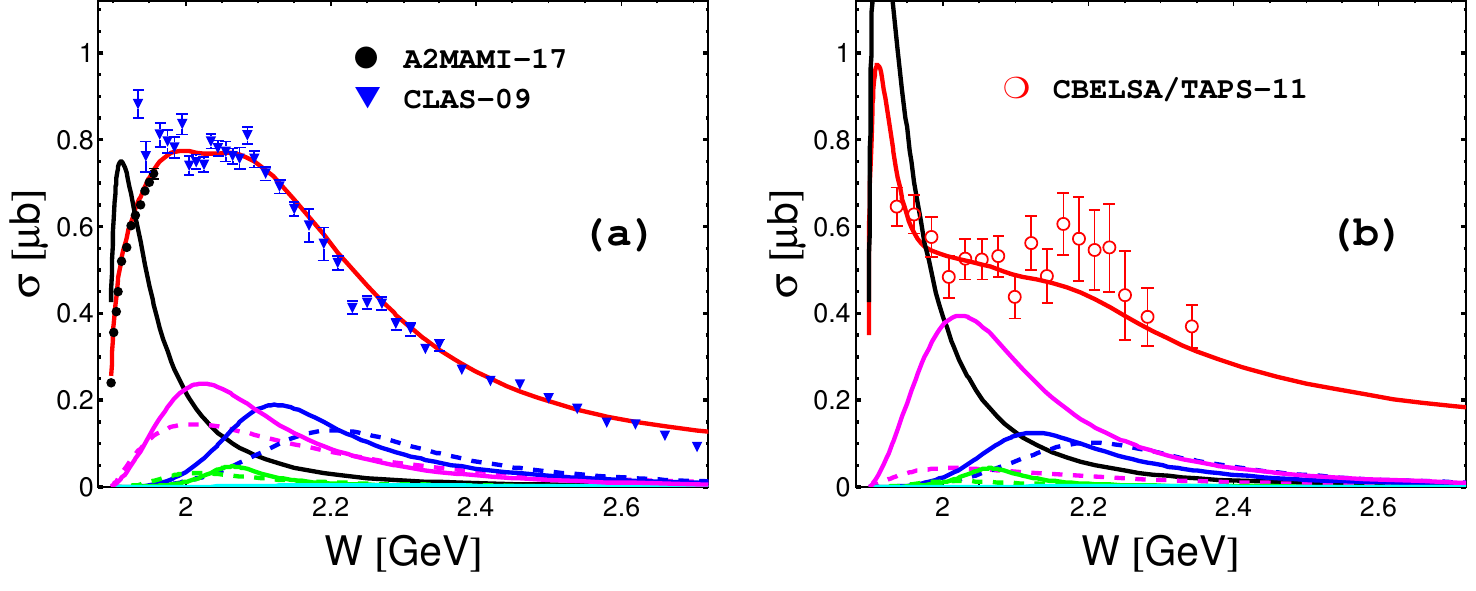}}
\caption{Partial contributions of the resonances to the total cross
section for $(\gamma,\eta^\prime)$ on protons (a) and neutrons (b).
The solid red lines show our full EtaMAID solution. The other curves
show resonance contributions of $S_{11}(1895)$: black solid,
$P_{11}(1880,2100)$: magenta solid, $P_{13}(1900)$: magenta dashed,
$D_{13}(1875)$: green solid, $D_{15}(2000)$: green dashed,
$F_{15}(2000)$: blue solid, and $F_{17}(1990)$: blue dashed.}
\label{fig:tcs_resonances_etapr}
\end{center}
\end{figure*}

\begin{figure*}[!ht]
\begin{center}
\resizebox{0.7\textwidth}{!}{\includegraphics{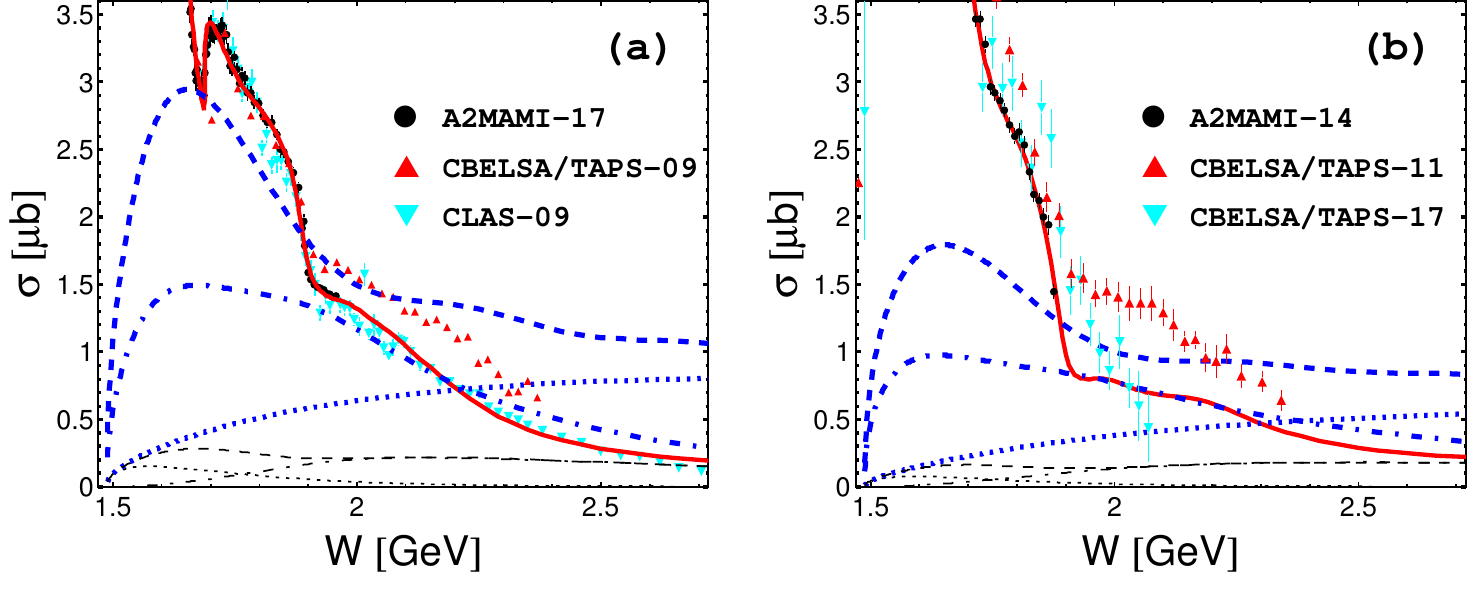}}
\caption{Partial contributions of the background to the total cross
section for $(\gamma,\eta)$ on protons (a) and neutrons (b). The
solid red lines show our full EtaMAID solution. The wide blue
dotted, dash-dotted, and dashed lines show Born, Regge, and
Born+Regge, respectively, without damping factors. The thin black
dotted, dash-dotted, and dashed lines show the same, when damping
factors are applied. The CBELSA/TAPS data have not been used in our
fit. } \label{fig:tcs_bgr_eta}
\end{center}
\end{figure*}

\begin{figure*}[!ht]
\begin{center}
\resizebox{0.7\textwidth}{!}{\includegraphics{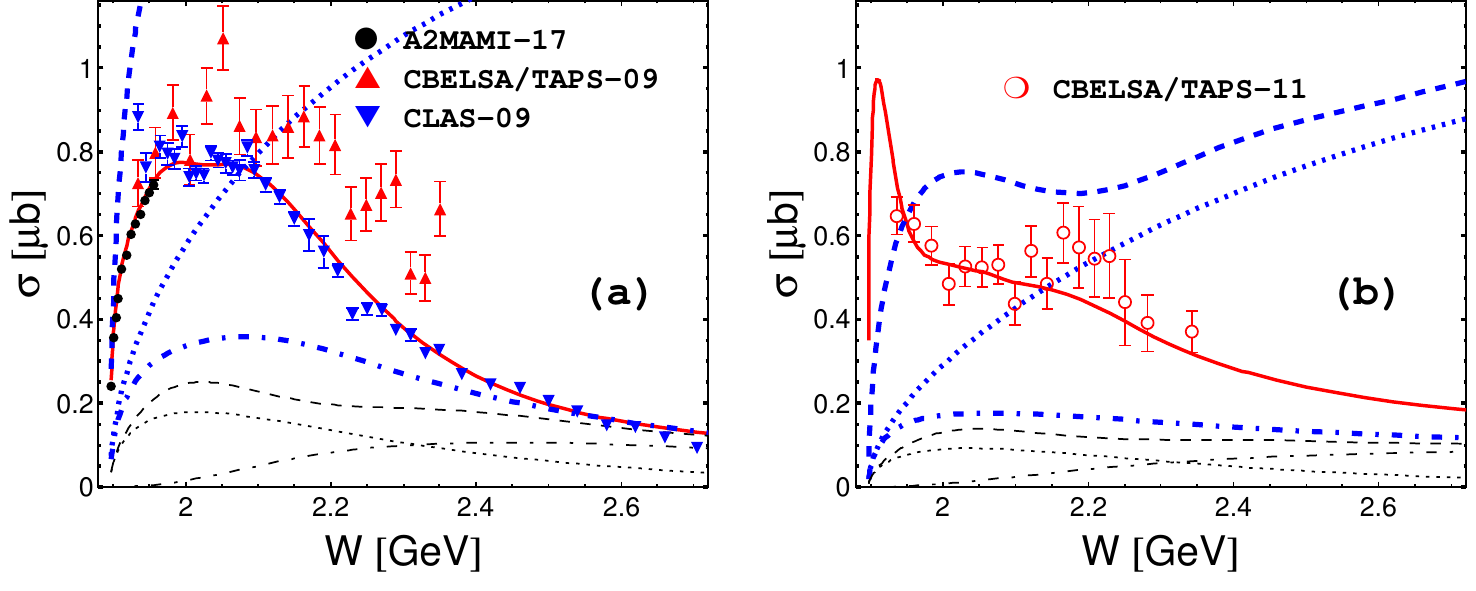}}
\caption{Partial contributions of the background to the total cross
section for $(\gamma,\eta^\prime)$ on protons (a) and neutrons (b).
Notation of curves are as in Fig.~\ref{fig:tcs_bgr_eta}. }
\label{fig:tcs_bgr_etapr}
\end{center}
\end{figure*}

\begin{figure*}[!ht]
\begin{center}
\resizebox{0.7\textwidth}{!}{\includegraphics{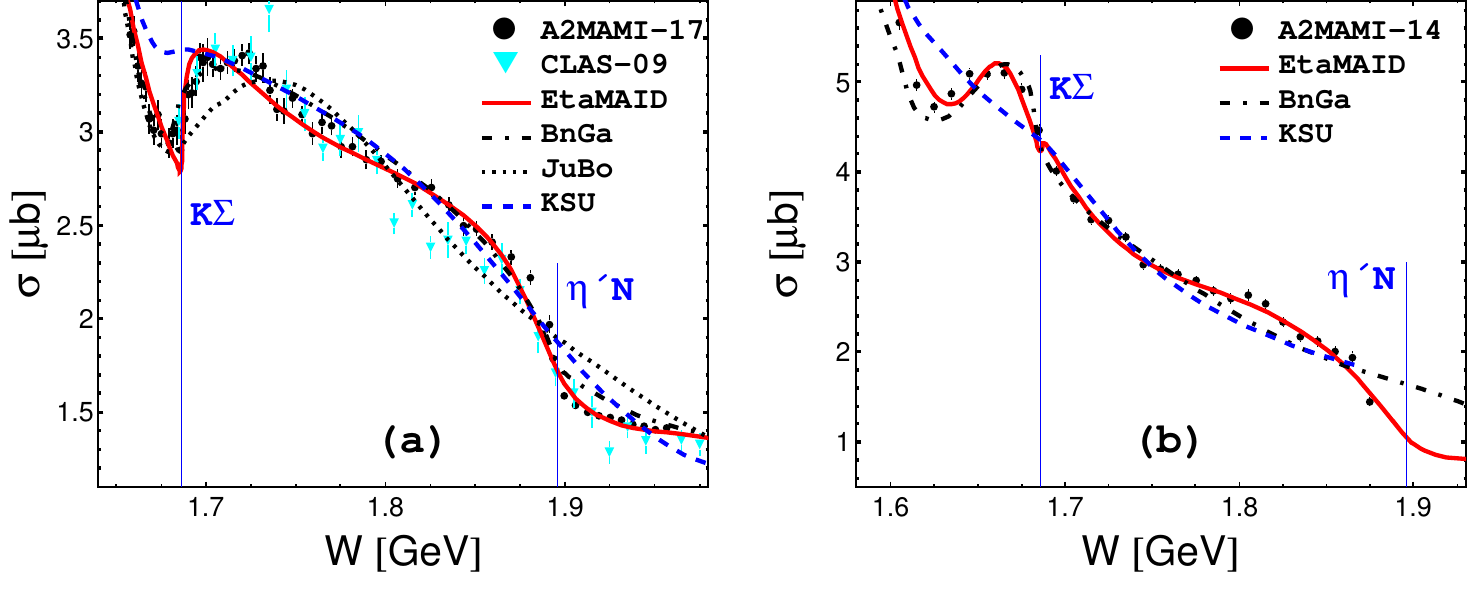}}
\caption{Total cross section for $(\gamma,\eta)$ on protons (a) and
neutrons (b) in comparison with other newly updated PWA. The solid
red lines show our full EtaMAID solution. The black dash-dotted,
dotted and blue dashed curves are obtained from the newly updated
BnGa~\cite{Anisovich:2018}, J\"uBo~\cite{Ronchen:2018}
and KSU~\cite{KSU2018} partial wave analyses. Near $\eta$ threshold,
below 1.6~GeV, all solutions are practically identical. Vertical
lines correspond to thresholds of $K\Sigma$, and $\eta^\prime N$
photoproduction. } \label{fig:tcs_4models_eta}
\end{center}
\end{figure*}

The data in Figs.~\ref{fig:tcs} - \ref{fig:tcs_4models_eta} are from
A2 Collaboration at MAMI: A2MAMI-17~\cite{Kashevarov:2017} and
A2MAMI-14~\cite{Werthmueller:2014}; CBELSA/TAPS Collaboration:
CBELSA/TAPS-09~\cite{Crede:2009zzb}, CBELSA/TAPS-11 for $\eta
n$~\cite{Jaegle2:2011} and for $\eta^\prime n$~\cite{Jaegle:2011},
and CBELSA/TAPS-17~\cite{Witthauer:2017bonn}. In the case of the
CLAS-09 data, we show data points that were obtained in a Legendre
fit to the differential cross sections from CLAS
collaboration~\cite{Williams:2009yj} and are affected by additional
uncertainties due to a limited angular range of the data especially
in forward direction. The total cross section data shown here have
not been used in our fit, only the differential cross sections were
fitted.

The fit results for the total cross sections are presented in
Fig.~\ref{fig:tcs} together with corresponding experimental data.

In Fig.~\ref{fig:tcs}~(a), there are very interesting features
visible at energies $W\approx 1680$~MeV and $W\approx 1890$~MeV,
which can be explained by cusp effects due to the opening of new
strong channels in the $S$-wave.

The cusp in the $\eta p$ total cross section, in connection with the
steep rise of the $\eta^\prime p$ from its threshold,
Fig.~\ref{fig:tcs}~(b), is explained by a strong coupling of the
$S_{11}(1895)$ resonance to both channels, see also
Figs.~\ref{fig:tcs_resonances_eta_log} - \ref{fig:tcs_pw_eta_log}.
Unfortunately, there are no data for the $\eta^\prime n$ channel
near threshold and only one data point exists in the cusp region for
the $\eta n$ channel, Fig.~\ref{fig:tcs}~(a). Nevertheless our
solution demonstrates also a strong coupling of the $S_{11}(1895)$
for these neutron channels.

Other interesting structures are observed as a dip in $\gamma p
\rightarrow \eta p$ and a bump in $\gamma n \rightarrow \eta n$
around $W \approx 1680$ MeV, Fig.~\ref{fig:tcs}~(a). Both structures
were observed experimentally many times and its existence is
unambiguous. However its nature is not yet fully understood. See for
more details Ref.~\cite{Krusche2014}. Our analysis shows that the
narrow bump in $\eta n$ and the dip in $\eta p$ channels have
different origin. The first is a result of an interference of few
resonances with a dominant contribution of the $P_{11}(1710)$, see
Fig.~\ref{fig:tcs_resonances_eta_log}~(b) and
Fig.~\ref{fig:tcs_pw_eta_log}~(b). The second one is mainly a sum of
$S_{11}(1535)$  and $S_{11}(1650)$ with opposite signs. However the
narrowness of this structure is explained by a cusp effect due to
the opening of the $K\Sigma$ decay channel of the $S_{11}(1650)$
resonance, see Fig.~\ref{fig:tcs_resonances_eta_log}~(a) and
Fig.~\ref{fig:tcs_pw_eta_log}~(a).

In Fig.~\ref{fig:tcs_resonances_eta_log} -
\ref{fig:tcs_resonances_etapr} we show partial resonance
contributions for $\eta$ and $\eta^\prime$ photoproduction in four
channels. In Fig.~\ref{fig:tcs_resonances_eta_log} we concentrate on
the most important $S_{11}$ resonances $N(1535)\frac{1}{2}^-$,
$N(1650)\frac{1}{2}^-$ and $N(1895)\frac{1}{2}^-$. The
$S_{11}(1535)$ completely dominates both proton and neutron
channels. And, as a side remark, due to the large branchings into
the $\pi N$ and $\eta N$ channels, this resonance produces a very
significant cusp effect in the cross sections of pion
photoproduction~\cite{Althoff:1979mb,Ahrens:2006gp}. The second
$S_{11}(1650)$ exhibits visible cusp effects due to the opening of
the $K\Lambda$ and $K\Sigma$ channels. Also the third $S_{11}(1895)$
shows a visible cusp at $\eta^\prime N$ threshold. In the full
solution the $K\Lambda$ cusp remains hidden under the strong
$S_{11}(1535)$ contribution, also the $K\Sigma$ cusp becomes
invisible in the neutron channel. But in the proton channel this
cusp appears as a very pronounced dip with even a kind of a bump
afterwards. The $\eta^\prime N$ cusps due to the third
$S_{11}(1895)$ resonance are visible in both proton and neutron
channels, and in case of the proton the cusp is very well supported
by the high-precision data of A2-MAMI.

The cusp structures are even better visible in
Fig.~\ref{fig:tcs_pw_eta_log}, where all resonances within the same
partial wave are summed up. In the cases of $P_{11}$ and $D_{13}$
these are sums over even four $N^*$ resonances. From this figure it
becomes very clear that the bump structures at $W\approx 1680$~MeV
is a cusp effect of the $S_{11}(1650)$ in the proton channel and a
resonance effect of the $P_{11}(1710)$ in the neutron channel.

The largest $N^*$ resonance contributions in $(\gamma,\eta)$ total
cross sections are from $S_{11}(1535,1650,1895)$, $P_{11}(1710)$,
$P_{13}(1720,1900)$, and $D_{13}(1700,1875)$.

Fig.~\ref{fig:tcs_resonances_etapr} shows the partial
contributions of the $N^*$ resonances to the total cross sections
for $(\gamma,\eta^\prime)$ on proton and neutron. The largest
resonance contributions in the total cross sections for
$(\gamma,\eta^\prime)$ are from $S_{11}(1895)$, $P_{11}(1880)$,
$P_{11}(2100)$, $F_{15}(2000)$ and $F_{17}(1990)$. It is interesting
to note, that the first two of them have Breit-Wigner masses below
threshold but appear as resonance bumps above threshold due to phase
space factors.

In both channels the $S_{11}$ resonance dominates near threshold and
the second largest peak arises from $P_{11}$, followed by large
contributions from $F$-wave resonances. This is different in the
most recent BnGa analysis~\cite{Anisovich:2017}, where the
$P_{13}(1900)$ plays a dominant role and $F$-waves are practically
negligible. Such ambiguities in the PWA can be expected when only
two observables are measured as in the $\eta^\prime$ proton channel.
For the neutron channel, there is even only the differential cross
section measured. Such an incompleteness in the polarization
observables naturally leads to large ambiguities in the partial wave
analysis.

In Fig.~\ref{fig:tcs_bgr_eta} and
Fig.~\ref{fig:tcs_bgr_etapr} we show the background contributions
for $\eta$ and $\eta^\prime$ photoproduction in four channels. The
blue dotted, dash-dotted and dashed lines are obtained by Born
terms, $t$-channel vector meson exchanges in Regge parametrization
and the sum of both, respectively. The Born terms rise very strongly
already near threshold in $\eta^\prime$ photoproduction and appear
also very large in $\eta$ photoproduction for energies above 2~GeV.
The $t$-channel Regge contributions are also quite large in the
resonance region below 2.5~GeV and dominate the cross section for
energies above 2~GeV. The double counting of Regge and resonances
becomes quite obvious. Therefore, as explained in sect.~III before,
we have introduced damping factors for the background contributions,
Eqs.~(\ref{eq:Born_damp}) and (\ref{eq:Regge_damp}), yielding to the
black dotted, dash-dotted and dashed lines for Born, Regge and total
background, respectively.

Finally, in Fig.~\ref{fig:tcs_4models_eta} we compare our
EtaMAID2018 solution with the new 2018 updates of Bonn-Gatchina
(BnGa)~\cite{Anisovich:2018}, J\"ulich-Bonn
(J\"uBo)~\cite{Ronchen:2018} and Kent-State University
(KSU)~\cite{KSU2018}.

While EtaMAID has analyzed all four channels up to $W=4.5$~GeV
($E\approx 10$~GeV), BnGa analyzed three, $\eta p$, $\eta n$, and
$\eta^\prime p$ up to $W=2.5$~GeV, KSU analyzed two, $\eta p$ up to
$W=2.0$~GeV and $\eta n$ up to $W=1.9$~GeV and J\"uBo analyzed only
the $\eta p$ channel up to $W=2.4$~GeV. J\"uBo and KSU, which did
not include the latest A2MAMI-17 data, differ significantly from the
data in the dip region around $W\approx 1680$~MeV and around the
$\eta^\prime$ threshold, the BnGa solution describes the data much
better. The best description of the data is obtained with EtaMAID.

In appendix~\ref{app:BG-BW} we list all background and resonance
parameters of our model. For a few selected very important $S_{11}$
and $P_{11}$ resonances we also give an error analysis for
Breit-Wigner parameters based on MINUIT-MINOS in
table~\ref{tab:bw-with-errors}. We also have calculated effective
$\eta^\prime N$ branching ratios by integrating the decay spectrum
above $\eta^\prime N$ threshold according to
Ref.~\cite{Beringer:1900zz}. For the $N(1880)1/2^+$ and
$N(1895)1/2^-$ we obtained $(6.3\pm 2)\%$ and $(19.5\pm 5)\%$,
respectively. A complete resonance analysis, especially with pole
positions and residues will be published in a following paper.

\begin{table*}
\caption{\label{tab:bw-with-errors} Breit-Wigner parameters for
selected resonances: mass M$_{BW}$, total width $\Gamma_{BW}$,
branching ratio $\beta_{\eta N}$ to $\eta$N, and helicity amplitudes
$A_{1/2}^{p(n)}$ for proton (neutron). The first row for each
resonance gives a parameter set of the presented EtaMAID solution.
The parameters indicated without errors were fixed during the fit.
The second row indicate an overall status of the resonance and lists
the corresponding parameters estimated by PDG~\cite{Tanabashi:2018}
(NE means "No Estimates" given by PDG.). The effective $\eta^\prime
N$ branching ratios according to Ref.~\cite{Beringer:1900zz} for the
$N(1880)1/2^+$ and $N(1895)1/2^-$ are $(6.3\pm 2)\%$ and $(19.5\pm
5)\%$, respectively.} \label{tab:bwpar}
\begin{center}
\begin{tabular*}{16.5cm}
{@{\hspace{0.1cm}}c @{\hspace{0.1cm}}|
@{\hspace{0.1cm}}c
@{\hspace{0.3cm}}c
@{\hspace{0.3cm}}c
@{\hspace{0.3cm}}c
@{\hspace{0.5cm}}c
@{\hspace{0.5cm}}c
@{\hspace{0.5cm}}c
@{\hspace{0.5cm}}c }
\hline\hline\noalign{\smallskip}
Resonance $J^P$ &
$M_{BW}$ [MeV] &
$\Gamma_{BW}$ [MeV] &
$\beta_{\eta N}$ $[\%]$ &
$A_{1/2}^p$ $[10^{-3}$GeV$^{-1/2}]$ &
$A_{1/2}^n$ $[10^{-3}$GeV$^{-1/2}]$ & \\
\noalign{\smallskip}\hline\noalign{\smallskip}
$N(1535)1/2^-$ &$1522\pm8$  &$175\pm25$ &$34\pm5$  &$+115$      &$-102\pm8$  \\
 ****          &$1530\pm15$ &$150\pm25$ &$42\pm13$ &$+105\pm15$ &$-75\pm20$  \\
\noalign{\smallskip}\hline\noalign{\smallskip}
$N(1650)1/2^-$ &$1626^{+10}_{-5}$  &$133\pm20$ &$19\pm6$  &$+55$      &$-25\pm20$        \\
 ****          &$1650\pm15$        &$125\pm25$ &$25\pm10$ &$+45\pm10$ &$-10^{+40}_{-30}$ \\
\noalign{\smallskip}\hline\noalign{\smallskip}
$N(1710)1/2^+$ &$1670\pm20$ &$63^{+55}_{-18}$  &$12\pm4$ &$5.5$ &$-42^{+16}_{-12}$ \\
 ****          &$1710\pm30$ &$140\pm60$       &$30\pm20$ &NE    &NE                \\
\noalign{\smallskip}\hline\noalign{\smallskip}
$N(1880)1/2^+$ &$1882\pm24$ &$90^{+70}_{-30}$ &$43^{+10}_{-20}$ &$60$ &$-7^{+60}_{-60}$ \\
 ***           &$1880\pm50$ &$300\pm100$      &NE              &NE   &NE               \\
\noalign{\smallskip}\hline\noalign{\smallskip}
$N(1895)1/2^-$ &$1894.4^{+5}_{-15}$ &$71^{+25}_{-13}$  &$3.3\pm1.5$       &$-32$ &$+43^{+30}_{-50}$ \\
 ****          &$1895\pm25$         &$120^{+80}_{-40}$ &$25^{+15}_{-10}$  &NE    &NE                \\
\noalign{\smallskip}\hline\hline
\end{tabular*}
\end{center}
\end{table*}

\subsection{\boldmath Comparison with the data of $d\sigma/d\Omega, \Sigma, T, F, E$
for $\gamma\,p \rightarrow \eta\,p$}

\begin{figure*}[!ht]
\begin{center}
\resizebox{0.7\textwidth}{!}{\includegraphics{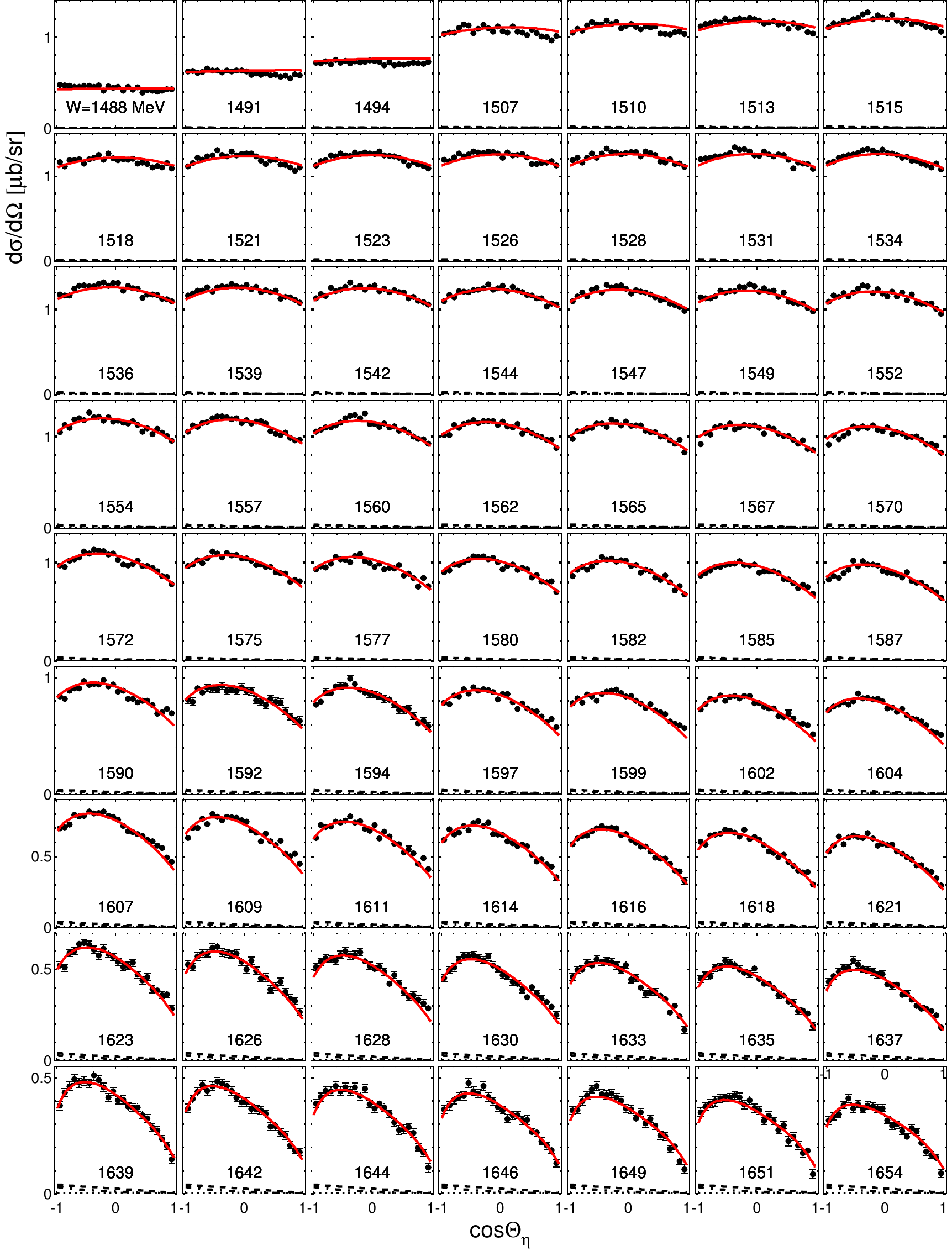}}
\caption{Differential cross section for $(\gamma,\eta)$ on the
proton for 1488~MeV~$\le W \le$~1654~MeV as function of cosine of
the c.m. scattering angle. The solid red lines show our full
solutions, whereas the black dotted, dash-dotted, and dashed lines
are Born terms, Regge, and full background, respectively. The data
are from A2MAMI~\cite{Kashevarov:2017}. }
\label{fig:eta-p_dcs1_mami}
\end{center}
\end{figure*}

\begin{figure*}[!ht]
\begin{center}
\resizebox{0.7\textwidth}{!}{\includegraphics{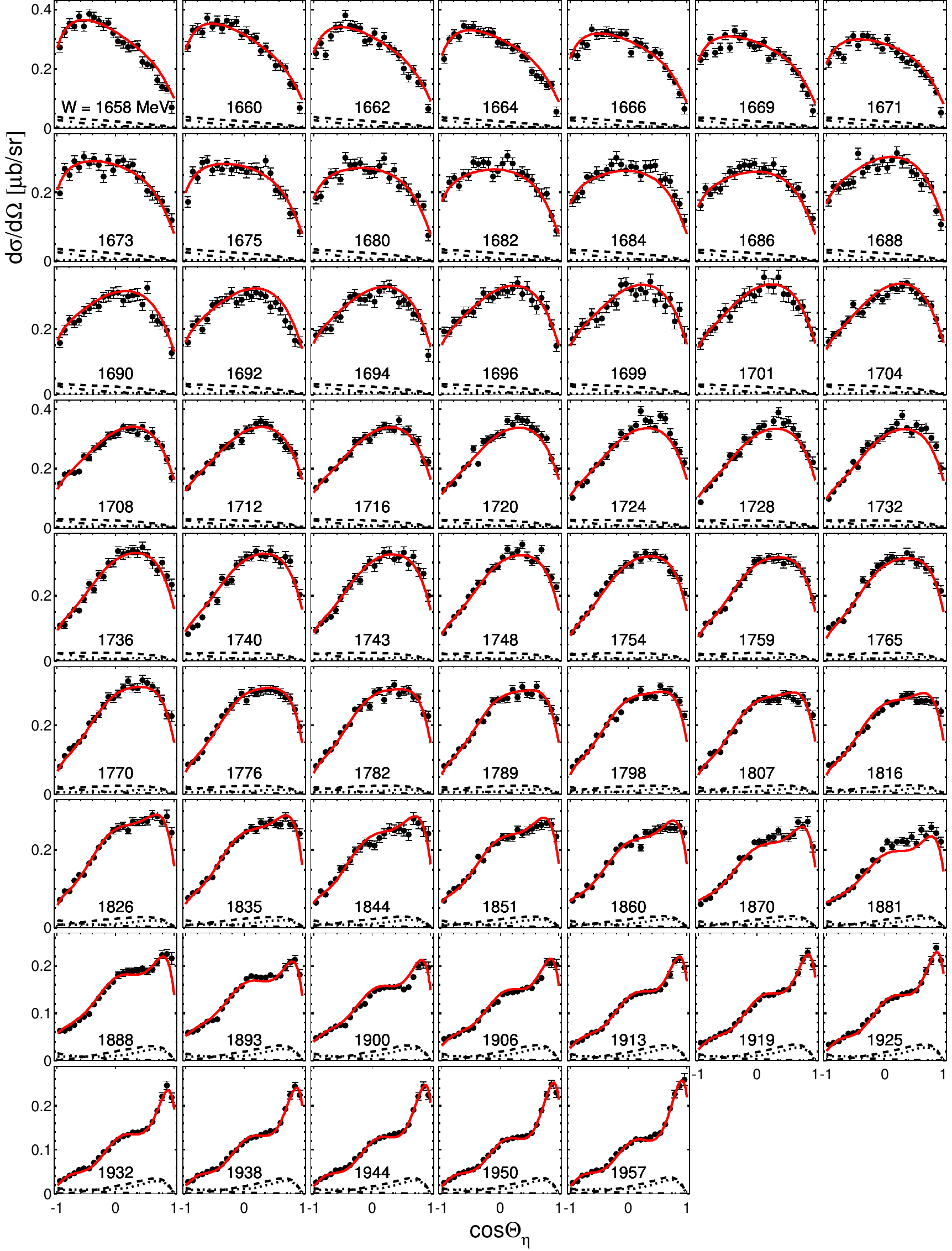}}
\caption{Same as in Fig.~\ref{fig:eta-p_dcs1_mami} for c.m. energies
1658~MeV~$\le W \le 1957$~MeV. } \label{fig:eta-p_dcs2_mami}
\end{center}
\end{figure*}

\begin{figure*}[!ht]
\begin{center}
\resizebox{0.7\textwidth}{!}{\includegraphics{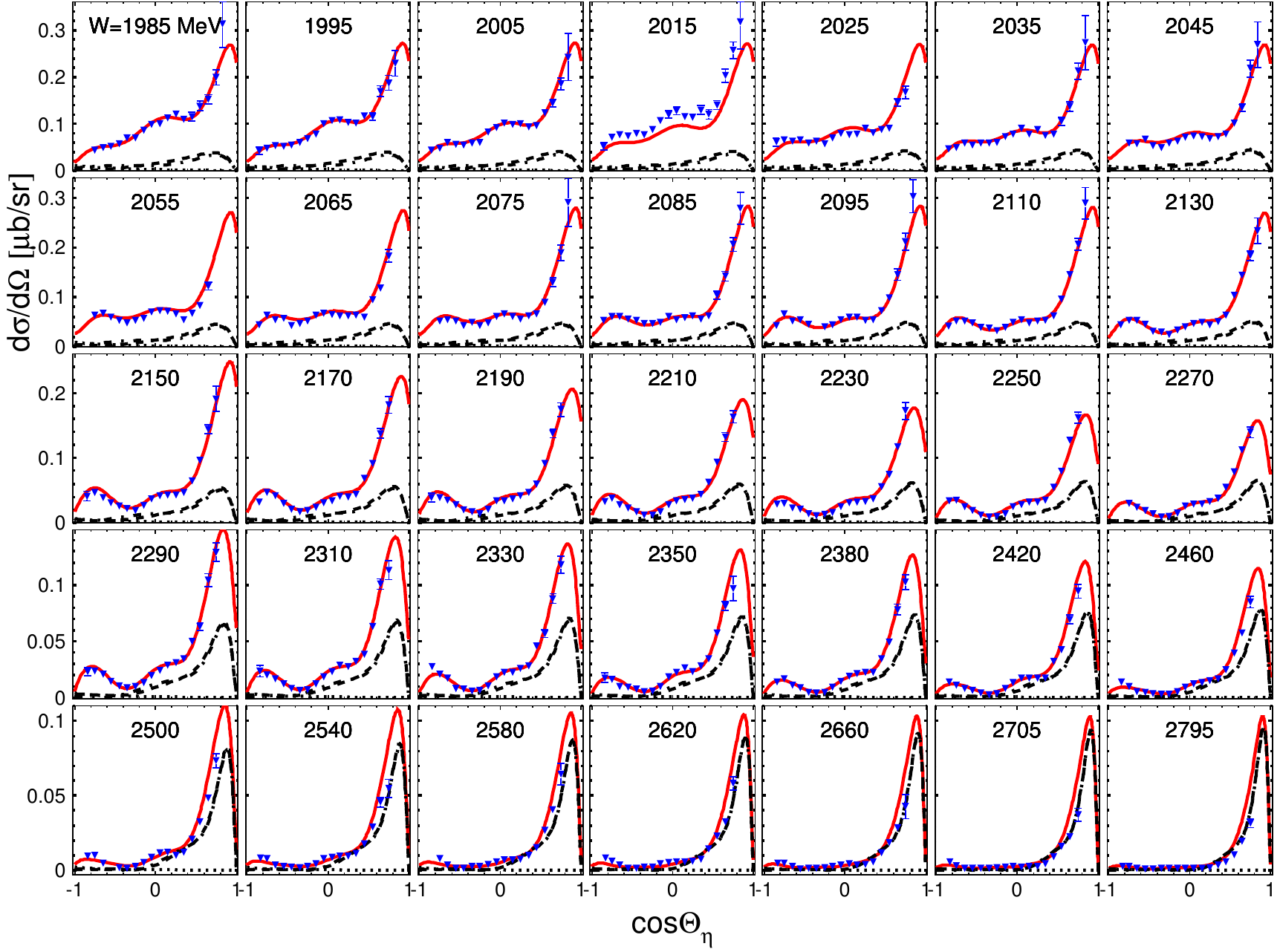}}
\caption{Same as in Fig.~\ref{fig:eta-p_dcs1_mami} for c.m. energies
1985~MeV~$\le W \le 2795$~MeV. The data are from
CLAS~\cite{Williams:2009yj}.} \label{fig:eta-p_dcs_clas}
\end{center}
\end{figure*}

In this subsection we turn to differential cross sections and
polarization observables for $\eta$ production on the proton target.
Figs.~\ref{fig:eta-p_dcs1_mami} - \ref{fig:eta-p_dcs_clas} display
the differential cross section for the reaction $\gamma p\to\eta p$
as function of the cosine of the c.m. scattering angle in comparison
to the full solution (solid red curves). We point out that our full
solution provides an excellent description of the data over the
whole energy and angular range, including the $K\Sigma$ and
$\eta^\prime$ cusp regions, $W\approx1680$~MeV and
$W\approx1890$~MeV, respectively. It is informative to observe the
impact of the background contributions, Born (dotted curves), Regge
(dash-dotted), and Born + Regge (dashed) contributions. Throughout
the whole energy range of MAMI data~\cite{Kashevarov:2017} and well
into the CLAS~\cite{Williams:2009yj} energy range, for $W\leq
2200$~MeV the background contributions are quite small, although
background-resonance interference may be non-negligible. This
observation is of importance to assess the issue of double counting
mentioned in the introduction in view of using a modified Regge
amplitude in the resonance region. We interpret the small relative
impact of the background amplitudes for $W\leq 2200$~MeV as an
indication that double counting does not pose problems for those
energies, and only the two highest resonances of our analysis,
$N(2190)\frac{7}{2}^-$ and $N(2250)\frac{9}{2}^-$, may be severely
affected by that problem. Above $W\approx2500$~MeV the Regge
contribution becomes dominant. We postpone a detailed study of the
contribution of the modified Regge background to resonant partial
waves in the transition region 2200~MeV~$\le W \le$~2500~MeV and
extraction of higher resonance parameters to the upcoming work.

\begin{figure*}[!ht]
\begin{center}
\resizebox{0.8\textwidth}{!}{\includegraphics{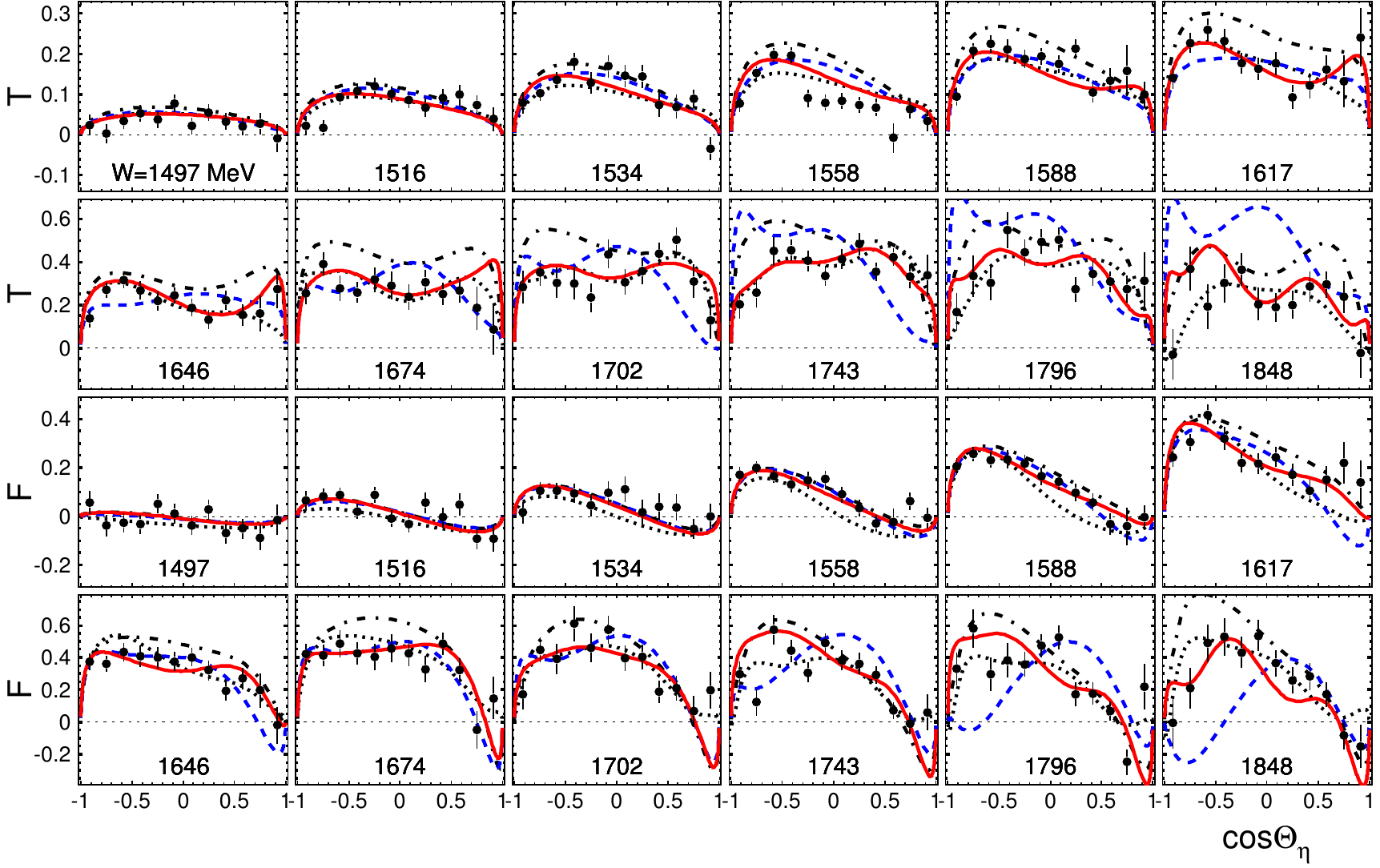}}
\caption{Target polarization $T$ (upper panels) and beam-target
polarization $F$ (lower panels) asymmetries for $(\gamma,\eta)$ on
the proton. The solid red lines show our full solution. Results of
other PWA are shown by the black dash-dotted
(BnGa~\cite{Anisovich:2018}), the black dotted
(J{\"u}Bo~\cite{Ronchen:2018}), and the blue dashed
(KSU~\cite{KSU2018}) lines. The data points are from
A2MAMI~\cite{Akondi:2014}.} \label{fig:eta-p_tf}
\end{center}
\end{figure*}

Polarization observables are much more critical tests for
models than total and differential cross sections as they are
sensitive to real and imaginary parts of interferences of
amplitudes. Fig.~\ref{fig:eta-p_tf} shows a comparison of
EtaMAID2018, BnGa, J\"{u}Bo and KSU models to data on target
polarization $T$ and beam-target polarization $F$ asymmetries from
A2MAMI for 1497~MeV~$\le W \le$~1848~MeV as function of cosine of
the c.m. scattering angle. It is seen that our solution describes
the data nicely for all energies and in the full angular range,
whereas other models show considerable deviations from data for
$W\geq1600$~MeV. We observe a significant spread between data points
in some neighboring angular bins, so more precise and
self-consistent data on this observable will help discriminating
among the models.

\begin{figure*}[!ht]
\begin{center}
\resizebox{0.7\textwidth}{!}{\includegraphics{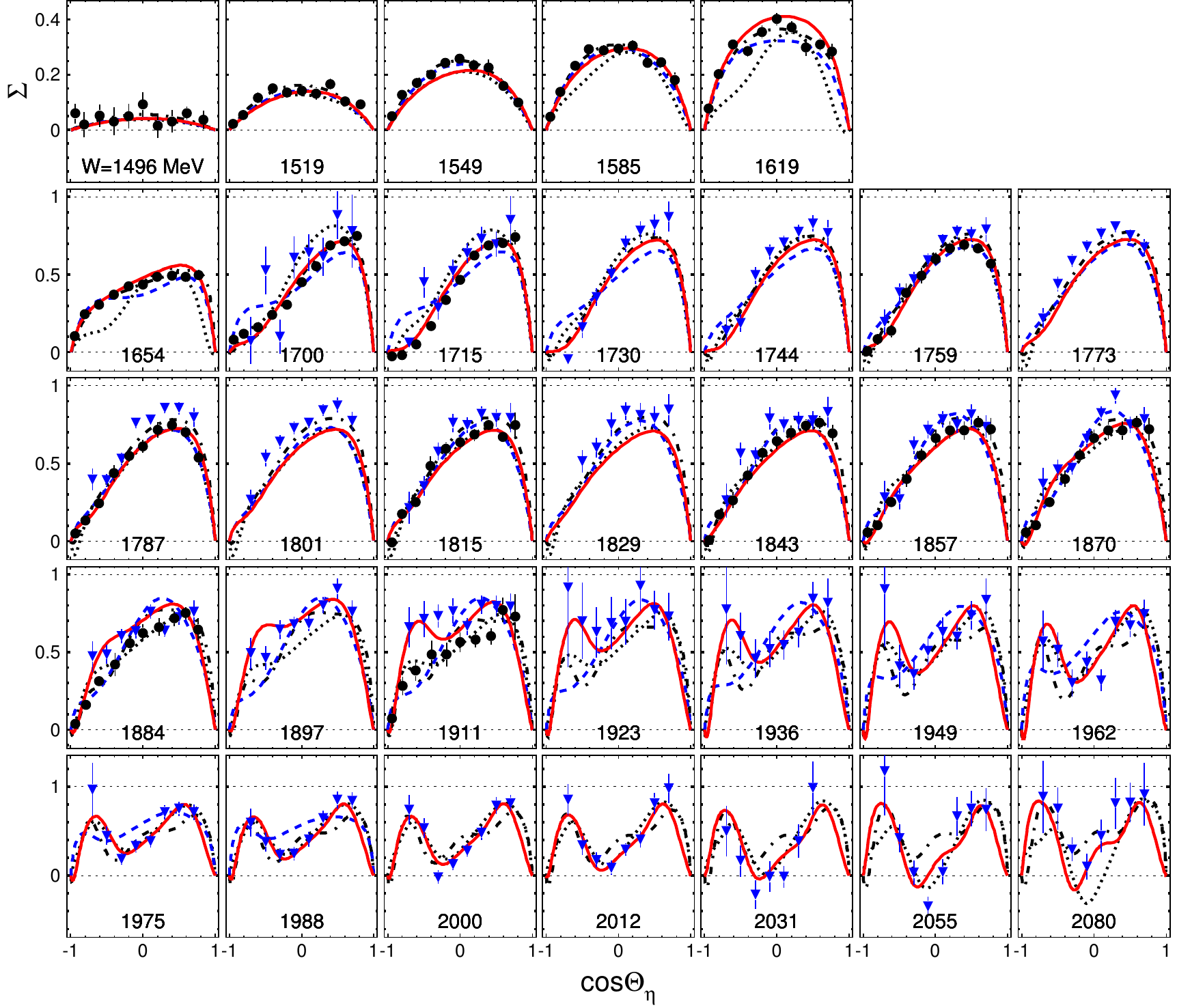}}
\caption{Photon beam asymmetry $\Sigma$ for $(\gamma,\eta)$ on the
proton. The black circles and blue triangles are data from
GRAAL~\cite{Bartalini:2007fg} and CLAS~\cite{Collins:2017}
respectively. Notation of the curves are as in
Fig.~\ref{fig:eta-p_tf}. } \label{fig:eta-p_sigma}
\end{center}
\end{figure*}

In Fig.~\ref{fig:eta-p_sigma}, data on the photon beam asymmetry
$\Sigma$ from GRAAL~\cite{Bartalini:2007fg} and
CLAS~\cite{Collins:2017} in comparison with models are shown. The
two data sets show a disagreement in several energy bins in the
overlap region 1700~MeV~$\le W \le$~1900~MeV which makes it
difficult to judge the quality of the model description of the data.
The J\"{u}Bo model fails to reproduce the high quality GRAAL data at
lower energies, especially at backward angles, while all other
models describe that energy region successfully. At highest
energies, this asymmetry shows a peculiar shape, peaks at forward
and backward angles and a dip around $90^\circ$. Since EtaMAID2018,
J\"{u}Bo and BnGa models deviate somewhat from each other, better
statistics data on $\Sigma$ at these energies will be helpful.

%

\begin{figure*}[!ht]
\begin{center}
\resizebox{0.7\textwidth}{!}{\includegraphics{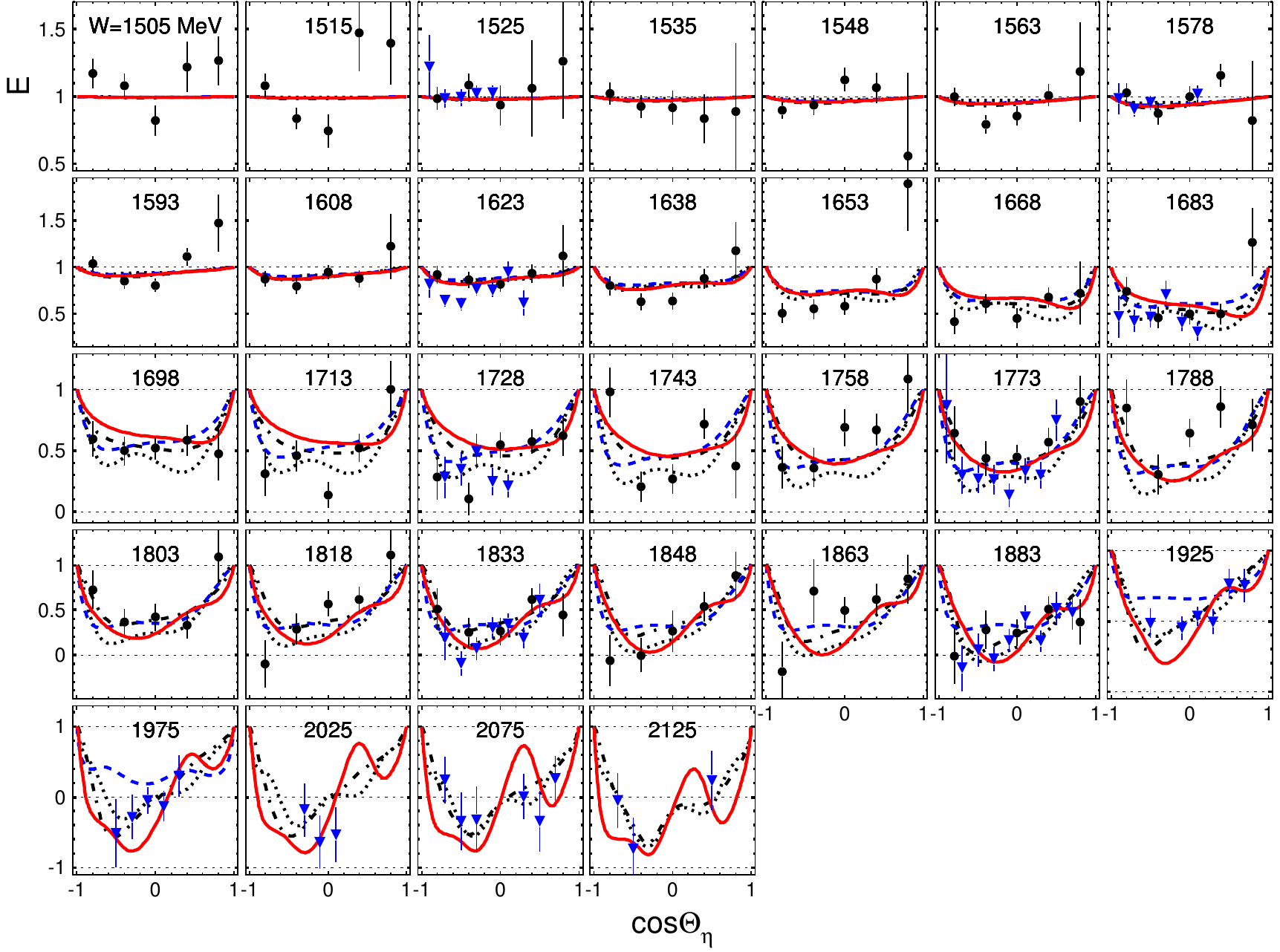}}
\caption{Beam-target polarization asymmetry $E$ for $(\gamma,\eta)$
on the proton. The black circles and blue triangles are data from
MAMI~\cite{Witthauer:2017} and CLAS~\cite{Senderovich:2016}
respectively. Notation of the curves are as in
Fig.~\ref{fig:eta-p_tf}. } \label{fig:eta-p_e}
\end{center}
\end{figure*}

For the beam-target polarization asymmetry $E$,
Fig.~\ref{fig:eta-p_e}, the situation is similar: All models give
very similar results for $W\leq1700$~MeV but start deviating above.
Current quality of the data does not permit to draw firm conclusions
from this comparison.

\begin{figure*}[!ht]
\begin{center}
\resizebox{0.7\textwidth}{!}{\includegraphics{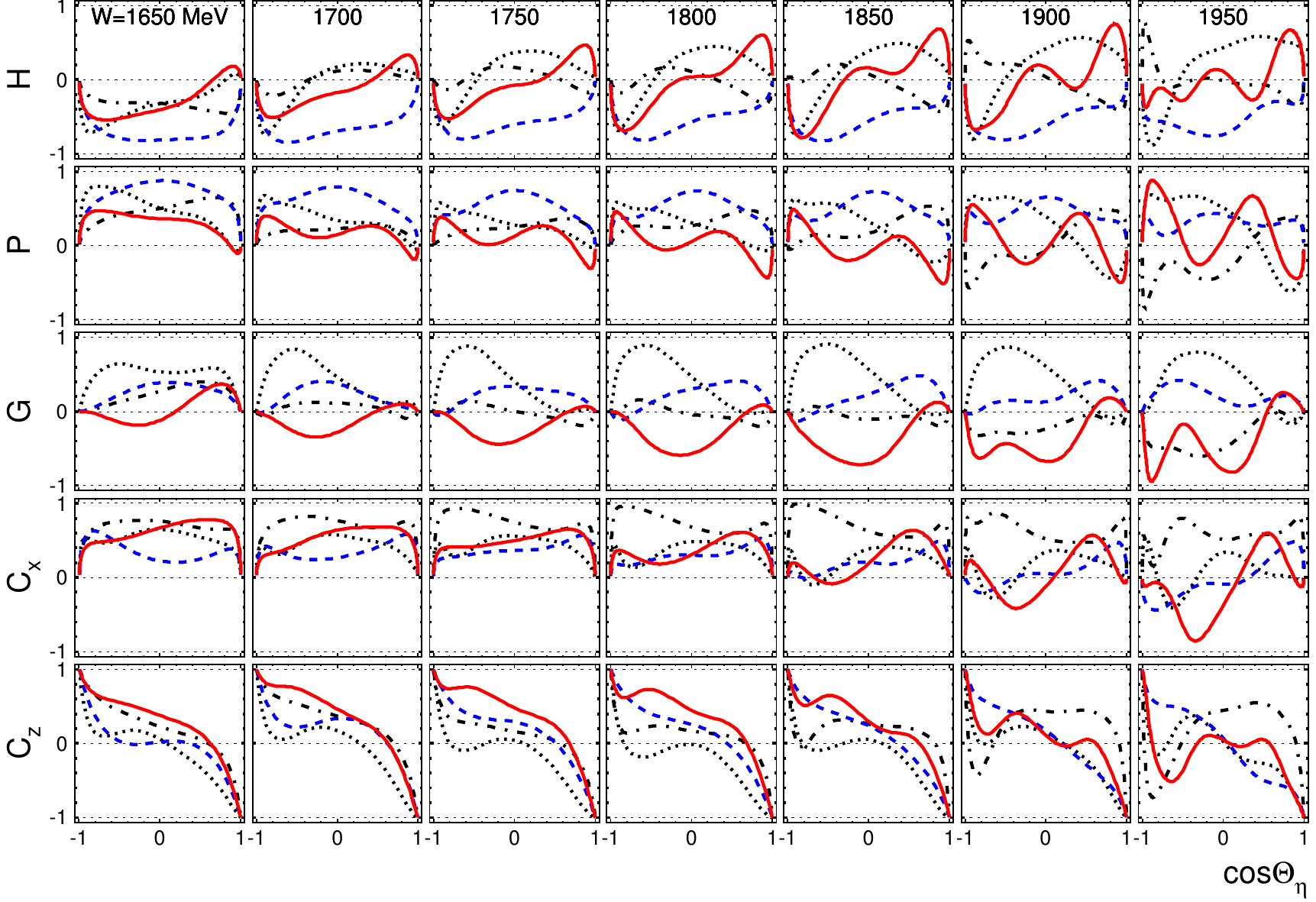}}
\caption{ Predictions for P, H, G, $C_x$, and $C_z$ observables for
$(\gamma,\eta)$ on the proton. Notations of the curves are as in
Fig.~\ref{fig:eta-p_tf}. } \label{fig:eta-p_predict}
\end{center}
\end{figure*}

In view of this general sensitivity of polarization
observables  to models, in Fig.~\ref{fig:eta-p_predict} we plot
predictions of the four models for the observables $P$, $H$, $G$,
$C_x$, and $C_z$ for $(\gamma,\eta)$ on the proton, for which no
data exist. All these observables look very promising for
discriminating the models, especially at higher energies.

It is interesting to note that in some energy regions, the
observables $P$ and $H$ are almost identical up to a sign. As can be
seen from $P+H$, Eq.~(A9) together with the multipole expansion,
Eq.~(\ref{eq:multipoles}), all $S$-wave contributions cancel exactly
and the leading terms are imaginary parts of $P-D$ interferences. In
EtaMAID a sizable deviation between $P$ and $-H$ is only seen at
higher energies in Fig.~\ref{fig:eta-p_predict}, while BnGa and
J\"uBo exhibit larger differences.

\subsection{\boldmath Comparison with the data of $d\sigma/d\Omega, \Sigma, E$
for $\gamma\,n \rightarrow \eta\,n$}

\begin{figure*}[!ht]
\begin{center}
\resizebox{0.7\textwidth}{!}{\includegraphics{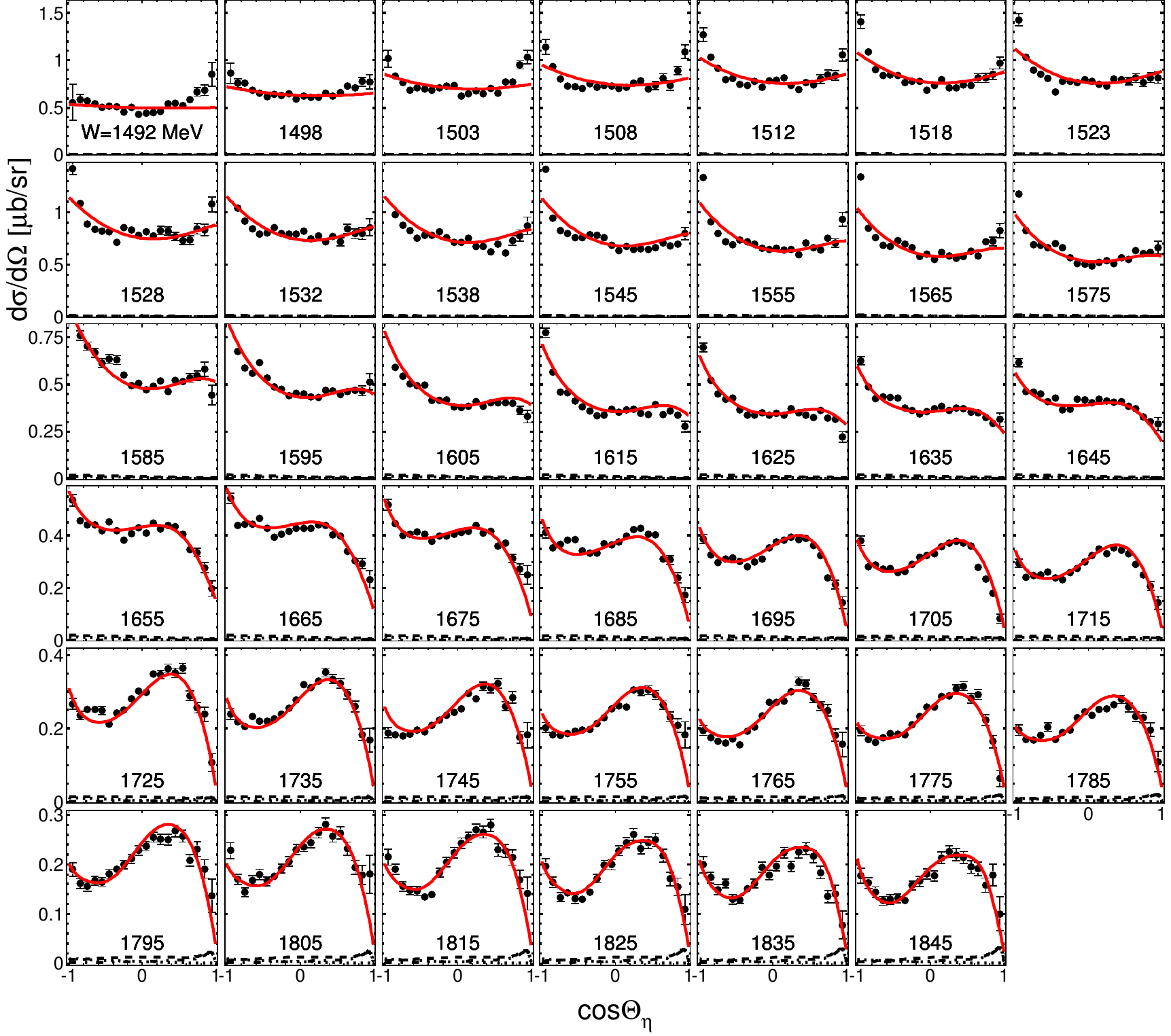}}
\caption{Differential cross section for $(\gamma,\eta)$ on the
neutron. The red solid lines show our full solutions, whereas the
black dotted, dash-dotted and dashed lines are Born terms, Regge,
and full background, respectively. The data are from
A2MAMI~\cite{Werthmueller:2014}. } \label{fig:eta-n_dcs}
\end{center}
\end{figure*}

\begin{figure*}[!ht]
\resizebox{0.7\textwidth}{!}{\includegraphics{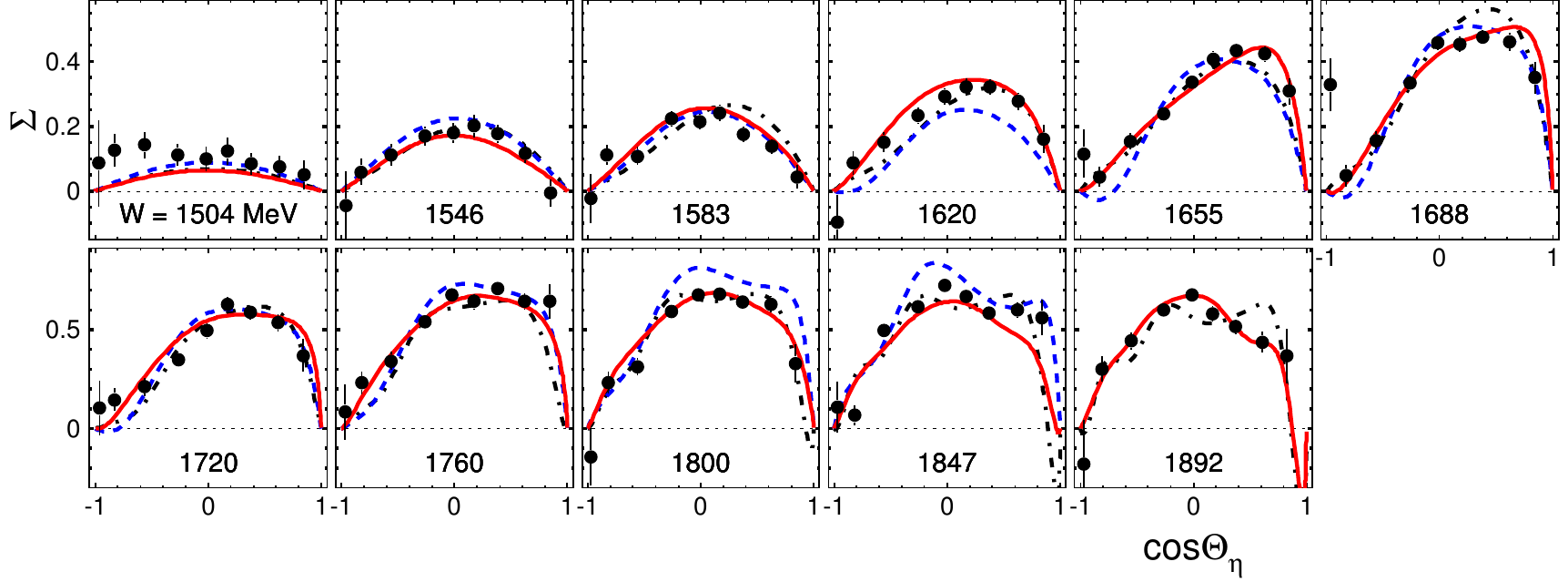}}
\caption{Photon beam asymmetry $\Sigma$ for $(\gamma,\eta)$ on the
neutron. The data are from GRAAL~\cite{Fantini:2008}. The solid red
lines show our full solution. Results of other PWA analyses are
shown by the black dotted (BnGa~\cite{Anisovich:2017}), and blue
dashed (KSU~\cite{KSU2018}) lines. } \label{fig:eta-n_sigma}
\end{figure*}

\begin{figure*}[!ht]
\resizebox{0.7\textwidth}{!}{\includegraphics{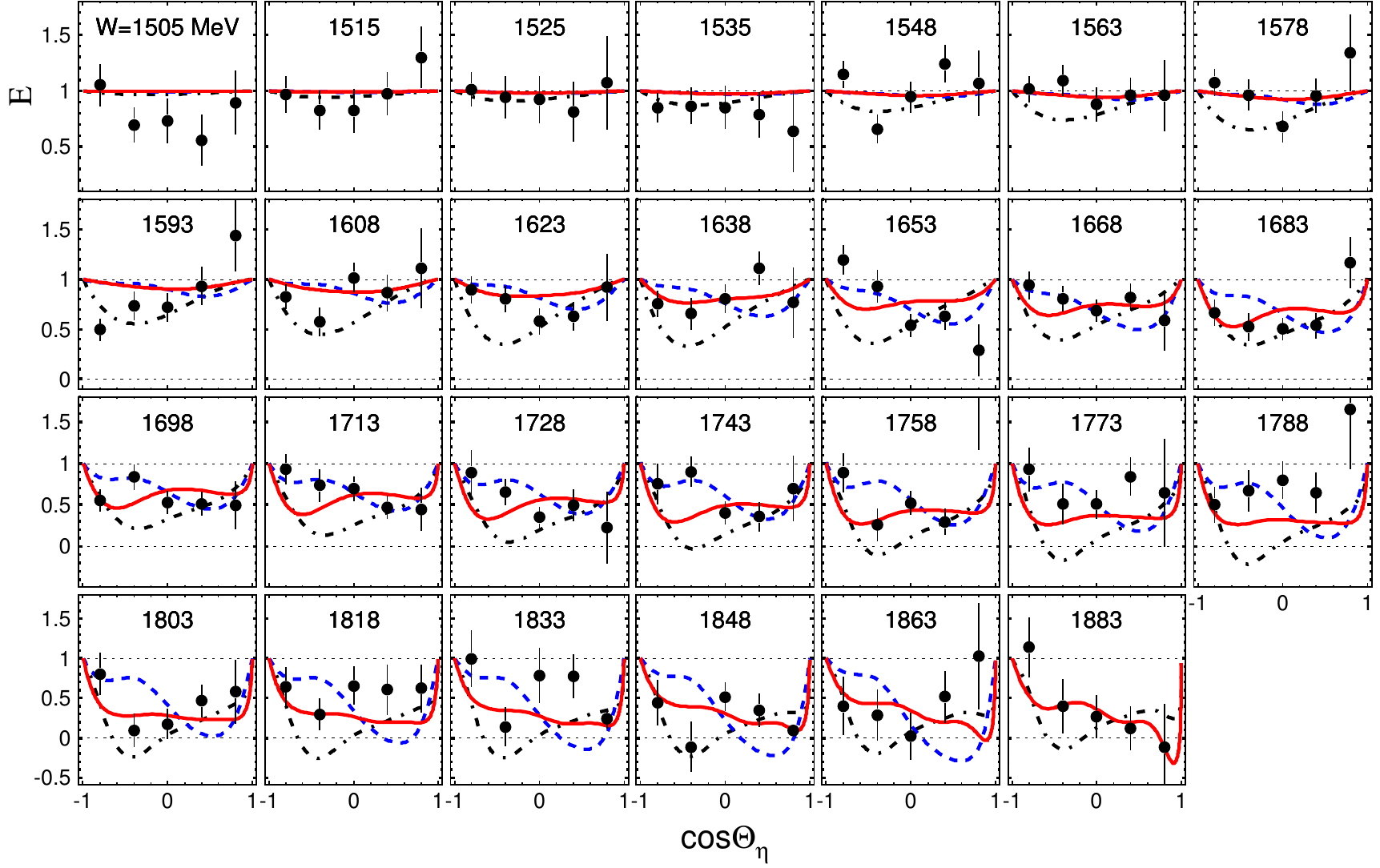}}
\caption{Beam-target polarization asymmetry $E$ for $(\gamma,\eta)$
on the neutron. The data are from A2MAMI~\cite{Witthauer:2017}.
Notation of the curves are as in Fig.~\ref{fig:eta-n_sigma}. }
\label{fig:eta-n_e}
\end{figure*}

Results for $\gamma\,n \rightarrow \eta\,n$ reaction are shown in
Figs.~\ref{fig:eta-n_dcs} - \ref{fig:eta-n_e}. Similar to the proton
target, we observe a very good description of the differential cross
section data in the full energy range where very precise A2MAMI data
are available, with the mere exception of some very backward or very
forward (in the c.m. frame) data points at low energies where
nuclear effects may lead to some systematic effects that were not
fully accounted for~\cite{Krusche2014}. This lack of strength in the
extreme backward and forward kinematics is however not reflected in
the description of the total cross section, see Fig.~\ref{fig:tcs}.

Polarization observables $\Sigma$ and $E$ are shown in
Figs.~\ref{fig:eta-n_sigma} and \ref{fig:eta-n_e}, respectively. Our
model describes the data nicely, although at present the
uncertainties of the data do not allow for a definitive comparison
of the models and will help to remove ambiguities, which are still
visible in the partial wave analysis, see sect.~\ref{sec:pwa}.

%

\begin{figure*}[!ht]
\begin{center}
\resizebox{0.7\textwidth}{!}{\includegraphics{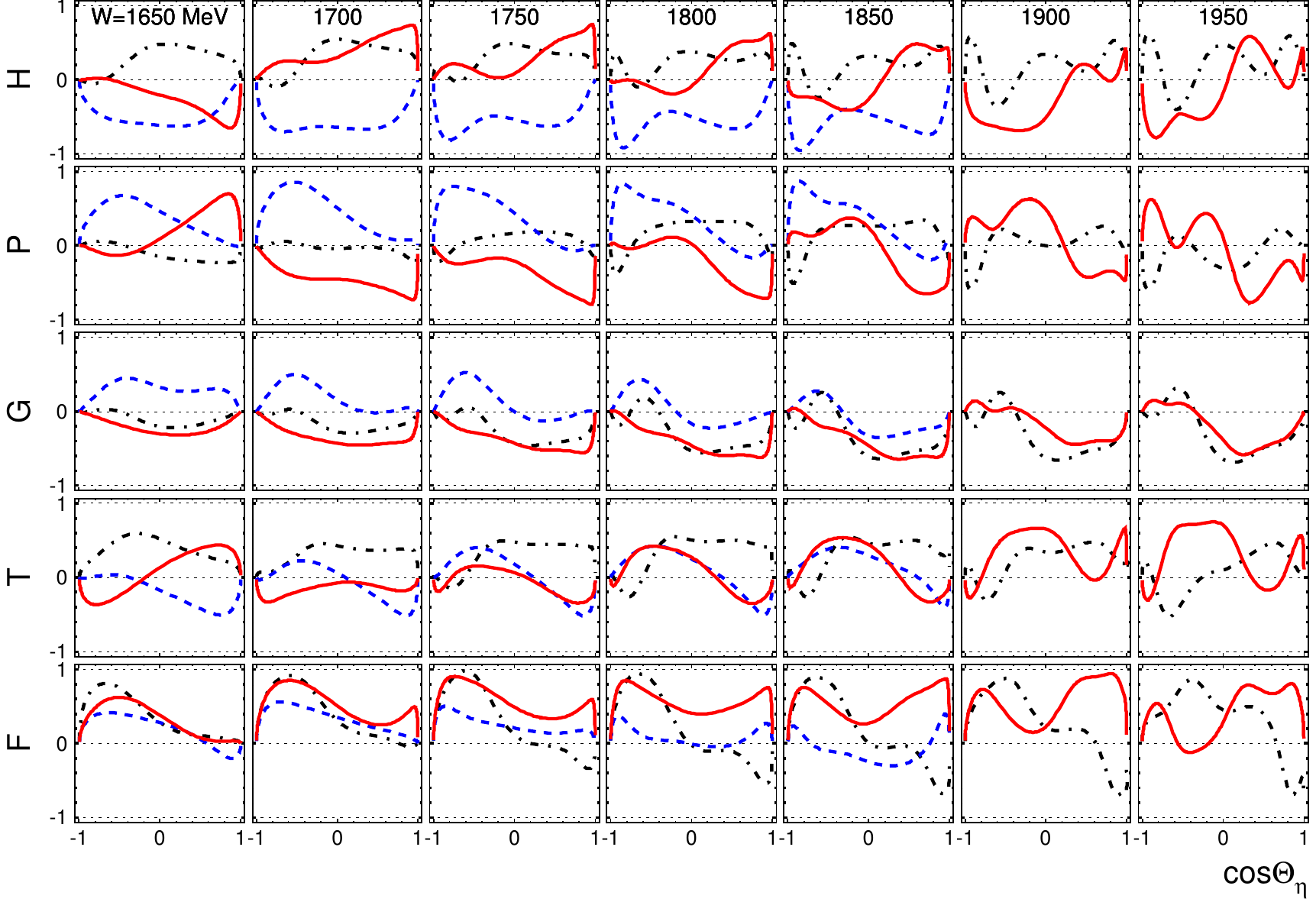}}
\caption{ Predictions for P, H, G, T and F observables for
$(\gamma,\eta)$ on the neutron. Notation of the curves are as in
Fig.~\ref{fig:eta-n_sigma}. } \label{fig:eta-n_predict}
\end{center}
\end{figure*}

In Fig. \ref{fig:eta-n_predict} we plot the polarization
observables $P$, $H$, $G$, $T$, and $F$ for $(\gamma,\eta)$ on the
neutron. As for the proton target, data on these observables will
yield a crucial test for our understanding of the models. As also
discussed for the proton, the symmetry between the $P$ and $H$
observables is again much more pronounced for EtaMAID than for other
solutions, a signature for a stronger $S$-wave dominance in EtaMAID.

\subsection{\boldmath Comparison with the data of $d\sigma/d\Omega$ and $\Sigma$ for
$\gamma\,p \rightarrow \eta^\prime\,p$ and $\gamma\,n \rightarrow
\eta^\prime\,n$}

\begin{figure*}[!ht]
\begin{center}
\resizebox{0.7\textwidth}{!}{\includegraphics{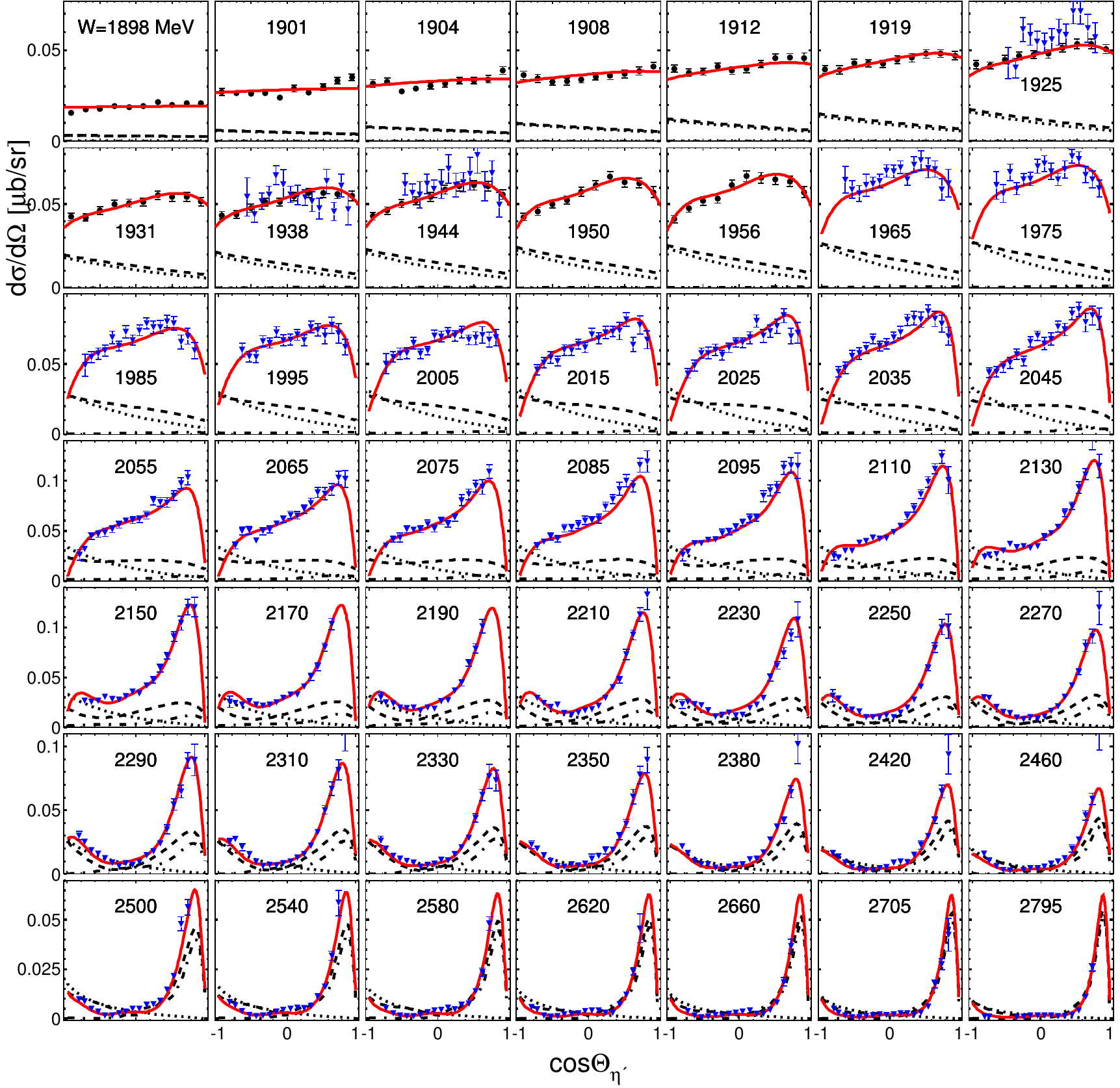}}
\end{center}
\caption{Differential cross section for $(\gamma,\eta^\prime)$ on
the proton. The red solid lines show our full solutions, whereas the
black dotted, dash-dotted, and dashed lines are Born terms, Regge,
and full background, respectively. The black circles are data from
A2MAMI~\cite{Kashevarov:2017} and blue triangles from
CLAS~\cite{Williams:2009yj}. } \label{fig:etapr-p_dcs}
\end{figure*}

\begin{figure*}[!ht]
\resizebox{0.6\textwidth}{!}{\includegraphics{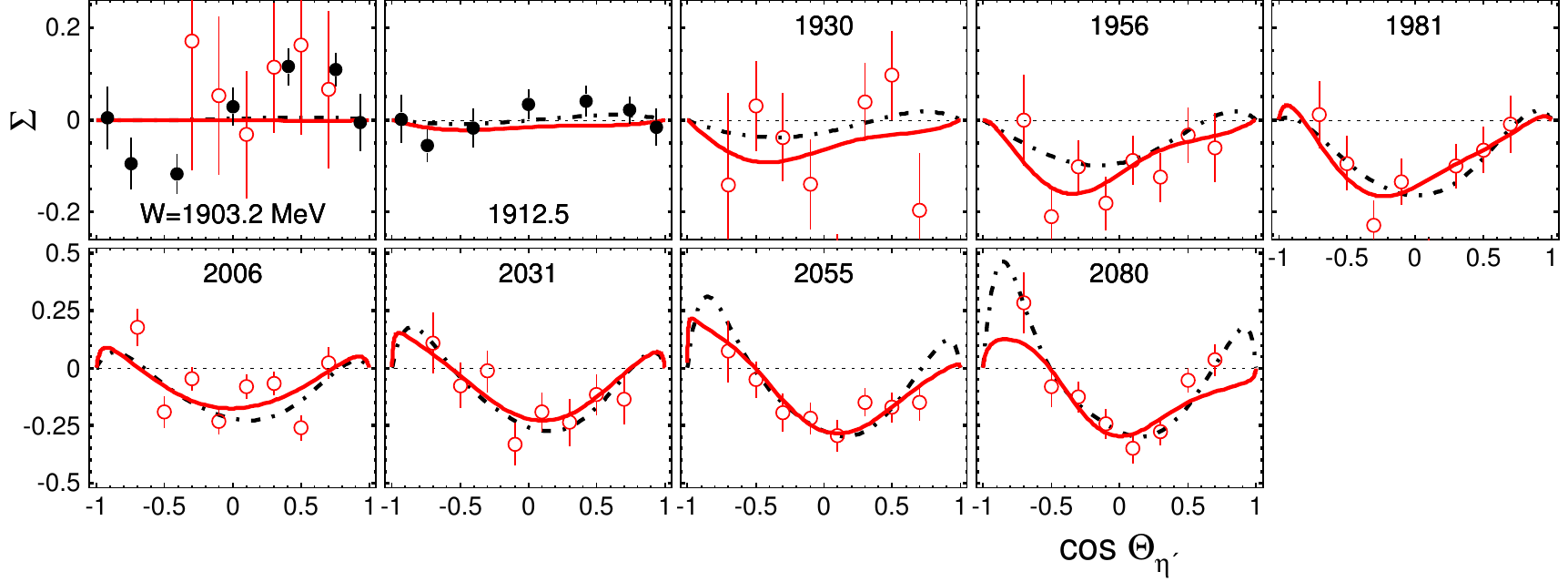}}
\caption{Photon beam asymmetry $\Sigma$ for $(\gamma,\eta^\prime)$
on the proton. The black full and red opened circles are data from
GRAAL~\cite{Sandri:2015} and CLAS~\cite{Collins:2017}, respectively.
Notation of the curves as in Fig.~\ref{fig:eta-p_tf}. }
\label{fig:etapr-p_sigma}
\end{figure*}

Results for $\gamma\,p \rightarrow \eta^\prime\,p$ are presented in
Figs.~\ref{fig:etapr-p_dcs} and ~\ref{fig:etapr-p_sigma}. For the
differential cross sections, we show the impact of the background
contributions, Born (dotted), Regge (dash-dotted), and Born + Regge
(dashed) contributions. Because of the higher threshold for
$\eta^\prime$ production, the background has much more relative
impact than for $\eta$ production. In particular, we observe that
the Born contribution, more precisely, the $u$-channel Born diagram,
gives a very sizable contribution at backward angles for
$W\geq$1912~MeV. We note in this respect that a reggeization of the
$u$-channel nucleon exchange is worthwhile to study in the upcoming
work. With this in mind, we notice a very good description of the
data in the full energy and angular range by our solution. Apparent
disagreement between A2MAMI and CLAS data, where the two data sets
overlap, play little role at present due to a much better statistics
of Mainz data.

The new solution reproduces all data for this reaction quite well,
with the exception of the first two energy bins for $\Sigma$, where
GRAAL data show a clear $\sim\sin\theta$ structure, see
Fig.~\ref{fig:etapr-p_sigma}. CLAS data are presently too uncertain
to confirm or disprove this behavior. The models do not show any
flexibility in that energy bin to possibly describe such a sine
structure. This will be further discussed in
section~\ref{sect:NarrowResonances}.
%

\begin{figure*}[!ht]
\begin{center}
\resizebox{0.6\textwidth}{!}{\includegraphics{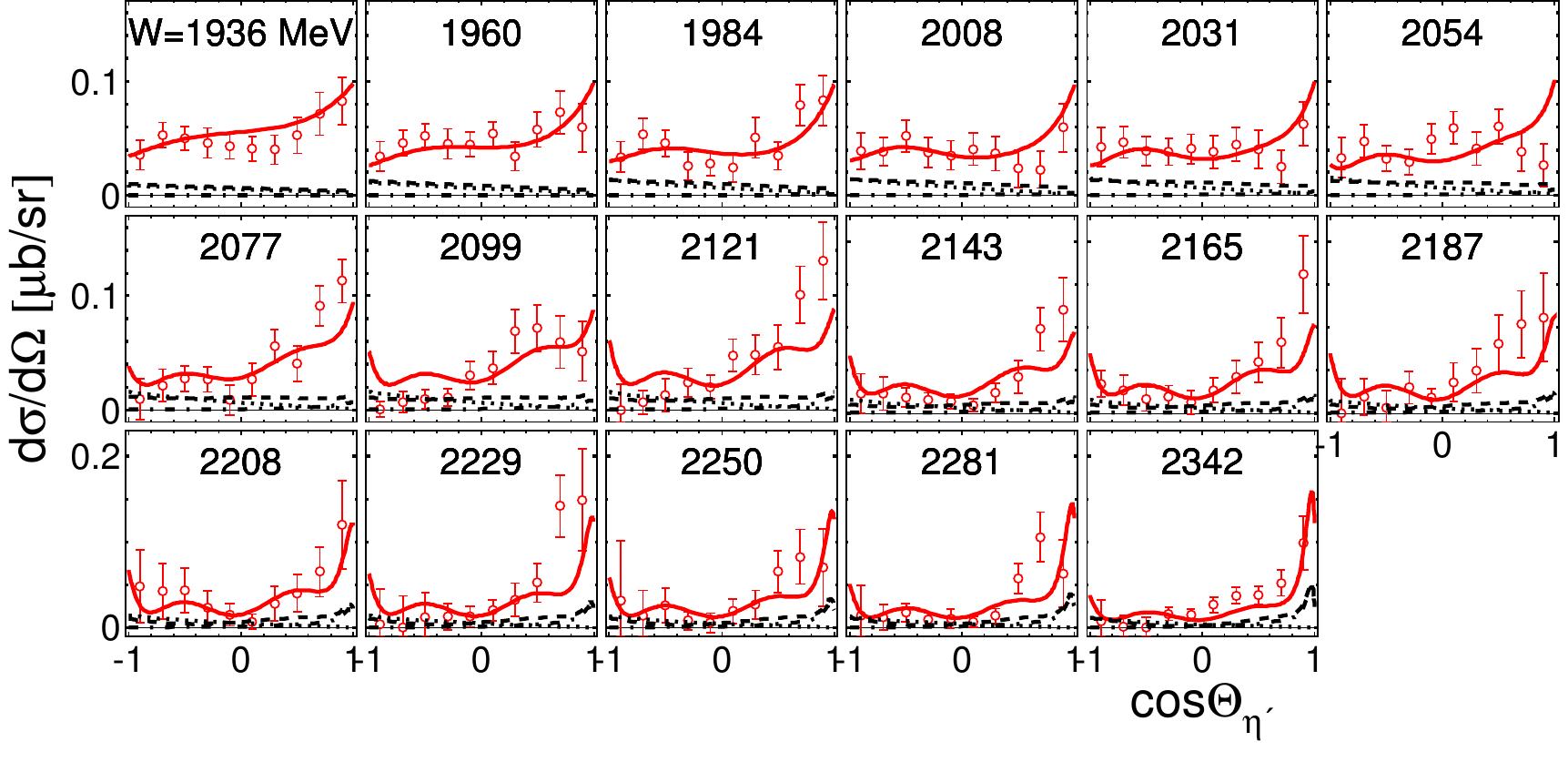}}
\caption{Differential cross section for $(\gamma,\eta^\prime)$ on
the neutron. The solid red lines show our full solutions, whereas
the black dotted, dash-dotted, and dashed lines are Born terms,
Regge, and full background, respectively. The data points are from
CBELSA/TAPS~\cite{Jaegle:2011}. } \label{fig:etapr-n_dcs}
\end{center}
\end{figure*}

For $\gamma\,n \rightarrow \eta^\prime\,n$ only one data set exists,
the unpolarized cross sections measured by the CBELSA/TAPS
Collaboration~\cite{Jaegle:2011}. The data together with our full
solution are presented in Fig~\ref{fig:etapr-n_dcs}. There is some
disagreement in the range $W = 2077 - 2121$ MeV. This channel has
not been analyzed by other PWA groups.

\section{\boldmath Narrow resonances in $\eta$ and $\eta^\prime$
photoproduction}\label{sect:NarrowResonances}

Around 2005, in $\eta$ photoproduction on the neutron a bump in the
total cross section in the vicinity of $W=1685$~MeV was observed
that was especially pronounced in the cross section ratio
$\sigma_n/\sigma_p$~\cite{Kuznetsov:2006kt,Kuznetsov:2010as,Werthmuller:2013rba}.
Many attempts were made to explain this effect, some explanations
introduced a new narrow $N(1685)$ resonance, where, however, the
quantum numbers were not uniquely
determined~\cite{Arndt:2003ga,Anisovich:2013sva}. Mostly a $P_{11}$
resonance was assumed, which also matched the position of a
predicted non-strange pentaquark state~\cite{Diakonov:1997mm}. The
range of the width was determined as $15-45$~MeV. Due to further
lack of evidence and more conventional explanations of the bump
structure in terms of interferences of $S_{11}$ resonances, in 2016
PDG has decided to remove this state from the listings. For further
reading see Ref.~\cite{Krusche2014}.

Here we want to discuss a further attempt to study possible
consequences from a narrow $N(1900)$ state, a few MeV above
$\eta^\prime$ threshold. Anisovich et
al.~\cite{Anisovich:2018-narrow} have shown that a narrow
$N(1900)\frac{3}{2}^-$ $D_{13}$ resonance with a mass $M_R=1900\pm
1$~MeV and a total width of less than 3~MeV can explain the
unexpected energy and angular dependence of the differential cross
section $d\sigma/d\Omega$ from A2MAMI and of the beam asymmetry
$\Sigma$ from GRAAL. In our EtaMAID analysis we can confirm the
possibility for an explanation with a narrow resonance, however, in
EtaMAID we would obtain a narrow $S_{11}$ resonance with quantum
numbers $\frac{1}{2}^-$, mass $M_R=1902.6\pm 1.0$~MeV and width
$\Gamma_R=2.1\pm 0.5$~MeV.

As it was pointed out before, the photon beam asymmetry $\Sigma$
measured at GRAAL exhibits a very unexpected behavior. First of all,
it shows a nodal structure with a sinus-type shape in the angular
distribution, which is a sign of higher partial wave content
compared to the beam asymmetry in threshold $\eta$ photoproduction.
Second, it appears with a strong energy behavior, changing the
magnitude of the beam asymmetry significantly within only a few MeV.
And third, it appears very close to the $\eta^\prime$ threshold and
decreases strongly within only a few MeV. Naturally, in this region
an effect would increase in magnitude rather than decrease, when the
energy rises.

The first issue can be easily investigated by the partial wave
series of the beam asymmetry. Expanded into partial waves up to $F$
waves ($L_{max}=3$), the beam asymmetry observable $\check{\Sigma}$
(see appendix \ref{app:BG-BW}) can be expressed in its angular dependence up to
$x^4$ with $x=\mbox{cos}\,\theta$
\begin{equation}
\check{\Sigma}=\sigma_0(x)\Sigma(x)=(1-x^2)\sum_{k=0}^4 a_k\, x^k\,,
\end{equation}
where the observed nodal structure arises from the coefficient
$a_1$, which can be separated into $S-F$, $P-D$ and $D-F$
interferences of partial waves, $a_1 = a_1^{SF} + a_1^{PD} +
a_1^{DF}$. Using Eq.~(A2) and the partial wave expansion of the CGLN
amplitudes, we get in details
\begin{eqnarray}\label{eq:a1_interferences}
a_1^{S_{11}-F_{15}}&=&15\mbox{Re}\{E_{0+}^*(E_{3-}+M_{3-})\}\,,\\
a_1^{S_{11}-F_{17}}&=&15\mbox{Re}\{E_{0+}^*(E_{3+}-M_{3+})\}\,,\nonumber\\
a_1^{P_{11}-D_{15}}&=&15\mbox{Re}\{M_{1-}^*(M_{2+}-E_{2+})\}\,,\nonumber\\
a_1^{P_{13}-D_{13}}&=&18\mbox{Re}\{E_{1+}^*E_{2-}+M_{1+}^*M_{2-}\}\,,\nonumber\\
a_1^{P_{13}-D_{15}}&=&3\mbox{Re}\{-9E_{1+}^*E_{2+}+M_{1+}^*(5E_{2+}+4M_{2+})\}\,,\nonumber\\
a_1^{D_{13}-F_{15}}&=&-3\mbox{Re}\{E_{2-}^*(4E_{3-}-5M_{3-})-9M_{2-}^*M_{3-}\}\,,\nonumber\\
a_1^{D_{13}-F_{17}}&=&-15\mbox{Re}\{E_{2-}^*(5E_{3+}+M_{3+})+6M_{2-}^*M_{3+}\}\,,\nonumber\\
a_1^{D_{15}-F_{15}}&=&-\frac{189}{2}\mbox{Re}\{E_{2+}^*E_{3-}+M_{2+}^*M_{3-}\}\,.\nonumber
\end{eqnarray}
Interferences of $P_{11}-D_{13}$ and $D_{15}-F_{17}$ do not
contribute.

\begin{figure}    
\begin{center}
\resizebox{0.5\textwidth}{!}{\includegraphics{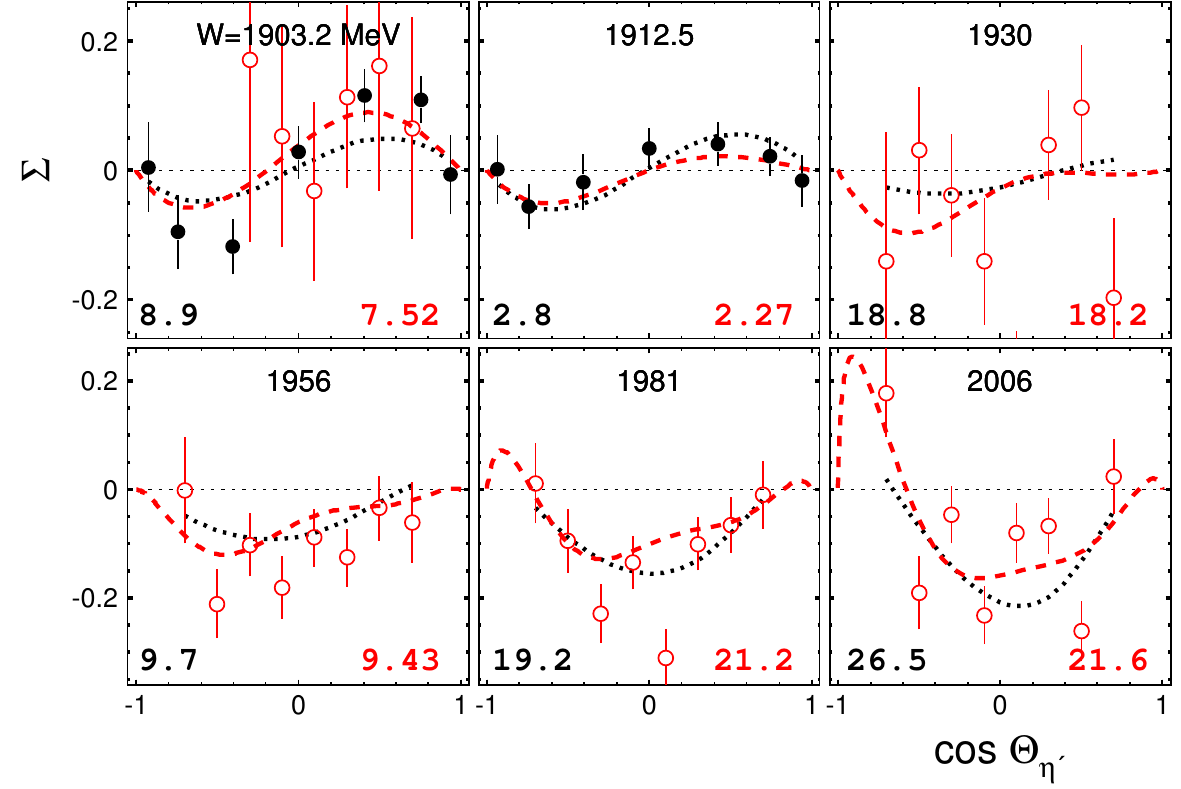}}
\caption{Photon beam asymmetry $\Sigma$ for $(\gamma,\eta^\prime)$
on the proton for selected energy bins. The black full and red open
circles are data from GRAAL~\cite{Sandri:2015} and
CLAS~\cite{Collins:2017}, respectively. The dashed red lines show
our solution with a narrow $S_{11}(1900)$ resonance and
corresponding $\chi^2$ in the lower right corner for each panel and
the black dotted lines BnGa~\cite{Anisovich:2018-narrow} with a
narrow $D_{13}(1900)$ resonance and $\chi^2$ on the left. }
\label{fig:etapr-p_sigma_narrow}
\end{center}
\end{figure}

In Fig.~\ref{fig:etapr-p_sigma_narrow} we show our result with a
narrow $S_{11}(1900)$ and the BnGa solution with a narrow
$D_{13}(1900)$ for $\eta^\prime$ photoproduction on the proton. Both
solutions can describe the GRAAL data similarly well, whereas
without a narrow resonance both solutions predict an almost zero
value for the threshold beam asymmetry, see
Fig.~\ref{fig:etapr-p_sigma}. According to the multipole expansion
of the $a_1$ coefficient, Eq.~(\ref{eq:a1_interferences}), the nodal
structure of the angular dependence of the beam asymmetry is
explained with a $S_{11}-F_{15}$ interference in EtaMAID and with a
$P_{13}-D_{13}$ interference in BnGa.

\begin{figure}    
\begin{center}
\resizebox{0.5\textwidth}{!}{\includegraphics{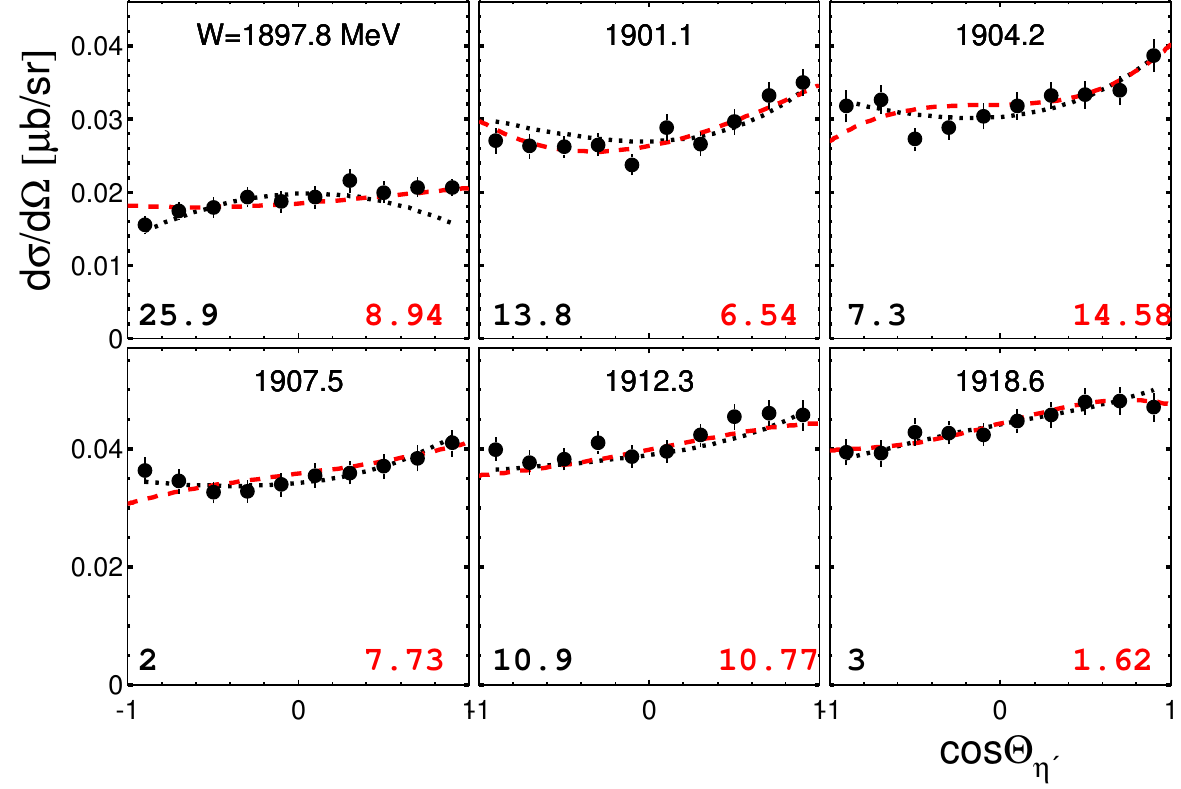}}
\caption{Differential cross section for $(\gamma,\eta^\prime)$ on
the proton for selected energy bins. The black circles are data from
A2MAMI~\cite{Kashevarov:2017}. Notations are as in
Fig.~\ref{fig:etapr-p_sigma_narrow}. }
\label{fig:etapr-p_dcs_narrow}
\end{center}
\end{figure}

Besides the beam asymmetry, also the differential cross section
exhibits small unexplained structures in the standard solutions, see
Fig.~\ref{fig:etapr-p_dcs}. This is also much improved with the
inclusion of a narrow resonance as shown in
Fig.~\ref{fig:etapr-p_dcs_narrow}.

\begin{figure*}[!ht]
\begin{center}
\resizebox{0.65\textwidth}{!}{\includegraphics{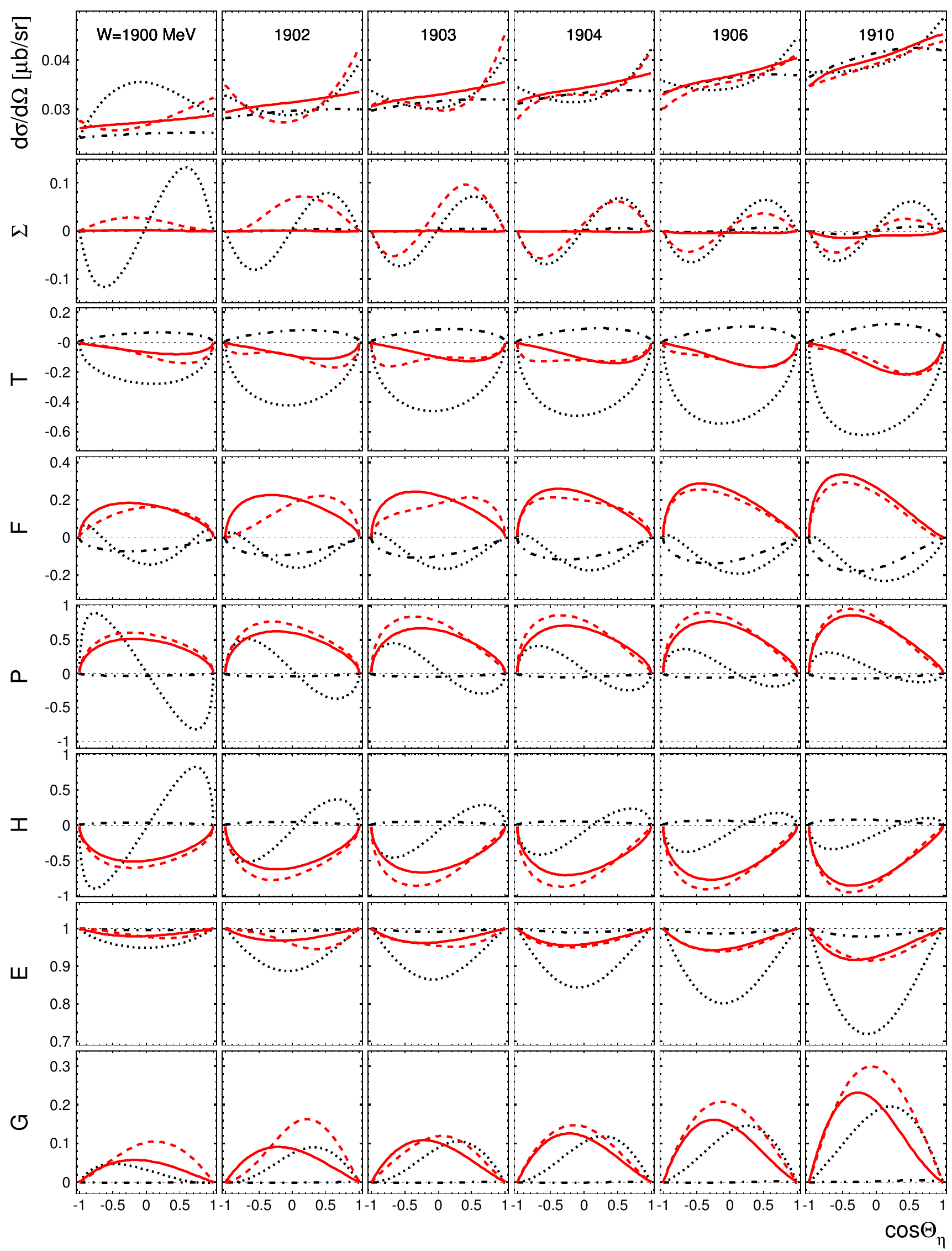}}
\caption{ Predictions for all 8 single- and beam-target double
polarization observables for $(\gamma,\eta^\prime)$ on the proton.
The red solid and black dash-dotted lines are the 2018 standard
solutions of EtaMAID and BnGa without narrow resonances. The red
dashed lines show the predictions of our EtaMAID solution with a
narrow $S_{11}(1900)$ resonance, while the black dotted lines are
obtained with the BnGa solution and a narrow $D_{13}(1900)$
resonance~\cite{Anisovich:2018-narrow}. }
\label{fig:etapr-p_predict}
\end{center}
\end{figure*}

With the two energy bins of the GRAAL beam asymmetry and the lowest
energy bins of the A2MAMI differential cross sections, the evidence
for the existence of a narrow resonance is rather weak. Especially,
as with only two observables the quantum numbers of such a state
cannot uniquely be determined. Therefore, we investigate the effects
of such narrow resonances on further not yet measured polarization
observables using beam and target polarization. In
Fig.~\ref{fig:etapr-p_predict} we show the standard solutions and
the addition of narrow resonances from EtaMAID and BnGa on the full
set of 8 polarization observables that could be measured with beam-
and target-polarization techniques, without recoil polarization
detection.

For such narrow resonances small energy bins are certainly needed.
The differential cross section, which can be expected with highest
statistics, should be re-measured and analyzed in finer energy bins.
Most important, due to the nodal structure change, is a new
measurement of the photon beam asymmetry, aiming for a similar
precision as in the GRAAL measurement. $P$ and $H$ observables,
which are almost identical up to a sign, are sensitive to a narrow
$D_{13}$ resonance, but almost independent of a narrow $S_{11}$
state. Also $T$ and $F$ observables are less sensitive but could be
obtained at MAMI with high accuracy.

\section{Partial wave amplitudes}\label{sec:pwa}

Compared to pion photoproduction, a comparison of partial waves from
different PWA is not straightforward in $\eta$ or $\eta^\prime$
photoproduction. First of all, different conventions for isospin
matrix elements are used in the literature, which appear as $+1$ in
the BnGa, J\"uBo and KSU analysis and $-1$ in the MAID and SAID
analysis. This overall sign or phase convention is denoted as
$C_{\eta N}$ in our BW ansatz of Eq.~(\ref{eq:BWres}). Second, for
$\eta$ and $\eta^\prime$ photoproduction no such convenient
unitarity constraints as the Watson Theorem exist, that determine
the phases in the low-energy regime. The only, somewhat weaker
constraints arise from channel couplings, which is more advanced in
coupled-channels approaches as BnGa, J\"uBo and KSU. In EtaMAID we
introduce coupling to pion channels only via the Breit-Wigner ansatz
and the parametrization of the energy dependent widths. E.g. the
$N(1535)\frac{1}{2}^-$ provides a very strong constraint because of
its large branchings of about $50\%$ for $\pi N$ and $40\%$ for
$\eta N$. For other partial waves, such BW constraints are much less
effective.

Therefore, even if complete experiments were performed, final
ambiguities would remain, which could not be resolved by
experimental observables. All physical observables are sums of
bi-linear products of amplitudes and conjugated amplitudes, e.g.
Re~$\{H_i(W,\theta)\,H_j^*(W,\theta)\}$, and are therefore invariant
under an overall energy- and angle-dependent phase $\phi(W,\theta)$.
This phase depends very much on the models and on couplings with
other channels, which finally will always be incomplete.

For a better comparison between the different newly updated 2018
PWA, that all use practically the same database, we have performed a
phase rotation of all amplitudes to our EtaMAID2018 phase,
\begin{equation}\label{eq:phaserotation}
H_i^{BG}\rightarrow \tilde{H}_i^{BG} = H_i^{BG}\,\cdot\, e^{i(
\phi_{H1}^{MD}(W,\theta) - \phi_{H1}^{BG}(W,\theta) )}\,,\;
i=1,\ldots,4\,,
\end{equation}
where MD stands for the EtaMAID model and BG for any other PWA, as
BnGa, J\"uBo, and KSU. For a detailed discussion of angle-dependent
phase ambiguities, see Ref.~\cite{Svarc:2017yuo,Svarc:2018aay}.

\begin{figure}[!ht]
\begin{center}
\resizebox{0.5\textwidth}{!}{\includegraphics{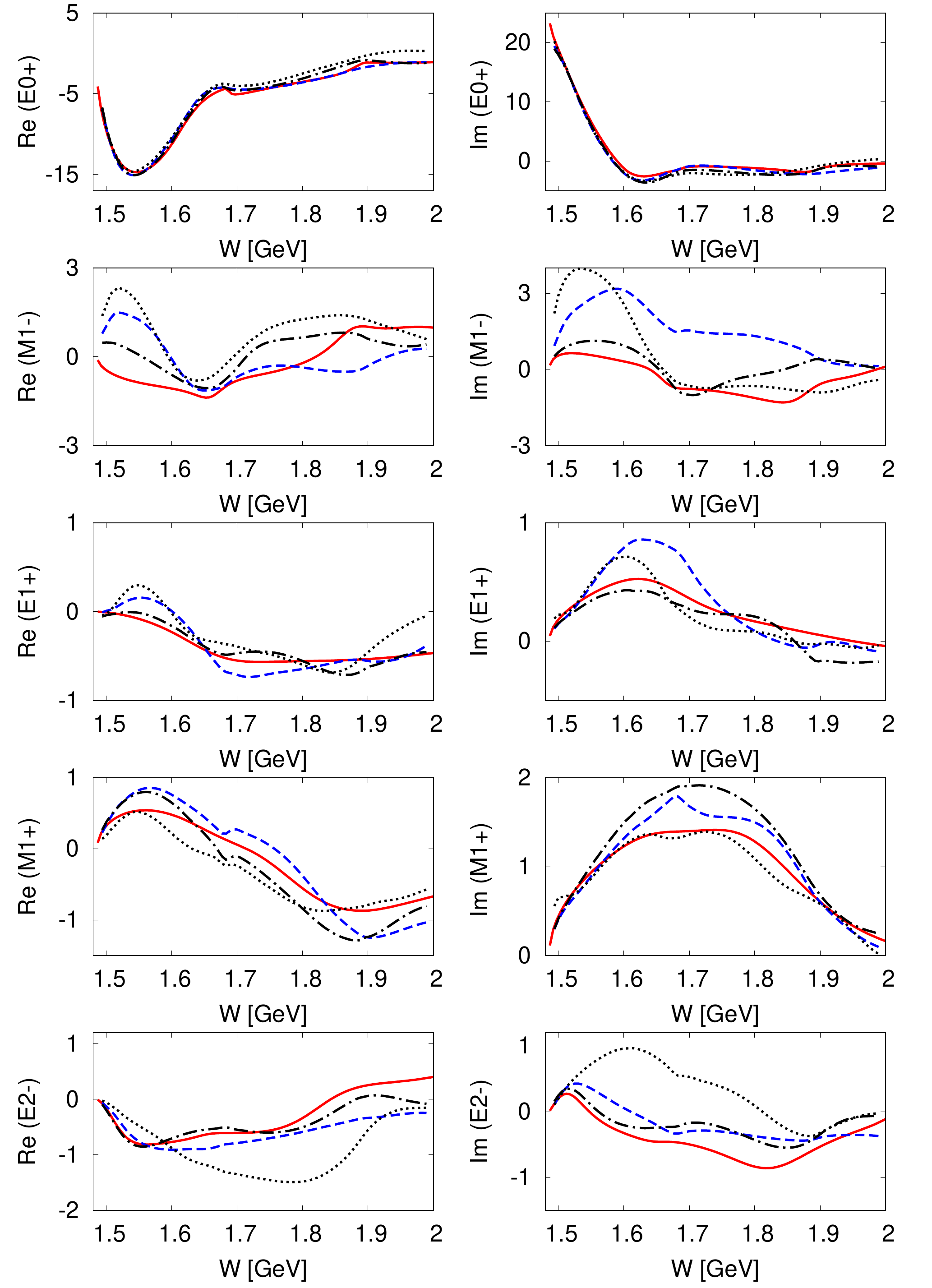}}
\caption{Comparison of $E_{0+}(S_{11})$, $M_{1-}(P_{11})$,
$E_{1+}(P_{13})$, $M_{1+}(P_{13})$, and $E_{2-}(D_{13})$ multipoles
for $\gamma p \rightarrow \eta p$, obtained from rotated helicity
amplitudes (see Eq.~(\ref{eq:phaserotation})) of different PWA. The
solid red lines show our EtaMAID2018 solution. Results of other PWA
analyses are shown by the black dash-dotted
(BnGa~\cite{Anisovich:2018}), the black dotted
(J{\"u}Bo~\cite{Ronchen:2018}), and the blue dashed
(KSU~\cite{KSU2018}) lines. The multipoles are given in units of
mfm.} \label{fig:mult1_rotated}
\end{center}
\end{figure}

\begin{figure}[!ht]
\begin{center}
\resizebox{0.5\textwidth}{!}{\includegraphics{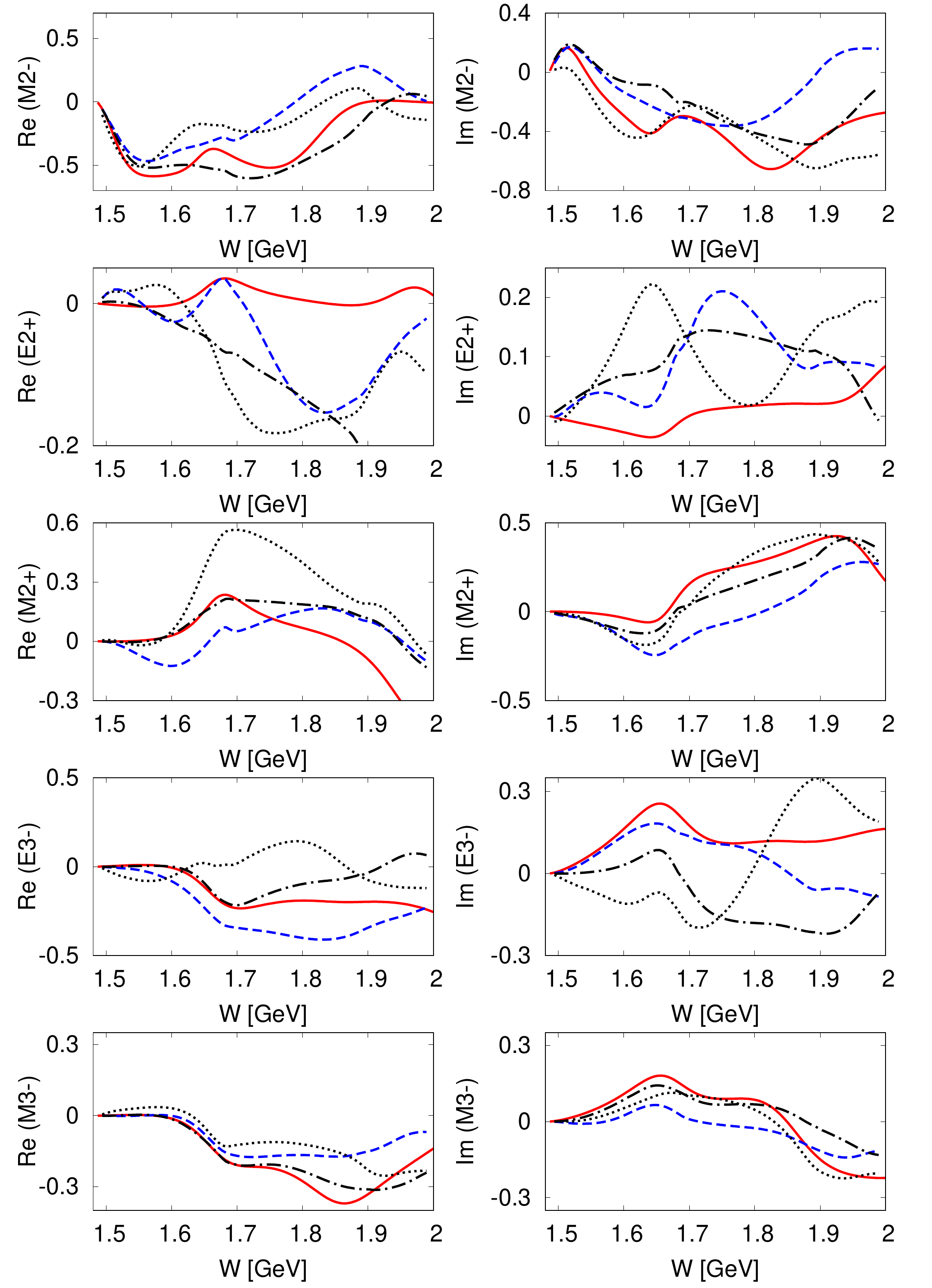}}
\caption{Comparison of $M_{2-}(D_{13})$, $E_{2+}(D_{15})$,
$M_{2+}(D_{15})$, $E_{3-}(F_{15})$, and $M_{3-}(F_{15})$ multipoles
from different PWA. Further details as in
Fig.~\ref{fig:mult1_rotated}} \label{fig:mult2_rotated}
\end{center}
\end{figure}

In Figs.~\ref{fig:mult1_rotated} and \ref{fig:mult2_rotated} we
compare the multipoles from rotated helicity amplitudes of EtaMAID,
BnGa, J\"uBo, and KSU. While the $S$ wave is practically identical
among all solutions, all other partial waves show deviations from
small up to huge. Moderate deviations we can see in $E_{1+}, M_{1+},
E_{2-}$, and $M_{3-}$, those we can already expect from different
fits to the measured data, as can be seen in
sect.~\ref{sect:Results}. Other partial waves as $M_{1-}$ and
especially $E_{2+}$ show very large deviations, which are most
likely due to the incompleteness of the database, where such
ambiguities must be expected.

A possible solution of this problem could be obtained along the
lines of Ref.~\cite{Osmanovic:2017fwe} by using constraints from
fixed-$t$ analyticity. But in addition also improvements of the
database with further observables and higher statistics would be
very helpful.


\section{Summary and Conclusions}\label{sec:conclusions}

Here we present a new update of EtaMAID for $\eta$ and $\eta^\prime$
photoproduction with four channels, $\eta p$, $\eta n$, $\eta^\prime
p$, $\eta^\prime n$. A large amount of data has been measured during
the last decade, mostly from A2MAMI, CBELSA and CLAS. Some of the
new polarization observables showed large discrepancies with our
previous solutions EtaMAID2001 and EtaMAID2003, and gave therefore a
lot of insight in further details of the partial wave analysis. In a
new approach, the high-energy regime $W>2.5$~GeV was first described
with a Regge approach, and the resonance regime from threshold up to
$W<2.5$~GeV with 21 $N^*$ resonances for $(\gamma,\eta)$ and 12
$N^*$ resonances for $(\gamma,\eta^\prime)$. All known $N^*$ states
listed by PDG have been investigated and, except for only 2 cases,
an improvement in our fit was found. Resonances found to be
insignificant for our analysis are $N(2040)\frac{3}{2}^+$ (a
one-star state only seen in $J/\Psi$ decays) and
$N(2220)\frac{9}{2}^+$ (a four-star high spin state mainly seen in
$\pi N$). In order to avoid or at least strongly reduce the double
counting from Regge plus resonances, we introduced damping factors
for Born and $t$-channel exchange contributions.

We obtained very good fits to almost all data, except for some
cases, where data from MAMI and CBELSA were in conflict and it did
not make sense to use both in the database for our fit. In these
cases we decided to use the MAMI data. From all $N^*$ resonances
that were significantly improving our fits, we found the largest
contributions in $(\gamma,\eta)$ from: $N(1535)\frac{1}{2}^-$,
$N(1650)\frac{1}{2}^-$, $N(1895)\frac{1}{2}^-$,
$N(1710)\frac{1}{2}^+$, $N(1720)\frac{3}{2}^+$,
$N(1900)\frac{3}{2}^+$, $N(1520)\frac{3}{2}^-$,
$N(1700)\frac{3}{2}^-$, and $N(1875)\frac{3}{2}^-$. For
$(\gamma,\eta^\prime)$ these are $N(1895)\frac{1}{2}^-$,
$N(1880)\frac{1}{2}^+$, $N(2100)\frac{1}{2}^+$,
$N(2000)\frac{5}{2}^+$, and $N(1990)\frac{7}{2}^+$. While
$N(1700)\frac{3}{2}^-$ and $N(1710)\frac{1}{2}^+$ are practically
neutron resonances, $N(1650)\frac{1}{2}^-$ and
$N(1880)\frac{1}{2}^+$ are much larger in the proton channel. Other
resonances contribute about equally in the proton and neutron
channels, see also photon couplings in table~\ref{tab:BWelectro} of
appendix~\ref{app:BG-BW}.

Generally, in a Breit-Wigner resonance analysis, the resonance
parameters are subject to model dependence. This could be rather
weak for prominent resonances with widths $\Gamma\lesssim 120$~MeV,
but for broad resonances with widths of several hundred MeV, the
model dependence can be very large.
In an upcoming work we plan to perform a
detailed resonance analysis with a search of $t$-matrix poles and
residues. In an application of the L+P method, successfully applied
in pion elastic scattering and pion photoproduction, we can expect
to reduce the model dependence for the resonance properties
considerably.

The new solution EtaMAID2018 is online available on the MAID web
pages~\cite{MAID}.


\begin{acknowledgments}
We want to thank Deborah R\"onchen of the J\"ulich-Bonn, Victor
Nikonov of the Bonn-Gatchina, and Mark Manley of the
Kent-State-University collaborations for providing us with their
most recent partial wave results and Andrey Sarantsev for many
helpful discussions. This work was supported by the Deutsche
Forschungsgemeinschaft (SFB 1044).
\end{acknowledgments}

\clearpage
\begin{appendix}

\section{Observables expressed in CGLN amplitudes}\label{app:obs}

Here we give the differential cross section, the three single-spin
asymmetries and the eight beam-target and beam-recoil
double-polarization observables expressed in CGLN amplitudes. In
addition we give the combination $\check{P}+\check{H}$, where most
of the terms cancel. A full list of all polarization observables
including also target-recoil polarization expressed in CGLN and in
helicity amplitudes can be found in Ref.~\cite{Osmanovic:2017fwe}.
In the literature, the sign definitions of double-polarization
observables is not unique. For an overview of the conventions see
Ref.~\cite{Sandorfi:2011nv}. Here we follow the conventions by
Barker~\cite{Barker:75}, SAID~\cite{Arndt:2002} and
MAID~\cite{MAID07}.

\begin{eqnarray}
\begin{split}
\sigma_{0} =& \,\mbox{Re}\,\left\{ \fpf{1}{1} + \fpf{2}{2} + \sin^{2}\theta\,(\fpf{3}{3}/2 + \fpf{4}{4}/2 \right. \\ 
            & \mbox{} \left. + \fpf{2}{3} + \fpf{1}{4} 
     + \cos\theta\,\fpf{3}{4}) - 2\cos\theta\,\fpf{1}{2} \right\} \rho \nonumber\\
\check{\Sigma} =& -\sin^{2}\theta\;\mbox{Re}\,\left\{\left(\fpf{3}{3} +\fpf{4}{4}\right)/2 +\fpf{2}{3} +\fpf{1}{4} \right.\\ 
               &\mbox{} \left.+ \cos\theta\,\fpf{3}{4}\right\}\rho \\
\check{T}      =& \sin\theta\;\mbox{Im}\,\left\{\fpf{1}{3}-\fpf{2}{4}+\cos\theta\,(\fpf{1}{4}-\fpf{2}{3}) \right. \\
               &\mbox{} \left. - \sin^{2}\theta\,\fpf{3}{4}\right\}\rho \\
\check{P}      =&  -\sin\theta\;\mbox{Im}\,\left\{ 2\fpf{1}{2} + \fpf{1}{3} - \fpf{2}{4} \right. \\ 
               &\mbox{} \left. + \cos\theta\,(\fpf{1}{4} -\fpf{2}{3}) - \sin^{2}\theta\,\fpf{3}{4}\right\}\rho \\
\check{E}      =& \,\mbox{Re}\,\left\{ \fpf{1}{1} + \fpf{2}{2} - 2\cos\theta\,\fpf{1}{2} \right. \\
               &\mbox{} \left. + \sin^{2}\theta\,(\fpf{2}{3} + \fpf{1}{4}) \right\}\rho \\
\check{F}      =& \sin\theta\;\mbox{Re}\,\left\{\fpf{1}{3} - \fpf{2}{4} - \cos\theta\,(\fpf{2}{3} - \fpf{1}{4})\right\}\rho \\
\check{G}      =& \sin^{2}\theta\;\mbox{Im}\,\left\{\fpf{2}{3} + \fpf{1}{4}\right\}\rho \\
\check{H}      =& \sin\theta\;\mbox{Im}\,\left\{2\fpf{1}{2} + \fpf{1}{3} - \fpf{2}{4} \right. \\
               &\mbox{} \left. + \cos\theta\,(\fpf{1}{4} - \fpf{2}{3})\right\}\rho \\
\check{P}+\check{H} &= \sin^3\theta\;\mbox{Im}\,\left\{ \fpf{3}{4}\right\}\rho \\
\check{C}_{x'} =& \sin\theta\;\mbox{Re}\,\left\{\fpf{1}{1} -\fpf{2}{2} - \fpf{2}{3} + \fpf{1}{4} \right. \\
               &\mbox{} \left. - \cos\theta\,(\fpf{2}{4} - \fpf{1}{3})\right\}\rho \\
\check{C}_{z'} =&  \,\mbox{Re}\,\left\{2\fpf{1}{2} - \cos\theta\,(\fpf{1}{1} + \fpf{2}{2}) \right. \\
               &\mbox{} \left. + \sin^{2}\theta\,(\fpf{1}{3} + \fpf{2}{4})\right\}\rho \\
\check{O}_{x'} =& \sin\theta\;\mbox{Im}\,\left\{\fpf{2}{3} - \fpf{1}{4} + \cos\theta\,(\fpf{2}{4} - \fpf{1}{3})\right\}\rho \\
\check{O}_{z'} =&  - \sin^{2}\theta\;\mbox{Im}\,\left\{\fpf{1}{3} + \fpf{2}{4}\right\}\rho \,.
\end{split}
\end{eqnarray}
with $\check{\Sigma}={\Sigma}\,\sigma_0$ and $\rho=q/k$.

\section{Expansion of CGLN amplitudes in terms of invariant
amplitudes}\label{app:FtoA}

The CGLN amplitudes are obtained from the invariant amplitudes $A_i$
by the following equations \cite{Den61}:
\begin{eqnarray}\label{eq:CGLN1}
\begin{split}
{F}_1 =& \frac{W-M_N}{8\pi\,W}\,\sqrt{(E_i+M_N)(E_f+M_N)}\big[ A_1  \\
       &+(W-M_N)\,A_4 - \frac{2M_N\nu_B}{W-M_N}\,(A_3-A_4)\big]\,,\nonumber \\
{F}_2 =& \frac{W+M_N}{8\pi\,W}\,|{\bold q}|\,\sqrt{\frac{E_i-M_N}{E_f+M_N}}\big[-A_1 + (W+M_N)\,A_4\\ 
       &- \frac{2M_N\nu_B}{W+M_N}\,(A_3-A_4)\big]\,, \nonumber \\
{F}_3 =& \frac{W+M_N}{8\pi\,W}\,|{\bold q}|\,\sqrt{(E_i-M_N)(E_f+M_N)}\big[(W-M_N)\,A_2 \\
       &+ A_3-A_4\big]\,,  \nonumber \\
{F}_4 =& \frac{W-M_N}{8\pi\,W}\,{\bold q}^2\,\sqrt{\frac{E_i+M_N}{E_f+M_N}} \big[-(W+M_N)\,A_2 \\
       &+A_3 - A_4\big]\,,
\end{split}   
\end{eqnarray}
with $\nu_B=(t-m_{\eta}^2)/(4m_N)$.

\section{Background and Breit-Wigner resonance parameters}\label{app:BG-BW}
In this appendix we list all parameters used in our isobar model. In
table~\ref{tab:BWhadronic} we give the hadronic parameters for 21
$N^*$ resonances used in EtaMAID2018. For all of them we found
couplings to the $\eta N$ channel, and for 12 of them also to the
$\eta^\prime N$ channel. Table~\ref{tab:BWelectro} gives all photon
couplings for proton and neutron targets and the newly introduced
unitarization phases for all four channels. Finally,
table~\ref{tab:background} gives all background parameters for Born
terms and Regge amplitudes.
\begin{table*}[ht]
\caption{Hadronic Breit-Wigner parameters for nucleon resonances.
Masses $M_R$ and widths $\Gamma_R$ are given in MeV and the
branching ratios $\beta$ in $\%$. The coupling constants $g$ are
dimensionless. The damping parameters of the hadronic vertex
functions are fixed at $X=450$~MeV. For channel openings below
threshold, conventional branching ratios are not defined and are
marked with $-$. Further non-zero couplings are also found for
$N(1440)\frac{1}{2}^+$ with $g_{\eta N}=1.0$, for
$N(1650)\frac{1}{2}^-$ with $g_{K \Sigma}=1.21$ and for
$N(1710)\frac{1}{2}^+$ with $g_{\omega N}=0.907$.
\\}\label{tab:BWhadronic}
\begin{tabular}{|c|ccc|ccccccccc|c|}
\hline
 $N(\cdots)J^\pi$ & $\ell$ & $\zeta_{\eta N}$ &$\zeta_{\eta^\prime N}$ & $M_R$ & $\Gamma_R$ & $\beta_{\pi N}$ & $\beta_{\pi\pi N}$ & $\beta_{\eta N}$ & $\beta_{K
 \Lambda}$ & $\beta_{K \Sigma}$ & $\beta_{\omega N}$ & $\beta_{\eta^\prime N}$ & $g_{\eta^\prime N}$\\
\hline
$N(1440)\frac{1}{2}^+$ & 1 & $+1$ &      & 1430.0 & 350.0 & $65.0$ & $35.0$ & $-$    & $ -  $ & $ -  $ & $ -  $ & $ -  $ & $  0  $ \\
$N(1520)\frac{3}{2}^-$ & 2 & $+1$ &      & 1520.0 & 100.0 & $61.0$ & $38.9$ & $0.08$ & $ -  $ & $ -  $ & $ -  $ & $ -  $ & $  0  $ \\
$N(1535)\frac{1}{2}^-$ & 0 & $+1$ &      & 1521.7 & 174.7 & $52.0$ & $13.6$ & $34.7$ & $ -  $ & $ -  $ & $ -  $ & $ -  $ & $  0  $ \\
$N(1650)\frac{1}{2}^-$ & 0 & $-1$ &      & 1626.3 & 132.5 & $51.0$ & $27.2$ & $18.8$ & $ 3.0$ & $ -  $ & $ -  $ & $ -  $ & $  0  $ \\
$N(1675)\frac{5}{2}^-$ & 2 & $-1$ &      & 1680.0 & 100.0 & $41.0$ & $57.1$ & $0.94$ & $ 1.0$ & $ -  $ & $ -  $ & $ -  $ & $  0  $ \\
$N(1680)\frac{5}{2}^+$ & 3 & $+1$ &      & 1690.0 & 145.3 & $62.0$ & $37.8$ & $0.16$ & $ 0  $ & $ -  $ & $ -  $ & $ -  $ & $  0  $ \\
$N(1700)\frac{3}{2}^-$ & 2 & $+1$ &      & 1659.6 &  83.9 & $15.0$ & $80.8$ & $1.16$ & $ 3.0$ & $ 0  $ & $ -  $ & $ -  $ & $  0  $ \\
$N(1710)\frac{1}{2}^+$ & 1 & $+1$ &      & 1669.5 &  63.2 & $ 5.0$ & $68.2$ & $11.9$ & $15.0$ & $ 0  $ & $ -  $ & $ -  $ & $  0  $ \\
$N(1720)\frac{3}{2}^+$ & 1 & $+1$ &      & 1750.0 & 395.5 & $11.0$ & $79.7$ & $1.28$ & $ 8.0$ & $ 0  $ & $ -  $ & $ -  $ & $  0  $ \\
$N(1860)\frac{5}{2}^+$ & 3 & $-1$ & $+1$ & 1885.8 & 197.4 & $20.0$ & $76.5$ & $3.55$ & $ 0  $ & $ 0  $ & $ 0  $ & $ -  $ & $0.700$ \\
$N(1875)\frac{3}{2}^-$ & 2 & $+1$ & $-1$ & 1893.9 & 320.0 & $ 4.0$ & $46.0$ & $11.0$ & $ 4.0$ & $15.0$ & $20.0$ & $ -  $ & $0.168$ \\
$N(1880)\frac{1}{2}^+$ & 1 & $+1$ & $-1$ & 1882.1 &  90.0 & $ 6.0$ & $74.6$ & $0.44$ & $ 2.0$ & $17.0$ & $ 0  $ & $ -  $ & $0.400$ \\
$N(1895)\frac{1}{2}^-$ & 0 & $+1$ & $+1$ & 1894.4 &  70.7 & $ 2.5$ & $63.2$ & $3.27$ & $18.0$ & $13.0$ & $ 0  $ & $ -  $ & $0.405$ \\
$N(1900)\frac{3}{2}^+$ & 1 & $-1$ & $-1$ & 1898.7 & 450.0 & $ 3.0$ & $63.9$ & $3.06$ & $12.0$ & $ 5.0$ & $13.0$ & $0.03$ & $0.563$ \\
$N(1990)\frac{7}{2}^+$ & 3 & $+1$ & $+1$ & 2227.0 & 389.0 & $ 2.0$ & $89.9$ & $3.61$ & $ 0  $ & $ 0  $ & $ 0  $ & $ 4.5$ & $0.347$ \\
$N(2000)\frac{5}{2}^+$ & 3 & $-1$ & $+1$ & 2116.8 & 246.9 & $ 8.0$ & $87.3$ & $2.30$ & $ 0  $ & $ 0  $ & $ 0  $ & $ 2.4$ & $0.300$ \\
$N(2060)\frac{5}{2}^-$ & 2 & $+1$ & $-1$ & 1984.5 & 159.8 & $11.0$ & $84.1$ & $1.58$ & $ 0  $ & $ 3.0$ & $ 0  $ & $ 0.3$ & $0.130$ \\
$N(2100)\frac{1}{2}^+$ & 1 & $+1$ & $+1$ & 2010.0 & 260.0 & $16.0$ & $78.2$ & $1.69$ & $ 0  $ & $ 0  $ & $ 0  $ & $ 4.1$ & $0.300$ \\
$N(2120)\frac{3}{2}^-$ & 2 & $+1$ & $-1$ & 2061.3 & 101.9 & $ 5.0$ & $94.9$ & $0.05$ & $ 0  $ & $ 0  $ & $ 0  $ & $0.03$ & $0.021$ \\
$N(2190)\frac{7}{2}^-$ & 4 & $-1$ & $+1$ & 2250.0 & 591.2 & $16.0$ & $78.8$ & $4.54$ & $ 0.5$ & $ 0  $ & $ 0  $ & $0.18$ & $0.100$ \\
$N(2250)\frac{9}{2}^-$ & 4 & $+1$ & $-1$ & 2250.0 & 733.2 & $12.0$ & $84.4$ & $3.50$ & $ 0  $ & $ 0  $ & $ 0  $ & $0.10$ & $0.085$ \\
 \hline
\end{tabular}
\end{table*}

\begin{table*}[ht]
\caption{Electromagnetic Breit-Wigner parameters for nucleon
resonances. Photon couplings $_{N}\!A_{\lambda}$ are given in
$10^{-3}/\sqrt{\mbox{GeV}}$. Unitary phases $\phi$ are given in
degrees. The damping parameters of the electromagnetic vertex
functions are fixed at $X_\gamma=0$.
\\}\label{tab:BWelectro}
\begin{tabular}{|c|cccc|cccc|}
\hline
 $N(\cdots)J^\pi$ & $_pA_{1/2}$& $_pA_{3/2}$& $_nA_{1/2}$& $_nA_{3/2}$ & $\phi_{\eta p}$ & $\phi_{\eta n}$ & $\phi_{\eta^\prime p}$ & $\phi_{\eta^\prime n}$\\
\hline
$N(1440)\frac{1}{2}^+$ & $-60.0$ &   $0$   & $40.0$   & $0$     & $-0.4$ & $-89.0$ & $0$ & $0$  \\
$N(1520)\frac{3}{2}^-$ & $-39.7$ & $116.8$ & $-160.0$ & $-94.0$ & $55.3$ & $73.5$ & $0$ & $0$  \\
$N(1535)\frac{1}{2}^-$ & $115.0$ & $0$     & $-101.9$ & $0$     & $29.0$ & $28.2$ & $0$ & $0$  \\
$N(1650)\frac{1}{2}^-$ & $ 55.0$ & $  0  $ & $-25.4$ & $  0  $ & $  6.0$ & $ 15.5$ & $  0  $ & $  0  $   \\
$N(1675)\frac{5}{2}^-$ & $ 23.7$ & $ 20.0$ & $ -9.8$ & $ 43.2$ & $ 78.4$ & $ 59.1$ & $  0  $ & $  0  $   \\
$N(1680)\frac{5}{2}^+$ & $-29.4$ & $133.0$ & $129.7$ & $ 10.0$ & $ 64.6$ & $ 89.0$ & $  0  $ & $  0  $   \\
$N(1700)\frac{3}{2}^-$ & $ 15.2$ & $-14.0$ & $ 93.4$ & $-32.1$ & $ 60.9$ & $ 57.7$ & $  0  $ & $  0  $   \\
$N(1710)\frac{1}{2}^+$ & $  5.5$ & $  0  $ & $-42.2$ & $  0  $ & $-47.1$ & $-79.4$ & $  0  $ & $  0  $   \\
$N(1720)\frac{3}{2}^+$ & $100.0$ & $  7.7$ & $-64.9$ & $ 63.9$ & $ 87.8$ & $ 56.3$ & $  0  $ & $  0  $   \\
$N(1860)\frac{5}{2}^+$ & $-30.7$ & $ 29.0$ & $-24.5$ & $ 33.7$ & $-83.0$ & $-89.0$ & $-39.6$ & $-61.3$   \\
$N(1875)\frac{3}{2}^-$ & $ 18.0$ & $-35.4$ & $-32.0$ & $ 50.4$ & $ 34.6$ & $ 30.3$ & $-20.8$ & $ 86.2$   \\
$N(1880)\frac{1}{2}^+$ & $ 60.4$ & $  0  $ & $ -6.6$ & $  0  $ & $ 84.9$ & $ 89.0$ & $ 89.0$ & $ 60.7$   \\
$N(1895)\frac{1}{2}^-$ & $-32.0$ & $  0  $ & $ 42.9$ & $  0  $ & $ 51.5$ & $ 58.9$ & $ 57.8$ & $ 41.0$   \\
$N(1900)\frac{3}{2}^+$ & $-50.2$ & $-67.0$ & $-42.5$ & $ 17.9$ & $ 47.6$ & $ 89.0$ & $ 43.4$ & $ 89.0$   \\
$N(1990)\frac{7}{2}^+$ & $-12.4$ & $ 57.0$ & $-43.3$ & $-28.1$ & $  6.3$ & $  3.7$ & $ 11.8$ & $ -7.9$   \\
$N(2000)\frac{5}{2}^+$ & $-73.1$ & $-12.9$ & $ 12.8$ & $-59.2$ & $ 89.0$ & $ 51.5$ & $ 89.0$ & $ 50.8$   \\
$N(2060)\frac{5}{2}^-$ & $ 21.3$ & $ 62.0$ & $ 43.0$ & $  6.1$ & $ 70.6$ & $ 67.3$ & $ 89.0$ & $ 89.0$   \\
$N(2100)\frac{1}{2}^+$ & $ 63.9$ & $  0  $ & $-82.7$ & $  0  $ & $ 89.0$ & $ 14.5$ & $ 58.1$ & $ 36.3$   \\
$N(2120)\frac{3}{2}^-$ & $113.5$ & $160.0$ & $160.0$ & $100.0$ & $-26.2$ & $-89.0$ & $ 56.6$ & $ 24.3$   \\
$N(2190)\frac{7}{2}^-$ & $ 26.7$ & $ 60.0$ & $ 34.5$ & $ 18.7$ & $-89.0$ & $-89.0$ & $ 59.2$ & $  7.5$   \\
$N(2250)\frac{9}{2}^-$ & $-31.2$ & $-20.0$ & $ 24.1$ & $ 12.5$ & $ 82.8$ & $ 89.0$ & $ 89.0$ & $ 88.2$   \\
 \hline
\end{tabular}
\end{table*}

\begin{table*}[ht]
\caption{Background parameters for Born terms and Regge exchanges.
The Regge damping parameters $\Lambda_{R}$ for $\eta$ and
$\eta^\prime$ photoproduction are given in units of GeV, the
Regge-cut parameters $d_c$ in GeV$^{-2}$, all other parameters are
dimensionless. The Regge-cut parameters are the same for $\eta$ and
$\eta^\prime$ photoproduction.
\\}\label{tab:background}
\begin{tabular}{|c|c||c|c|}
\hline
$g_{\eta NN}^2/4\pi$ & 0.063 & $g_{\eta^\prime NN}^2/4\pi$ & 0.060 \\
$\alpha_{B,\eta}$    & 4.51  & $\alpha_{B,\eta^\prime}$    & 3.95  \\
$\Lambda_{R,\eta}$& 0.974 & $\Lambda_{R,\eta^\prime}$& 0.440 \\
$\lambda_{\eta \gamma}^{\rho}$ & 0.910 & $\lambda_{\eta^\prime \gamma}^{\rho}$ &  1.049  \\
$\lambda_{\eta \gamma}^{\omega}$ & 0.246 & $\lambda_{\eta^\prime \gamma}^{\omega}$ & 0.363   \\
$\lambda_{\eta \gamma}^{b_1}$ & 0.1 & $\lambda_{\eta^\prime \gamma}^{b_1}$ &  1  \\
$g_{\rho}^{v}$  &  2.71  & $g_{\rho}^{t}$  &  4.20 \\
$g_{\omega}^{v}$  & 14.2  & $g_{\omega}^{t}$  &  0 \\
$g_{h_1}/g_{b_1}$  & 0.667  & $g_{b_1}^{t}$  &  $-7.0$ \\
$c_{\rho \mathbb P}$  &  4.64 & $c_{\omega \mathbb P}$ & $-5.00$ \\
$c_{\rho f_2}$        &  3.10 & $c_{\omega f_2}$       & $1.11$ \\
${\tilde c}_{\rho\mathbb P}$  &  0 & ${\tilde c}_{\omega\mathbb P}$  &  0 \\
${\tilde c}_{\rho f_2}$  &  0.245 & ${\tilde c}_{\omega f_2}$  &  $-0.122$ \\
$d_{c,\rho\mathbb P}$   & 12.1 & $d_{c,\rho f_2}$   & 12.1 \\
$d_{c,\omega\mathbb P}$ & 2.09 & $d_{c,\omega f_2}$ & 2.09 \\
 \hline
\end{tabular}
\end{table*}

\end{appendix}

\clearpage



\end{document}